\documentclass{cernrep}
\usepackage{graphicx,here,rotate,epsf,epsfig}
 
\begin{document}

\title{SUPERSYMMETRY FOR ALP HIKERS}
 
\author{John Ellis}
 
\institute{CERN, Geneva, Switzerland}
%-----------------------------------------------------------------------
% If your printer does not reproduce dimensions exactly, it may be
% necessary to remove the % signs and adjust the dimensions in the
% following commands:
%
%     \setlength{\textheight}{24cm}
%     \setlength{\textwidth}{16cm}
%
% Similarly for the following, if you need to adjust the positioning
% on the paper:
%
%         \setlength{\topmargin}{-1cm}
%         \setlength{\oddsidemargin}{0pt}
%         \setlength{\evensidemargin}{0pt}
%------------------------------------------------------------------------
 
\newcommand{\mycomm}[1]{\hfill\break{ \tt===$>$ \bf #1}\hfill\break}

\def\ga{\mathrel{\raise.3ex\hbox{$>$\kern-.75em\lower1ex\hbox{$\sim$}}}}
\def\la{\mathrel{\raise.3ex\hbox{$<$\kern-.75em\lower1ex\hbox{$\sim$}}}}
\def\gev{{\rm \, Ge\kern-0.125em V}}
\def\tev{{\rm \, Te\kern-0.125em V}}
\def\beq{\begin{equation}}
\def\eeq{\end{equation}}
\def\st{\scriptstyle}
\def\ss{\scriptscriptstyle}
\def\mb{m_{\widetilde B}}
\def\msf{m_{\tilde f}}
\def\mst{m_{\tilde t}}
\def\mf{m_{\ss{f}}}
\def\mpar{m_{\ss\|}^2}
\def\mpl{M_{\rm Pl}}
\def\mchi{m_{\chi}}
\def\ohsq{\Omega_{\chi} h^2}
\def\msn{m_{\tilde\nu}}
\def\m12{m_{1\!/2}}
\def\mstpl{m_{\tilde t_{\ss 1}}^2}
\def\mstpr{m_{\tilde t_{\ss 2}}^2}

\def\ga{\mathrel{\raise.3ex\hbox{$>$\kern-.75em\lower1ex\hbox{$\sim$}}}}
\def\la{\mathrel{\raise.3ex\hbox{$<$\kern-.75em\lower1ex\hbox{$\sim$}}}}
\def\gyr{{\rm \, G\kern-0.125em yr}}
\def\gev{{\rm \, Ge\kern-0.125em V}}
\def\tev{{\rm \, Te\kern-0.125em V}}
\def\beq{\begin{equation}}
\def\eeq{\end{equation}}
\def\ss{\scriptscriptstyle}
\def\scs{\scriptstyle}
\def\mb{m_{\widetilde B}}
\def\mst{m_{\tilde\tau_R}}
\def\mstop{m_{\tilde t_1}}
\def\msl{m_{\tilde{\ell}_1}}
\def\stau{\tilde \tau}
\def\stop{\tilde t}
\def\sbot{\tilde b}
\def\mchi{m_{\tilde \chi}}
\def\mxi{m_{\tilde{\chi}_i^0}}
\def\mxj{m_{\tilde{\chi}_j^0}}
\def\mchari{m_{\tilde{\chi}_i^+}}
\def\mcharj{m_{\tilde{\chi}_j^+}}
\def\mgluino{m_{\tilde g}}
\def\msf{m_{\tilde f}}
\def\m12{m_{1\!/2}}
\def\mtb{\overline{m}_{\ss t}}
\def\mbb{\overline{m}_{\ss b}}
\def\mfb{\overline{m}_{\ss f}}
\def\mf{m_{\ss{f}}}
\def\gt{\gamma_t}
\def\gb{\gamma_b}
\def\gf{\gamma_f}
\def\thm{\theta_\mu}
\def\tha{\theta_A}
\def\thb{\theta_B}
\def\mgl{m_{\ss \tilde g}}
\def\cp{C\!P}
\def\ch{{\widetilde \chi}} 
\def\st{{\widetilde \tau}_{\scriptscriptstyle\rm 1}}
\def\sm{{\widetilde \mu}_{\scriptscriptstyle\rm R}}
\def\sel{{\widetilde e}_{\scriptscriptstyle\rm R}}
\def\sl{{\widetilde \ell}_{\scriptscriptstyle\rm R}}
\def\msn{m_{\ch}}
\def\tsq{|{\cal T}|^2}
\def\tcm{\theta_{\rm\scriptscriptstyle CM}}
\def\half{{\textstyle{1\over2}}}
\def\neq{n_{\rm eq}}
\def\qeq{q_{\rm eq}}
\def\slash#1{\rlap{\hbox{$\mskip 1 mu /$}}#1}%
\def\mw{m_W}
\def\mz{m_Z}
\def\mhb{m_{H}}
\def\mhl{m_{h}}
\newcommand\f[1]{f_#1}
\def\nl{\hfill\nonumber\\&&}

\def\gappeq{\mathrel{\rlap {\raise.5ex\hbox{$>$}}
{\lower.5ex\hbox{$\sim$}}}}

\def\lappeq{\mathrel{\rlap{\raise.5ex\hbox{$<$}}
{\lower.5ex\hbox{$\sim$}}}}

\def\Toprel#1\over#2{\mathrel{\mathop{#2}\limits^{#1}}}
\def\FF{\Toprel{\hbox{$\scriptscriptstyle(-)$}}\over{$\nu$}}

\newcommand{\Zee}{$Z^0$}

%%%%%%%%%%%%%%%%%%%%%%%%%%  my definitions  %%%%%%%%%%%%%%%%%%%%%%%%%%%

\def\Yi{\eta^{\ast}_{11} \left( \frac{y_{i}}{2} g' Z_{\chi 1} + 
        g T_{3i} Z_{\chi 2} \right) + \eta^{\ast}_{12} 
        \frac{g m_{q_{i}} Z_{\chi 5-i}}{2 m_{W} B_{i}}}

\def\Xii{\eta^{\ast}_{11} 
        \frac{g m_{q_{i}}Z_{\chi 5-i}^{\ast}}{2 m_{W} B_{i}} - 
        \eta_{12}^{\ast} e_{i} g' Z_{\chi 1}^{\ast}}

\def\Wi{\eta_{21}^{\ast}
        \frac{g m_{q_{i}}Z_{\chi 5-i}^{\ast}}{2 m_{W} B_{i}} -
        \eta_{22}^{\ast} e_{i} g' Z_{\chi 1}^{\ast}}

\def\Vi{\eta_{22}^{\ast} \frac{g m_{q_{i}} Z_{\chi 5-i}}{2 m_{W} B_{i}}
        + \eta_{21}^{\ast}\left( \frac{y_{i}}{2} g' Z_{\chi 1}
        + g T_{3i} Z_{\chi 2} \right)}

\def\zthree{\delta_{1i} [g Z_{\chi 2} - g' Z_{\chi 1}]}

\def\zfour{\delta_{2i} [g Z_{\chi 2} - g' Z_{\chi 1}]}

%%%%%%%%%%%%%%%%%%%%%%%%%%%%%%%%%%%%%%%%%%%%%%%%%%%%%%%%%%%%%%%%%%%%%%%%
 
\maketitle % this produces the title block

\begin{flushright}
hep-ph/0203114 \\
CERN-TH/2002-052
\end{flushright}
 
\begin{abstract}
These lectures provide a phenomenological introduction to supersymmetry,
concentrating on the minimal supersymmetric extension of the Standard
Model (MSSM). In the first lecture, {\it motivations} are provided for
thinking that supersymmetry might appear at the TeV scale, including the
naturalness of the mass hierarchy, gauge unification and the probable mass
of the Higgs boson. In the second lecture, simple {\it globally
supersymmetric field theories} are introduced, with the emphasis on
features important for model-building. {\it Supersymmetry breaking} and
{\it local supersymmetry (supergravity)} are introduced in the third 
lecture,
and the structure of sparticle mass matrices and mixing are reviewed.
Finally, the available {\it experimental and cosmological constraints} on
MSSM parameters are discussed and combined in the fourth lecture, and the
{\it prospects} for discovering supersymmetry in future experiments are
previewed.
\end{abstract}

\section{GETTING MOTIVATED}

\subsection{Defects of the Standard Model}

The Standard Model agrees with all confirmed experimental data from
accelerators, but is theoretically very unsatisfactory. It does not
explain the particle quantum numbers, such as the electric charge $Q$,
weak isospin $I$, hypercharge $Y$ and colour, and contains at least 19
arbitrary parameters. These include three independent gauge couplings and
a possible CP-violating strong-interaction parameter, six quark and three
charged-lepton masses, three generalized Cabibbo weak mixing angles and
the CP-violating Kobayashi-Maskawa phase, as well as two independent
masses for weak bosons.

As if 19 parameters were insufficient to appall you, at least nine more
parameters must be introduced to accommodate neutrino oscillations: three
neutrino masses, three real mixing angles, and three CP-violating phases,
of which one is in principle observable in neutrino-oscillation
experiments and the other two in neutrinoless double-beta decay
experiments. Even more parameters would be needed to generate neutrino
masses in a credible way, associated with a heavy-neutrino sector and/or
additional Higgs particles.

Eventually, one would like to include gravity in a unified theory along
with the other particle interactions, which involves introducing at least
two more parameters, Newton's constant $G_N = 1 / m_P^2: m_P \sim
10^{19}$~GeV that characterizes the strength of gravitational
interactions, and the cosmological constant $\Lambda$ or some time-varying
form of vacuum energy as seems to be required by recent cosmological data.
A complete theory of cosmology will presumably also need parameters to
characterize the early inflation of the Universe and to generate its
baryon asymmetry, which cannot be explained within the Standard Model.

The Big Issues in physics beyond the Standard Model are conveniently
grouped into three categories~\cite{StAnd}. These include the problem of
{\bf Unification}: is there a simple group framework for unifying all the
particle interactions, a so-called Grand Unified Theory (GUT), {\bf
Flavour}: why are there so many different types of quarks and leptons and
why do their weak interactions mix in the peculiar way observed, and {\bf
Mass}: what is the origin of particle masses, are they due to a Higgs
boson, why are the masses so small? Solutions to all these problems should
eventually be incorporated in a Theory of Everything (TOE) that also
includes gravity, reconciles it with quantum mechanics, explains the
origin of space-time and why it has four dimensions, etc. String theory,
perhaps in its current incarnation of M theory, is the best (only?)
candidate we have for such a TOE~\cite{TOE}, but we do not yet understand
it well enough to make clear experimental predictions.

Supersymmetry is thought to play a r\^ole in solving many of these
problems beyond the Standard Model. As discussed later, GUT predictions
for the unification of gauge couplings work best if the effects of
relatively light supersymmetric particles are included~\cite{GUTs}. Also,
the hierarchy of mass scales in physics, and particularly the fact that
$m_W \ll m_P$, appears to require relatively light supersymmetric
particles: $M \lappeq 1$~TeV for its stabilization~\cite{hierarchy}.
Finally, supersymmetry seems to be essential for the consistency of string
theory~\cite{GSW}, although this argument does not really restrict the
mass scale at which supersymmetric particles should appear.

Thus there are plenty of good reasons to study supersymmetry, and we
return later to examine in more detail the motivations provided by
unification and the mass hierarchy problem.

\subsection{The Electroweak Vacuum}

Generating particle masses within the Standard Model requires breaking its
gauge symmetry, and the only consistent way to do this is by breaking the
symmetry of the electroweak vacuum:
\begin{equation}
m_{W,Z} \; \ne \; 0 \; \; \leftrightarrow \; \; < 0 | X_{I, {I_3}} | 0 > 
\; \ne 
\; 0
\label{break}
\end{equation}
where the symbols $I, I_3$ denote the weak isospin quantum numbers of
whatever object $X$ has a non-zero vacuum expectation value. There are a
couple of good reasons to think that $X$ must have (predominantly) isospin
$I = 1/2$. One is the ratio of the $W$ and $Z$ boson masses~\cite{RV}:
\begin{equation}
\rho \; \equiv \; {m_W^2 \over m_Z^2 \cos^2 \theta_W} \; \simeq \; 1,
\label{rho}
\end{equation}
and the other reason is to provide non-zero fermion masses. Since 
left-handed fermions $f_L$ have $I = 1/2$, right-handed fermions $f_R$ 
have 
$I = 0$ and fermion mass terms couple them together: $m_f {\bar f_L} f_R$, 
we must break isospin symmetry by 1/2 a unit:
\begin{equation}
m_f \; \ne \; 0 \; \; \leftrightarrow \; \; < 0 | X_{1/2, \pm 1/2} | 0 > 
\; \ne \; 0.
\label{fermionmass}
\end{equation}

The next question is, what is the nature of $X$? Is it elementary or composite?
In the initial formulation of the Standard Model, it was assumed that 
$X$ should be an elementary Higgs-Brout-Englert~\cite{H,BE} field $H$: $< 
0 | H^0 | 0 > 
\ne 0$, which would have a physical excitation that manifested itself as a 
neutral scalar Higgs boson~\cite{H}. However, as discussed in more detail 
later, 
an elementary Higgs field has problems with quantum (loop) corrections.
Those due to Standard Model particles
are quadratically divergent, resulting in a large cutoff-dependent 
contribution to the physical masses of the Higgs boson, $W,Z$ bosons and 
other particles:
\begin{equation}
\delta m_H^2 \; \simeq \; {\cal O}({\alpha \over \pi}) \Lambda^2,
\label{quadratic}
\end{equation}
where $\Lambda$ represents the scale at which new physics appears. 

The sensitivity (\ref{quadratic}) disturbs theorists, and one of the
suggestions to avoid it was to postulate replacing an elementary
Higgs-Brout-Englert field $H$ by a composite field such as a condensate of
fermions:  $< 0 | {\bar F} F | 0 > \ne 0$. This possibility was made more
appealing by the fact that fermion condensates are well known in
solid-state physics, where Cooper pairs of electrons are responsible for
conventional superconductivity, and in strong-interaction physics, where
quarks condense in the vacuum: $< 0 | {\bar q} q | 0 > \ne 0$.

In order to break the electroweak symmetry at a large enough scale,
fermions with new interactions that become strong at a higher mass scale
would be required. One suggestion was that the Yukawa interactions of the
top quark might be strong enough to condense them: $< 0 | {\bar t} t | 0 >
\ne 0$~\cite{topcolour}, but this would have required the top quark to 
weigh more than
200~GeV, in simple models. Alternatively, theorists proposed the existence
of new fermions held together by completely new interactions that became
strong at a scale $ \sim 1$~TeV, commonly called {\it Technicolour}
models~\cite{TC}.

Specifically, the technicolour concept was to clone the QCD
quark-antiquark condensate
\begin{equation}
< 0 | {\bar q_L} q_R | 0 > \sim
\Lambda_{QCD}^3: \; \; \Lambda_{QCD} \sim 1 {\rm GeV}, 
\label{QCDqqbar}
\end{equation}
on a much larger
scale, postulating a condensate of new massive fermions $< 0 | {\bar Q_L}
Q_R | 0 > \sim \Lambda_{TC}^3$ where $\Lambda_{TC} \sim 1$~TeV. Assigning
the techniquarks to the same weak representations as conventional quarks,
$I_L = 1 / 2, I_R = 0$, the technicondensate breaks electroweak symmetry
in just the way required to reproduce the relation (\ref{rho}). Just as
QCD with two massless flavours would have three massless pions and a
massive scalar meson, this simplest version of technicolour would predict
three massless technipions that are eaten by the $W^\pm$ and $Z^0$ to
provide their masses via the Higgs-Brout-Englert mechanism, leaving over a
single physical scalar state weighing about 1~TeV, that would behave in
some ways like a heavy Higgs boson.

Unfortunately, this simple technicolour picture must be complicated, in 
order to cancel triangle anomalies and to give masses to 
fermions~\cite{ETC}, so the 
minimal default option has become a model with a single technigeneration:
\begin{eqnarray}
\pmatrix{
\nu \cr
\ell }
\pmatrix{
u \cr
d}_{1,2,3}
\longrightarrow
\pmatrix{
N \cr
L}_{1,\ldots,N_{TC}}
\pmatrix{ U \cr
D}_{1,\ldots,N_{TC};1,2,3}
\label{TCgeneration}
\end{eqnarray}
One may then study models with different numbers $N_{TC}$ of technicolours, 
and also different numbers of techniflavours $N_{TF}$ if one wishes. The 
single-technigeneration model (\ref{TCgeneration}) above has $N_{TF} = 
4$, corresponding to the technilepton doublet $(N, L)$ and the three 
coloured techniquark doublets $(U, D)_{1,2,3}$.

The absence of any light technipions is already a problem for this
scenario~\cite{Tpi}, as is the observed suppression of flavour-changing
neutral interactions~\cite{DE}. Another constraint is provided by
precision electroweak data, which limit the possible magnitudes of
one-loop radiative corrections due to virtual techniparticles. These may
conveniently be parameterized in terms of three combinations of vacuum
polarizations: for example~\cite{STU}
\begin{equation}
T \equiv {\epsilon_1 \over \alpha} \equiv {\Delta \rho \over \alpha},
\label{Trho}
\end{equation}
where
\begin{equation}
\Delta \rho = {\Pi_{ZZ} (0) \over m_Z^2} - {\Pi_{WW} (0) \over m_W^2} - 2 
\tan \theta_W {\Pi_{\gamma Z} (0) \over m_Z^2},
\label{Delrho}
\end{equation}
leading to the following approximate expression:
\begin{equation}
T = {3 \over 16 \pi \sin^2 \theta_W \cos^2 \theta_W} ({m_t^2 \over m_Z^2}) 
-
{3 \over 16 \pi \cos^2 \theta_W} {\rm ln} ({m_H^2 \over m_Z^2}) + \dots 
\label{TSM}
\end{equation}
There are analogous expressions for the other two combinations of vacuum 
polarizations:
\begin{eqnarray}
S & \equiv & {4 \sin^2 \theta_W \over \alpha} \epsilon_3 =  
{1 \over 12 \pi} {\rm ln} ({m_H^2 \over m_Z^2}) + \dots \\
U & \equiv & - {4 \sin^2 \theta_W \over \alpha} \epsilon_2  
\label{TU}
\end{eqnarray}
The electroweak data may then be used to constrain $\epsilon_{1,2,3}$ 
(or, equivalently, $S,T,U$), and thereby extensions of the Standard Model 
with the same $SU(2) \times U(1)$ gauge group and extra matter particles 
that do not have important other interactions with ordinary matter.
This approach does not include vertex corrections, so the most important 
one, that for $Z^0 \to {\bar b} b$, is treated by introducing 
another parameter $\epsilon_b$.

\begin{figure}%[t]
\begin{center}
\includegraphics[height=3in]{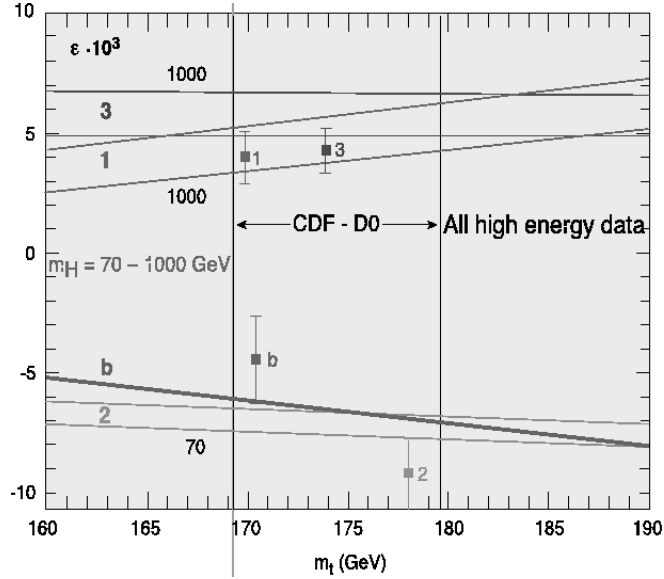}
\caption[]{The ranges of the vacuum-polarization parameters 
$\epsilon_{1,2,3,b}$ allowed by the precision electroweak 
data~\cite{epsilons} are compared with the predictions of the 
Standard Model, as functions of the masses of the $t$ quark and 
the Higgs boson.}
\label{fig:epsilons}
\end{center}
\end{figure}

\begin{figure}%[t]
\begin{center}
\includegraphics[height=3in]{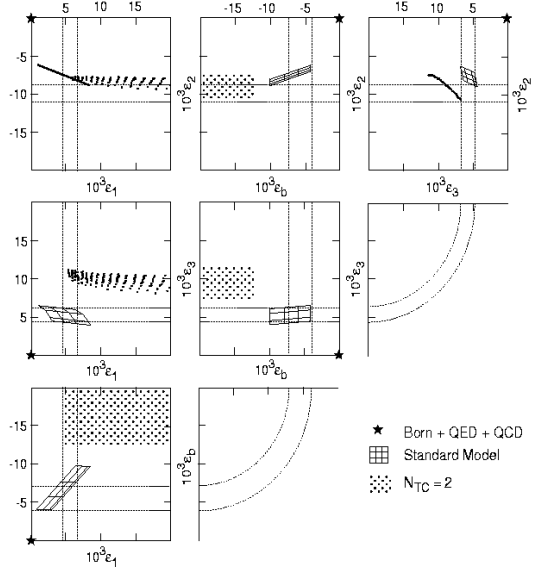}
\caption[]{Two-dimensional projections comparing the 
allowed ranges of the 
$\epsilon_i$ shown in Fig.~\ref{fig:epsilons} with the predictions of the 
Standard Model (hatched regions) and a minimal one-generation technicolor 
model (chicken-pox regions)~\cite{EFL}.}
\label{fig:TC}
\end{center}
\end{figure}

This simple parameterization is perfectly sufficient to provide big
headaches for the simple technicolour models described above.
Fig.~\ref{fig:epsilons} compares the values of the parameters $\epsilon_i$
extracted from the final LEP data with the values calculated in the
Standard Model for $m_t$ within the range measured by CDF and D0, and for
$113~{\rm GeV} < m_H < 1$~TeV. We see that the agreement is quite good,
which does not leave much room for new physics beyond the Standard Model
to contribute to the $\epsilon_i$.  Fig.~\ref{fig:TC} compares these
measured values also with the predictions of the simplest one-generation
technicolour model, with $N_{TC} = 2$ and other assumptions described
in~\cite{EFL}.

We see that the data seem to disagree quite strongly with these
technicolour predictions. Does this mean that technicolour is dead? Not
quite~\cite{Lane}, but it has motivated technicolour enthusiasts to pursue
epicyclic variations on the original idea, such as Œwalking
technicolour~\cite{walk}, in which the technicolour dynamics is not scaled
up from QCD in such a naive way.

\subsection{It Quacks like Supersymmetry}

Electroweak radiative corrections may be bad news for technicolour models,
but they do seem to provide hints for supersymmetry, as we now discuss.

\begin{figure}%[t]
%\centerline{
\begin{center}
\includegraphics[height=3in]{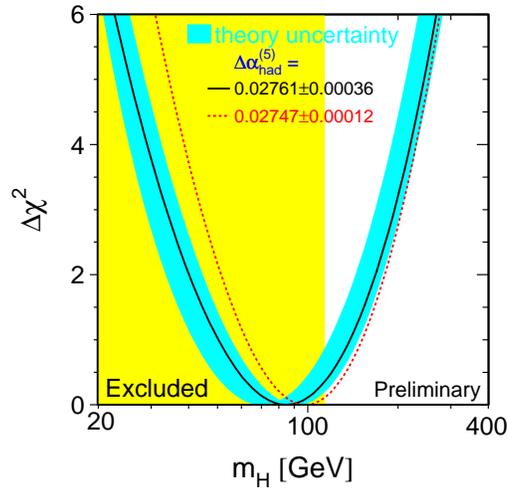}
\end{center}
\caption[]{The $\chi^2$ function for the mass of the Higgs boson
in the Standard Model provided by precision electroweak data, for two
different estimates of the hadronic contribution to the effective value of
$\alpha_{em}(m_Z)$. The shaded (blue) band covers other theoretical 
uncertainties~\cite{LEPEWWG}.}
\label{fig:mH}
\end{figure}

\begin{figure}%[t]
\begin{center}
\includegraphics[height=3in]{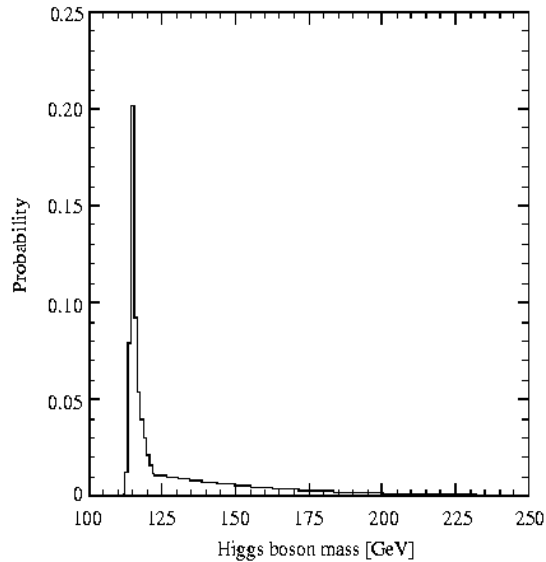}
\caption[]{The probability distribution for the mass of the Higgs boson
in the Standard Model obtained by combining precision electroweak data 
with the lower limit coming from direct experimental 
searches~\cite{Erler}.}
\label{fig:Erler}
\end{center}
\end{figure}

Fig.~\ref{fig:mH} summarizes the indirect information about the possible
mass of the Standard Model Higgs boson provided by fits to the precision
electroweak data, including the best available estimates of leading
multi-loop effects, etc. Depending on the estimate of the hadronic
contributions to $\alpha_{em}(m_Z)$ that one uses, the preferred value of
$m_H$ is around 100~GeV~\cite{LEPEWWG}. Including all the available 
electroweak data except the most recent NuTeV result on deep-inelastic 
$\nu$ scattering, and taking the value $\delta \alpha_{had} = 0.02747 \pm 
0.00012$ for the hadronic contribution to the effective value of 
$\alpha_{em}(m_Z)$, one finds~\cite{LEPEWWG}
\begin{equation}
m_H \; = \; 98^{+53}_{-36} \; {\rm GeV}.
\label{current mH}
\end{equation}
Also shown in Fig.~\ref{fig:mH} is the lower limit $m_H > 114.1$~GeV
provided by direct searches at LEP~\cite{LEPHWG}.  We see that the most
likely value of $m_H$ is 115~GeV, a point made graphically in
Fig.~\ref{fig:Erler}, where precision electroweak data are combined with
the lower limit coming from direct experimental searches~\cite{Erler}.
Values of the Higgs mass up to 199~GeV are allowed at the 99~\% confidence
level, so any assertion that LEP has excluded the majority of the range
allowed by the precision electroweak fit is grotesquely premature. On the
other hand, any resemblance between the most likely value and the mass
$m_H \sim 115$~GeV hinted by direct searches during the dying days of LEP
is surely coincidental (?).

\begin{figure}%[t]
\begin{center}
\hspace{0.01in}
%\rotate{\epsfig{file=HR.eps,height=3in}}
\rotate{\includegraphics[height=3in]{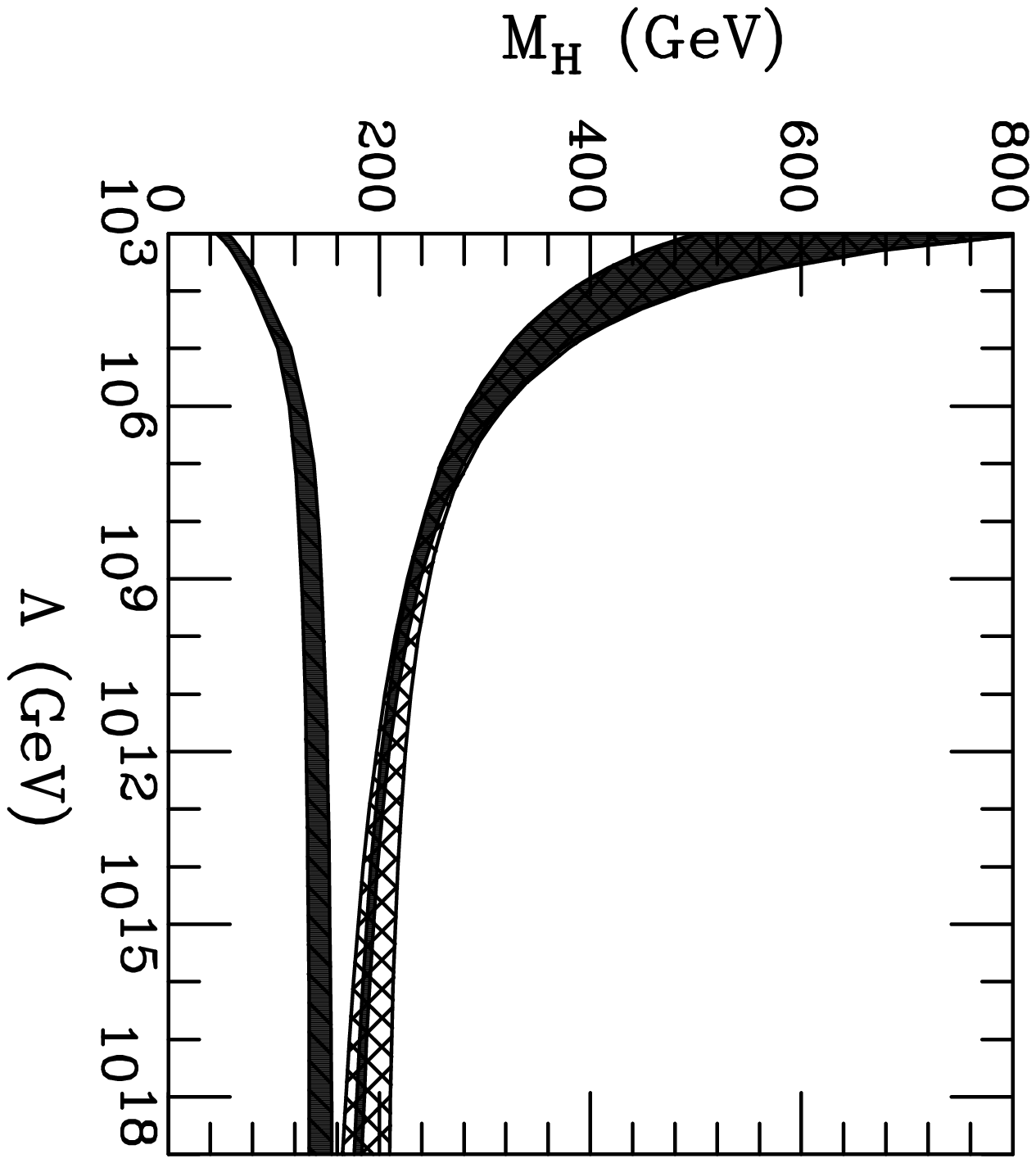}}
\end{center}
\caption[]{Range of $m_H$ allowed in the Standard Model if it is to remain 
valid up to a scale $\Lambda$~\cite{HR}. When $m_H$ is too large, 
renormalization of 
the Higgs self-coupling causes it to blow up at some scale below 
$\Lambda$. When $m_H$ is too small, renormalization of the effective Higgs 
potential by the $t$-quark Yukawa coupling drives it negative, rendering 
the present electroweak vacuum unstable.}
\label{fig:HR}
\end{figure}

If $m_H$ is indeed as low as about 115~GeV, this would be prima facie
evidence for physics beyond the Standard Model at a relatively low energy
scale~\cite{ER}, as seen in Fig.~\ref{fig:HR}. If $m_H$ is larger than the
central range marked in Fig.~\ref{fig:HR}~\cite{HR}, the large Higgs
self-coupling in the renormalization-group running of the effective Higgs
potential causes it to blow up at some scale below the corresponding scale
of $\Lambda$ marked on the horizontal axis. Conversely, if $m_H$ is below
the central band, the larger Higgs-top Yukawa coupling overwhelms the
relatively small Higgs self-coupling, driving the effective Higgs
potential negative at some scale below the corresponding value of
$\Lambda$. As a result, our present electroweak vacuum would be unstable,
or at least metastable with a lifetime that might be longer than the age
of the Universe~\cite{Strumia}.  In the special case $m_H \sim 115$~GeV,
this potential disaster could be averted only by postulating new physics
at some scale $\Lambda \lappeq 10^6$~GeV.

\begin{figure}%[t]
\centerline{\includegraphics[height=3in]{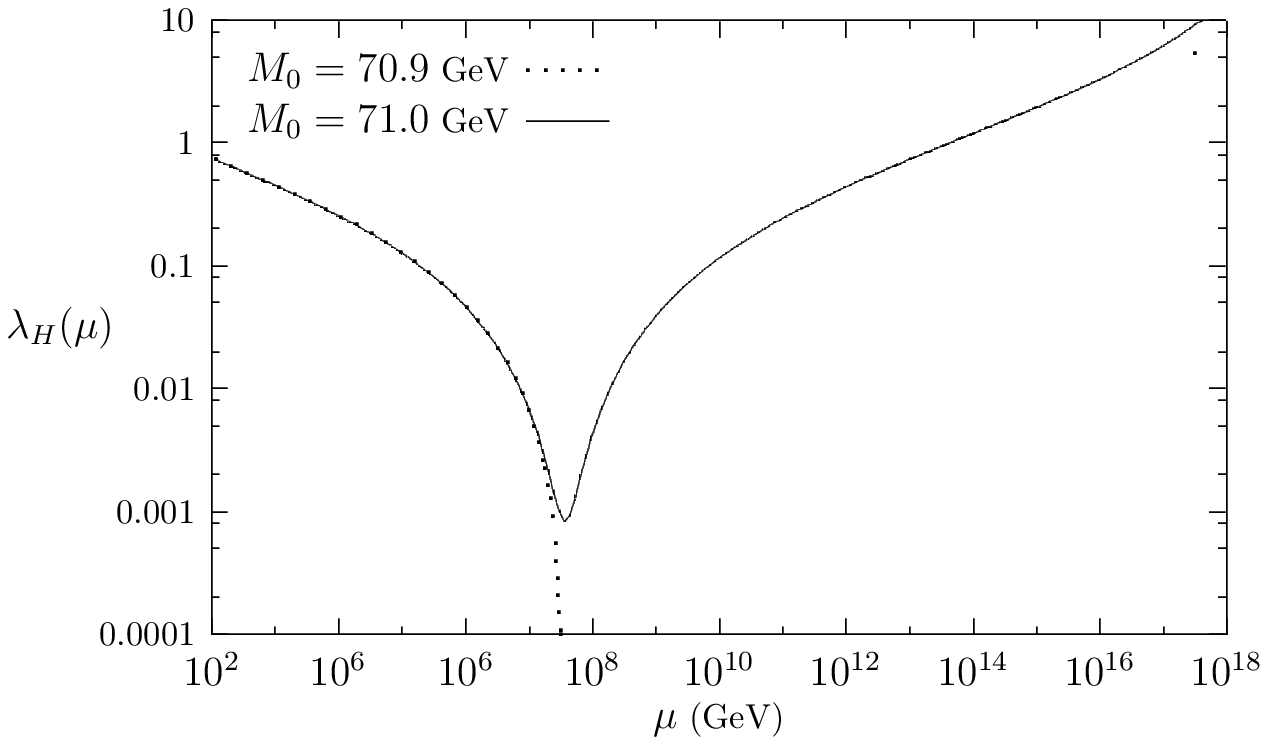}}
\caption[]{Renormalization of the effective Higgs self-coupling for 
different values of the coupling $M_0$ to new bosons $\phi$. It is seen 
that the coupled system must be tuned very finely in order for the 
potential not to collapse or blow up~\cite{ER}.}
\label{fig:ER1}
\end{figure}

\begin{figure}%[t]
\centerline{\includegraphics[height=3in]{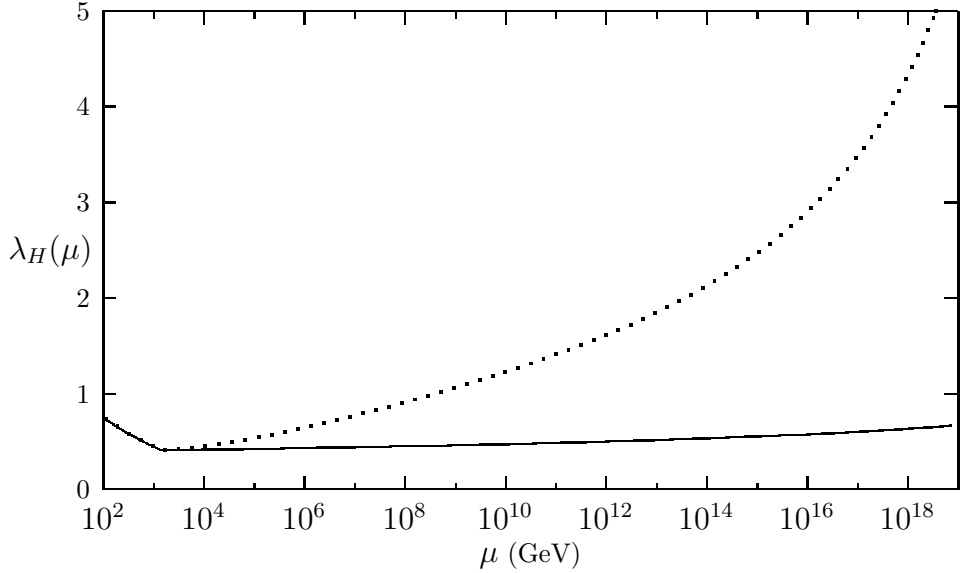}}
\caption[]{An example of the role played by fermionic superpartners
in the running of the Higgs self-coupling. The solid line corresponds to a 
supersymetric model, whereas the dotted line gives the running of the   
quartic
Higgs coupling when the contributions from fermionic Higgsino
and gaugino superpartners have been removed~\cite{ER}.}
\label{fig:ER2}
\end{figure}

This new physics should be bosonic~\cite{ER}, in order to counteract the 
negative
effect of the fermionic top quark. Let us consider introducing $N_I$
isomultiplets of bosons $\phi$ with isospin $I$, coupled to the
conventional Higgs boson by
\begin{equation}
\lambda_{22} |H|^2 |\phi|^2.
\label{Hphi}
\end{equation}
It turns out~\cite{ER} that the coupled renormalization-group equations 
for the $H,
\phi$ system are very sensitive to the chosen value of $\lambda_{22}$ in
(\ref{Hphi}). As seen in Fig.~\ref{fig:ER1}, if the coupling
\begin{equation}
M_0^2 \equiv \lambda_{22} <0|H|0>^2
\label{mzero}
\end{equation}
is too large, the effective Higgs potential blows
up, but it collapses if $M_0^2$ is too small, and the typical amount of
fine-tuning required is 1 in $10^3$! Radiative corrections may easily
upset this fine-tuning, as seen in Fig.~\ref{fig:ER2}. The fine-tuning is
maintained naturally in a supersymmetric theory, but is destroyed if one
has top quarks and their supersymmetric partners $\tilde t$, but not the
supersymmetric partners $\tilde H$ of the Higgs bosons.

If the new physics below $10^6$~GeV is not supersymmetry, it must quack
very much like it!

\subsection{Why Supersymmetry?}

The main theoretical reason to expect supersymmetry at an accessible
energy scale is provided by the {\it hierarchy problem}~\cite{hierarchy}:  
why is $m_W \ll m_P$, or equivalently why is $G_F \sim 1 / m_W^2 \gg G_N =
1 / m_P^2$? Another equivalent question is why the Coulomb potential in an
atom is so much greater than the Newton potential: $e^2 \gg G_N m^2 = m^2
/ m_P^2$, where $m$ is a typical particle mass?

Your first thought might simply be to set $m_P \gg m_W$ by hand, and 
forget about the problem. Life is not so simple, because, as already 
mentioned, quantum 
corrections to $m_H$ and hence $m_W$ are quadratically divergent in the 
Standard Model:
\begin{equation}
\delta m_{H,W}^2 \; \simeq \; {\cal O}({\alpha \over \pi}) \Lambda^2,
\label{Qdgt}
\end{equation}
which is $\gg m_W^2$ if the cutoff $\Lambda$, which represents the scale 
where new physics beyond the Standard Model appears, is comparable to the 
GUT or Planck scale. For example, if the 
Standard Model were to hold unscathed all the way up the Planck mass $m_P 
\sim 10^{19}$~GeV, the radiative correction (\ref{Qdgt}) would be 36 
orders of magnitude greater than the physical values of $m_{H,W}^2$! 

In principle, this is not a problem from the mathematical point of view of
renormalization theory. All one has to do is postulate a tree-level value
of $m_H^2$ that is (very nearly) equal and opposite to the `correction'
(\ref{Qdgt}), and the correct physical value may be obtained. However,
this strikes many physicists as rather unnatural: they would prefer a
mechanism that keeps the `correction' (\ref{Qdgt}) comparable at most to
the physical value.

This is possible in a supersymmetric theory, in which there are equal numbers of bosons and fermions with identical couplings. Since bosonic and fermionic loops have opposite signs, the residual one-loop correction is of the form
\begin{equation}
\delta m_{H,W}^2 \; \simeq \; {\cal O}({\alpha \over \pi}) (m_B^2 - 
m_F^2),
\label{susy}
\end{equation}
which is $\lappeq m_{H,W}^2$ and hence naturally small if the 
supersymmetric partner bosons $B$ and fermions $F$ have similar masses:
\begin{equation}
|m_B^2 - m_F^2| \; \lappeq \; 1~{\rm TeV}^2.
\label{natural}
\end{equation}
This is the best motivation we have for finding supersymmetry at 
relatively low energies~\cite{hierarchy}.

In addition to this first supersymmetric miracle of removing (\ref{susy})
the quadratic divergence (\ref{Qdgt}), many logarithmic divergences are
also absent in a supersymmetric theory. This is the underlying reason why
supersymmetry solves the fine-tuning problem of the effective Higgs
potential when $m_H \sim 115$~GeV, as advertised in the previous section.
Note that this argument is logically distinct from the absence of
quadratic divergences in a supersymmetric theory.

Many other arguments for supersymmetry were proposed before this
hierarchy/naturalness argument. Some of them remain valid, but none of
them fixed the scale at which supersymmetry should appear.

Back in the 1960's, there were many attempts to combine internal
symmetries such as flavour $SU(2)$ or $SU(3)$ with external Lorentz
symmetries, in groups such as $SU(6)$ and ${\tilde U}(12)$. However, it
was shown in 1967 by Coleman and Mandula~\cite{CM} that no non-trivial
combination of internal and external symmetries could be achived using
just bosonic charges. The first non-trivial extension of the Poincar\'e
algebra with fermionic charges was made by Gol¹fand and Likhtman in
1971~\cite{GL}, and in the same year Neveu and Schwarz~\cite{NS}, and
Ramond~\cite{R}, proposed two-dimensional supersymmetric models in
attempts to make fermionic string theories that could accommodate baryons.
Two years later, the first interesting four-dimensional supersymmetric
field theories were written down. Volkov and Akulov~\cite{VA} wrote down a
non-linear realization of supersymmetry with a massless fermion, that they
hoped to identify with the neutrino, but this identification was soon
found not to work, because the low-energy interactions of
neutrinos differed from those in the non-linear supersymmetric model.

Also in 1973, Wess and Zumino~\cite{WZ1,WZ2} started writing down
renormalizable four-dimensional supersymmetric field theories, with the
objective of describing mesons and baryons. Soon afterwards, together with
Iliopoulos and Ferrara, they were able to show that supersymmetric field
theories lacked many of the quadratic and other divergences found in
conventional field theories~\cite{noren}, and some thought this was an
attractive feature, but the physical application remained obscure for
several years. Instead, for some time, phenomenological interest in
supersymmetry was focused on the possibility of unifying fermions and
bosons, for example matter particles (with spin $1 / 2$) and force
particles (with spin $1$), or alternatively matter and Higgs particles, in
the same supermultiplets~\cite{Fayet}. With the discovery of local
supersymmetry, or supergravity, in 1976~\cite{sugra}, this hope was
extended to the unification of the graviton with lower-spin particles.
Indeed, for a short while, the largest supergravity theory was touted as
the TOE: in the words of Hawking~\cite{Hawk}, `Is the end in sight for 
theoretical physics?'.

These are all attractive ideas, and many play r\^oles in current theories,
but I reiterate that the only real motivation for expecting supersymmetry
at accessible energies $\lappeq 1$~TeV is the naturalness of the mass
hierarchy~\cite{hierarchy}.

\subsection{What is Supersymmetry?}

The basic idea of supersymmetry is the existence of fermionic charges
$Q_\alpha$ that relate bosons to fermions. Recall that all previous
symmetries, such as flavour $SU(3)$ or electromagnetic $U(1)$, have
involved scalar charges $Q$ that link particles with the same spin into
multiplets:
\begin{equation}
Q \; | {\rm Spin} J > \; = \; | {\rm Spin} J¹ >.
\label{scalarQ}
\end{equation}
Indeed, as mentioned above, Coleman and Mandula~\cite{CM} proved that it 
was
`impossible' to mix internal and Lorentz symmetries: $J_1 \leftrightarrow
J_2$. However, their `no-go' theorem assumed implicitly that the
prospective charges should have integer spins.

The basic element in their `proof' was the observation that the only
possible conserved tensor charges were those with no Lorentz indices,
i.e., scalar charges, and the energy-momentum vector $P_\mu$. To see how
their `proof' worked, consider two-to-two elastic scattering, $1 + 2 \to
3 + 4$, and imagine that there exists a conserved two-index tensor charge,
$\Sigma_{\mu \nu}$. By Lorentz invariance, its diagonal matrix elements
between single-particle states $| a >$ must take the general form:
\begin{equation}
< a | \Sigma_{\mu \nu} | a > \; = \; \alpha P^{(a)}_\mu P^{(a)}_\nu \; + \; \beta g_{\mu \nu},
\label{SigmaME}
\end{equation}
where $\alpha, \beta$ are arbitrary reduced matrix elements, and $g_{\mu
\nu}$ is the metric tensor. For $\Sigma_{\mu \nu}$ to be conserved in a
two-to-two scattering process, one must have
\begin{equation}
P^{(1)}_\mu P^{(1)}_\nu +  P^{(2)}_\mu P^{(2)}_\nu = 
P^{(3)}_\mu P^{(3)}_\nu + P^{(4)}_\mu P^{(4)}_\nu,
\label{trouble}
\end{equation}
where we assume that the symmetry is local, so that two-particle matrix
elements of $\Sigma_{\mu \nu}$ play no r\^ole. Since Lorentz invariance
also requires $ P^{(1)}_\mu + P^{(2)}_\mu = P^{(3)}_\mu + P^{(4)}_\mu$,
the only possible outcomes are $ P^{(1)}_\mu = P^{(3)}_\mu$ or $
P^{(4)}_\mu$. Thus the only possibilities are completely forward
scattering or completely backward scattering. This disagrees with
observation, and is in fact theoretically impossible in any local field
theory.

This rules out any non-trivial two-index tensor charge, and the argument
can clearly be extended to any higher-rank tensor with more Lorentz
indices. But what about a spinorial charge $Q_\alpha$? This can have no
diagonal matrix element:
\begin{equation}
< a | Q_\alpha | a > \; \ne \; 0,
\label{vanishes}
\end{equation}
and hence the Coleman-Mandula argument fails. 

So what is the possible form of a `supersymmetry' algebra that includes
such spinorial charges $Q^i_\alpha$~\footnote{In what follows, I shall
suppress the spinorial subscript $\alpha$ whenever it is not essential. 
The superscripts $i, j, ..., N$ denote different supersymmetry charges.}? 
Since 
the different $Q^i$ are supposed to generate symmetries, they must commute
with the Hamiltonian:
\begin{equation}
[ Q^i, H] \; = \; 0: \; i = 1,2, Š, N.
\label{commute}
\end{equation}
So also must the anticommutator of two spinorial charges:
\begin{equation}
[ \{Q^i, Q^j \}, H] \; = \; 0: \; i,j = 1,2, Š, N.
\label{anticommute}
\end{equation}
However, the part of the anticommutator $\{Q^i, Q^j \}$ that is symmetric
in the internal indices $i, j$ cannot have spin 0. Instead, as we
discussed just above, the only possible non-zero spin choice is $J = 1$,
so that
\begin{equation}
\{Q^i, Q^j \} \; \propto \delta^{ij} P_\mu + \dots \; Š: i,j = 1,2, Š, N.
\label{vector}
\end{equation}
In fact, as was proved by Haag, Lopuszanski and Sohnius~\cite{HLS}, 
the only allowed possibility is
\begin{equation}
\{Q^i, Q^j \} \; = \; 2 \delta^{ij} \gamma^\mu P_\mu {\cal C}+ \dotsŠ: i,j 
= 1,2, Š, N,
\label{HLS}
\end{equation}
where ${\cal C}$ is the charge-conjugation matrix discussed in more 
detail in Lecture 2, and the dots denote 
a possible `Œcentral charge' that is antisymmetric in the indices $i,j$, 
and hence can only appear when $N > 1$.

According to a basic principle of Swiss law, anything not illegal is
compulsory, so there MUST exist physical realizations of the supersymmetry
algebra (\ref{HLS}). Indeed, non-trivial realizations of the
non-relativistic analogue of (\ref{HLS}) are known from nuclear 
physics~\cite{Iachello},
atomic physics and condensed-matter physics. However, none of these is
thought to be fundamental.

In the relativistic limit, supermultiplets consist of massless particles 
with spins differing by half a unit. In the case of simple $N = 1$ 
supersymmetry, the basic building blocks are 
{\it chiral~supermultiplets}:
\begin{eqnarray}
\pmatrix{
{1 \over 2} \cr
0}
~e.g.,~
\pmatrix{
\ell~ (lepton) \cr
\tilde \ell~(slepton)} or
\pmatrix {q~(quark) \cr
\tilde q~(squark)}
\label{chiral}
\end{eqnarray}
{\it gauge supermultiplets}:

\begin{eqnarray}
\pmatrix{
1 \cr
{1 \over 2}}
~e.g.,~ 
\pmatrix{
\gamma~(photon)\cr
\tilde \gamma~(photino)}
~or~
\pmatrix{
g~(gluon)\cr
\tilde g~(gluino)}
\label{gaugesmultiplet}
\end{eqnarray}
and the {\it graviton supermultiplet} consisting of the spin-2 graviton 
and the spin-$3 / 2$ gravitino.

Could any of the known particles in the Standard Model be linked together
in supermultiplets? Unfortunately, none of the known fermions $q, \ell$
can be paired with any of the Œknown¹ bosons $\gamma, W^\pm Z^0, g, H$,
because their internal quantum numbers do not match~\cite{Fayet}. For
example, quarks $q$ sit in triplet representations of colour, whereas the
known bosons are either singlets or octets of colour. Then again, leptons
$\ell$ have non-zero lepton number $L = 1$, whereas the known bosons have
$L = 0$. Thus, the only possibility seems to be to introduce new
supersymmetric partners (spartners) for all the known particles: quark
$\to$ squark, lepton $\to$ slepton, photon $\to$ photino, Z $\to$ Zino, W
$\to$ Wino, gluon $\to$ gluino, Higgs $\to$ Higgsino, as suggested in
(\ref{chiral}, \ref{gaugesmultiplet}) above.

The best that one can say for supersymmetry is that it economizes on
principle, not on particles!

\subsection{(S)Experimental Hints}

By now, you may be wondering whether it makes sense to introduce so many
new particles just to deal with a paltry little hierarchy or naturalness
problem. But, as they used to say during the First World War, `if you know
a better hole, go to it.' As we learnt above, technicolour no longer seems
to be a viable hole, and I am not convinced that theories with large extra
dimensions really solve the hierarchy problem, rather than just rewrite
it. Fortunately, there are two hints from the high-precision electroweak
data that supersymmetry may not be such a bad hole, after all.

One is the fact, already advertised, that there probably exists a Higgs
boson weighing less than about 200~GeV~\cite{LEPEWWG}. This is perfectly 
consistent with
calculations in the minimal supersymmetric extension of the Standard Model
(MSSM), in which the lightest Higgs boson weighs less than about 
130~GeV~\cite{susyHiggs}, 
as we discuss in more detail in Lecture 3.

The other hint is provided by the strengths of the different gauge 
interactions, as measured at LEP~\cite{GUTs}. These may be run up to high 
energy scales 
using the renormalization-group equations, to see whether they unify as 
predicted in a GUT. The answer is no, if supersymmetry is not included in 
the calculations. In that case, GUTs would require
\begin{equation}
\sin^2 \theta_W \; = \; 0.214 \pm 0.004,
\label{GUT}
\end{equation}
whereas the experimental value of the effective neutral weak mixing 
parameter at the $Z^0$ peak is $\sin^2 \theta = 0.23149
\pm 0.00017$~\cite{LEPEWWG}. On the other hand, minimal supersymmetric 
GUTs predict
\begin{equation}
\sin^2 \theta_W \; \sim \; 0.232Š,
\label{susyGUT}
\end{equation}
where the error depends on the assumed sparticle masses, the preferred 
value being around 1~TeV, as suggested completely independently by the 
naturalness of the electroweak mass hierarchy. This argument is also 
discussed in more detail in Lecture 3.

\section{SIMPLE MODELS}

\subsection{Deconstructing Dirac}

In this Section, we tackle some unavoidable spinorology. The most familiar
spinors used in four-dimensional field theories are four-component Dirac
spinors $\psi$. You may recall that it is possible to introduce projection
operators
\begin{equation}
P_{L,R} \; \equiv \; {1 \over 2} ( 1 \mp \gamma_5 ),
\label{projections}
\end{equation}
where $\gamma_5 \equiv i \gamma^0 \gamma^1 \gamma^2 \gamma^3$, 
and the $\gamma_\mu$ can be written in the forms
\begin{eqnarray}
\gamma_\mu \; =  \; \pmatrix{
0 & \sigma_\mu \cr
\bar\sigma_\mu & 0}~,
\label{gamma}
\end{eqnarray}
where $\sigma_\mu \equiv (1, \sigma_i)$, ${\bar \sigma_\mu} \equiv 
(1, - \sigma_i)$. Then $\gamma_5$ can be written in the form 
diag$(- {\mathbf 1}, {\mathbf 1})$, where $- {\mathbf 1}, {\mathbf 1}$ 
denote $2 \times 2$ matrices. Next,we introduce the corresponding left- 
and right-handed spinors
\begin{equation}
\psi_{L,R} \; \equiv \; P_{L,R} \psi,
\label{LRspinors}
\end{equation}
in terms of which one may decompose the four-component spinor into a 
pair of two-component spinors:
\begin{eqnarray}
\psi = \pmatrix{ \psi_L \cr \psi_R}~.
\label{decomp}
\end{eqnarray}
These will serve as our basic fermionic building blocks.

Antifermions can be represented by adjoint spinors
\begin{equation}
{\bar \psi} \; \equiv \; \psi^\dagger \gamma^0 \; = \; 
( {\bar \psi_R}, {\bar \psi_L})
\label{adjoint}
\end{equation}
where the $\gamma^0$ factor has interchanged the left- and 
right-handed components $\psi_{L,R}$. We can now decompose in terms of 
these the conventional fermion kinetic term
\begin{equation}
{\bar \psi} \gamma_\mu \partial^\mu \psi \; = \; 
{\bar \psi_L} \gamma_\mu \partial^\mu \psi_L + 
{\bar \psi_R} \gamma_\mu \partial^\mu \psi_R
\label{kinetic}
\end{equation}
and the conventional mass term
\begin{equation}
{\bar \psi} \psi \; = \; {\bar \psi_R} \psi_L + {\bar \psi_L} \psi_R.
\label{massterm}
\end{equation}
We see that the kinetic term keeps separate the left- and right-handed 
spinors, whereas the mass term mixes them.

The next step is to introduce the charge-conjugation operator ${\cal C}$, 
which 
changes the overall sign of the vector current ${\bar \psi} \gamma^\mu \psi$. 
It transforms spinors into their conjugates:
\begin{eqnarray}
\psi^c \; \equiv {\cal C} \; {\bar \psi^T} \; = \; {\cal C} (\psi^\dagger 
\gamma^0)^T \; 
= \; 
\pmatrix{
\bar\psi_R\cr
\bar\psi_L}~,
\label{Cpsi}
\end{eqnarray}
and operates as follows on the $\gamma$ matrices:
\begin{equation}
{\cal C}^{-1} \gamma^\mu {\cal C} \; = \; - \gamma^{\mu T}.
\label{Cgamma}
\end{equation}
A convenient representation of ${\cal C}$ is:
\begin{equation}
{\cal C} \; = \; i \gamma^0 \gamma^2.
\end{equation}
It is apparent from the above that the conjugate of a left-handed 
spinor is right-handed:
\begin{eqnarray}
(\psi_L)^c \; = \; \pmatrix{ 0 \cr \bar\psi_L}~,
\label{LtoR}
\end{eqnarray}
so that the combination
\begin{equation}
\bar \psi^c_L \psi_L \; = \; \psi_L \sigma_2 \psi_L
\label{LLmass}
\end{equation}
mixes left- and right-handed spinors, and has the same form as a 
mass term (\ref{massterm}).

It is apparent from (\ref{LtoR}) that we can construct four-component 
Dirac spinors 
entirely out of two-component left-handed spinors and their conjugates:
\begin{eqnarray}
\psi = \pmatrix{
\psi_i \cr \psi^c_j}~,
\label{constructD}
\end{eqnarray}
a trick that will be useful later in our supersymmetric model-building. 
As examples, instead of working with left- and right-handed quark 
fields $q_L$ and $q_R$, or left- and right-handed lepton fields $\ell_L$ 
and $\ell_R$, we can write the theory in terms of left-handed antiquarks 
and antileptons: $q_R \to q^c_L$ and $\ell_R \to \ell^c_L$.

\subsection{Simplest Supersymmetric Field Theories}

Let us now consider a field theory~\cite{FF} containing just a single 
left-handed 
fermion $\psi_L$ and a complex boson $\phi$, without any interactions, 
as described by the Lagrangian
\begin{equation}
L_0 \; = \; i {\bar \psi_L} \gamma^\mu \partial_\mu \psi_L \; + \; 
| \partial \phi |^2.
\label{freetheory}
\end{equation}
We consider the simplest possible non-trivial transformation law for the 
free theory (\ref{freetheory}):
\begin{equation}
\phi \; \to \; \phi + \delta \phi, \; \; {\rm where} \; \delta \phi \; 
= \; \sqrt{2} {\bar E} \psi_L,
\label{Btransform}
\end{equation}
where $E$ is some constant right-handed spinor. In 
parallel with (\ref{Btransform}), we also consider the most general 
possible transformation law for the fermion $\psi$:
\begin{equation}
\psi_L \; \to \; \psi_L + \delta \psi_L, \; \; {\rm where} \; \delta \psi_L 
\; = \; - ~ a~i~\sqrt{2} (\gamma_\mu \partial^\mu \phi) E - F E^c,
\label{Ftransform}
\end{equation}
where $a$ and $F$ are constants to be fixed later, and 
we recall that $E^c$ is a 
left-handed spinor. We can now consider the resulting transformation of 
the full Lagrangian (\ref{freetheory}), which can easily be checked to 
take the form
\begin{equation}
\delta L_0 \; = \; \sqrt{2} \partial_\mu [ {\bar \psi} E \partial^\mu \phi 
+ {\bar E} \gamma^\mu \phi^* \gamma_\nu \partial^\nu \psi],
\label{Ltransform}
\end{equation}
if and only if we choose
\begin{equation}
a \; = \; 1 \; \; {\rm and} \; \; F \; = \; 0
\label{fixaF}
\end{equation}
in this free-field model.
With these choices, and the resulting total-derivative transformation 
law ({\ref{Ltransform}) for the free Lagrangian, the free action $A_0$ is 
invariant under the transformations (\ref{Btransform},\ref{Ftransform}), since
\begin{equation}
\delta A_0 \; = \; \delta \int d^4 x L_0 \; = \; 0.
\label{Atransform}
\end{equation}
Fine, you may say, but is this symmetry actually supersymmetry? 
To convince yourself that it is, consider the sequences of pairs
(\ref{Btransform},\ref{Ftransform}) of transformations starting from 
either the boson $\phi$ or the fermion $\psi$:
\begin{equation}
\phi \; \to \; \psi \; \to \; \partial \phi, \; \; \psi \; \to \; 
\partial \phi \; \to \; \partial \psi.
\label{transform2}
\end{equation}
In both cases, the action of two symmetry transformations is equivalent 
to a derivative, i.e., the momentum operator, corresponding exactly to the 
supersymmetry algebra. A free boson and a free fermion together realize 
supersymmetry: like the character in Moli\`ere, we have been talking 
prose all our lives without realizing it!

Now we look at interactions in a supersymmetric field theory~\cite{WZ1}.  
The most general interactions between spin-0 fields $\phi^i$ and spin-1/2
fields $\psi^i$ that are at most bilinear in the latter, and hence have a
chance of being renormalizable in four dimensions, can be written in the
form

\begin{equation}
L \; = \; L_0 \; - \; V(\phi^i, \phi^*_j) \; - \; {1 \over 2} M_{ij}(\phi, 
\phi^*){\bar \psi^{ci}} \psi^j
\label{interactions}
\end{equation}
where $V$ is a general effective potential, and $M_{ij}$ includes both
mass terms and Yukawa interactions for the fermions. Supersymmetry imposes
strong constraints on the allowed forms of $V$ and $M$, as we now see.
Suppose that $M$ depended non-trivially on the conjugate fields $\phi^*$: 
then the 
supersymmetric variation $\delta ( M {\bar \psi^c} \psi )$ would contain a 
term
\begin{equation}
{\partial M \over \partial \phi^*} \psi^* {\bar \psi^c} \psi
\label{Mbit}
\end{equation}
that could not be compensated by the variation of any other term. We
conclude that $M$ must be independent of $\phi^*$, and hence $M = M(\phi)$
alone. 

Another term in the variation of the last term in (\ref{interactions}) is
\begin{equation}
{\partial M_{ij} \over \partial \phi^k} {\bar E} \psi^k {\bar \psi^{ci}} 
\psi^j.
\label{Mbit2}
\end{equation}
This term cannot be cancelled by the variation of any other term, but can 
vanish by itself if $\partial M_{ij} / \partial \phi^k$ is completely 
symmetric in the indices $i, j, k$. This is possible only if
\begin{equation}
M_{ij} \; = \; {\partial W \over \partial \phi^i \partial \phi^j}
\label{symmetric}
\end{equation}
for some function $W(\phi)$ called the {\it superpotential}. If the theory
is to be renormalizable, $W$ can only be cubic. The trilinear term of $W$
determines the Yukawa couplings, and the bilinear part the mass terms.

We now re-examine the form of the supersymmetric transformation law 
(\ref{Ftransform}) itself. Yet another term in the variation of the second 
term in (\ref{interactions}) has the form
\begin{equation}
i M_{jk}{\bar \psi^{cj}} \gamma_\mu \partial^\mu \phi^k E + ({\rm Herm.
Conj.}).
\label{Mbit3}
\end{equation}
This can cancel against an $F$-dependent term in the variation of the
fermion kinetic term
\begin{equation}
- i  {\bar \psi_i} \gamma_\mu \partial^\mu F^i E^c +  ({\rm Herm. Conj.}),
\label{Fform}
\end{equation}
if the following relation between $F$ and $M$ holds: ${\partial F^*_i
\over \partial \phi^j} = M_{ij}$, which is possible if and only if
\begin{equation}
F^*_i \; = \; {\partial W \over \partial \phi^i}.
\label{formF}
\end{equation}
Thus the form of $W$ also determines the required form of the
supersymmetry transformation law.

The form of $W$ also determines the effective potential $V$, as we now
see. One of the terms in the variation of $V$ is
\begin{equation}
{\partial V \over \partial \phi^i} {\bar E} \psi^I \; + \; ({\rm Herm. Conj.}),
\label{Vvarn}
\end{equation}
which can only be cancelled by a term in the variation of $ M_{ij} {\bar
\psi^{ci}}\psi^j$, which can take the form $M_{ij} {\bar \psi^{ci}} F^j
E^c$ if
\begin{equation}
{\partial V \over \partial \phi^i} \; = \; M_{ij} F^i,
\label{Mvarn}
\end{equation}
which is in turn possible only if
\begin{equation}
V \; = \; | {\partial W \over \partial \phi^i} |^2 \; = \; | F^i|^2.
\label{finalV}
\end{equation}

We now have the complete supersymmetric field theory for interacting
chiral (matter) supermultiplets~\cite{WZ1}:
\begin{equation}
L \; = \; i {\bar \psi_i} \gamma_\mu \partial^\mu \psi^i + 
|\partial_\mu \phi^i|^2
- |{\partial W \over \partial \phi^i}|^2 - {1 \over 2}{\partial^2 W \over 
\partial \phi^i \partial^j} {\bar \psi^{ci}} \partial \psi^j + ({\rm Herm. 
Conj.}).
\label{finalL}
\end{equation}
This Lagrangian is invariant (up to a total derivative) under the 
supersymmetry transformations
\begin{equation}
\delta \phi^i \; = \; \sqrt{2} {\bar E} \psi^i, \; \delta \psi^i \; 
= \; -i \sqrt{2} \gamma_\mu \partial^\mu \phi^i E - F^i E^c: \; 
F^i \; = \; ({\partial W \over \partial \phi^i})^*.
\label{finalS}
\end{equation}
The simplest non-trivial superpotential involving a single superfield $\phi$ is
\begin{equation}
W \; = \; {\lambda \over 3} \phi^3 \; + \; {m \over 2} \phi^2.
\label{simpleW}
\end{equation}
It is a simple exercise for you to verify using the rules given
above that the corresponding Lagrangian is
\begin{equation}
L \; = \; i {\bar \psi} \gamma_\mu \partial^\mu \psi \; 
+ \; |\partial_\mu \phi |^2 \; - \; | m \phi + 
\lambda \phi^2|^2 \; - \; m {\bar \psi^c} \psi \; - \; \lambda \phi {\bar 
\psi^c} \psi.
\label{finalL2}
\end{equation}
We see explicitly that the bosonic component $\phi$ of the
supermultiplet has the same mass as the fermionic component $\psi$, and
that the Yukawa coupling $\lambda$ fixes the effective potential.

We now turn to the possible form of a supersymmetric gauge 
theory~\cite{WZ2}.
Clearly, it must contain vector fields $A^a_\mu$ and fermions $\chi^a$ in
the same adjoint representation of the gauge group. Once one knows the
gauge group and the fermionic matter content, the form of the Lagrangian
is completely determined by gauge invariance:
\begin{equation}
L \; = \; {i \over 2} {\bar \chi^a} \gamma^\mu D^\mu_{ab} \chi^b \; - \; 
{1 \over 4} F^a_{\mu \nu} F^{a, \mu \nu} \; [\; - \; {1 \over 2} (D^a)^2 
\; ].
\label{gaugeS}
\end{equation}
Here, the gauge-covariant derivative
\begin{equation}
D^\mu_{ab} \; \equiv \; \delta_{ab} \partial^\mu \; - \; g f_{abc} A_c^\mu,
\label{covderiv}
\end{equation}
and the gauge field strength is
\begin{equation}
F^a_{\mu \nu} \; \equiv \; \partial_\mu A^a_\nu \; - \; 
\partial_\nu A^a_\mu \; + \; g f^{abc} A^b_\mu A^c_\nu,
\label{fieldstrength}
\end{equation}
as usual. We return later to the $D$ term at the end of (\ref{gaugeS}). 
Yet another of the miracles of supersymmetry is that the Lagrangian 
(\ref{gaugeS}) is automatically supersymmetric, without any further 
monkeying around. The corresponding supersymmetry transformations may be 
written as
\begin{eqnarray}
\delta A^a_\mu \; & = & \; - {\bar E} \gamma_\mu \chi^a, \\
\delta \chi^a \; & = & \; - {i \over 2} F^a_{\mu \nu} \gamma^\mu 
\gamma^\nu E \; 
+ \; D^a E, \\
\delta D^a \; & = & \; - i {\bar E} \gamma_5 \gamma^\mu D_\mu^{ab}\chi^b.
\label{gaugetransf}
\end{eqnarray}
What about the $D$ term in (\ref{gaugeS})? It is a trivial consequence of
equations of motion derived from (\ref{gaugeS}) that $D^a = 0$. However,
this is no longer the case if matter is included. Then, it turns out, one
must add to (\ref{gaugeS}) the following:
\begin{equation}
\Delta L \; = \; - \sqrt{2} g \chi^a \phi^*_i (T^a)^i_j \psi^j \; 
+ \; ({\rm Herm. Conj.})
\; + \; g (\phi^*_i (T^a)^i_j \phi^j) D^a,
\label{gaugematterS}
\end{equation}
where $T^a$ is the group representation matrix for the matter fields
$\phi^i$. With this addition, the equation of motion for $D^a$ tells us
that
\begin{equation}
D^a \; = \; g \phi_i^* (T^a)^i_j \phi^j,
\label{formD}
\end{equation}
and we find a $D$ term in the full effective potential:
\begin{equation}
V \; = \; \Sigma_i | F_i|^2 \; + \; \Sigma_a {1 \over 2} (D^a)^2,
\label{fullV}
\end{equation}
where the form of $D^a$ is given in (\ref{formD}).

\subsection{Further Aspects of Supersymmetric Field Theories}

So far, we have taken a relatively unsophisticated approach to
supersymmetry. However, one of the reasons why theorists are so
enthusiastic about supersymmetry is because it is not just a new type of
symmetry, but extends the concept of space-time itself. Recall the basic
form of the supersymmetry algebra:
\begin{equation}
2 \delta^{ij} \gamma_\mu P^\mu {\cal C} \; = \; \{Q^i, Q^j\}.
\label{ysus}
\end{equation}
The reason this is written backwards here is to emphasize that one can
regard supersymmetric charges $Q^i$ as square roots of the translation
operator. Recall how the translation operator acts on a bosonic field:
\begin{equation}
\phi ( x + a) \; = \; e^{ia.P} \phi (x) e^{-ia.P},
\label{fintrans}
\end{equation}
where the momentum operator $P$ is the generator of infinitesimal translations:
\begin{equation}
i [ P_\mu, \phi (x) ] \; = \; \partial_\mu \phi (x).
\label{inftrans}
\end{equation}
Expanding the formula (\ref{fintrans}), we find the following expression
for a small finite translation:
\begin{equation}
\delta_a \phi (x) \; \equiv \; \phi (x + a ) - \phi (x) \; \simeq \; a_\mu 
\partial^\mu \phi (x) \; = \; ia_\mu [P^\mu, \phi (x)].
\label{smalltrans}
\end{equation} 
Following this deconstruction of translations, we now can see better how
the supersymmetric charge can be regarded, in some sense, as the square
root: `$Q \sim \sqrt{P}$', just as the Dirac equation can be regarded as
the square root of the Klein-Gordon equation. There is an exact
supersymmetric analogue of (\ref{smalltrans}):
\begin{equation}
\delta_{\bar E} \phi (x) \; = \; \sqrt{2} {\bar E} \psi (x) \; 
= \; i \sqrt{2} {\bar E} [ Q, \phi (x) ].
\label{infsusy}
\end{equation}
By analogy with (\ref{fintrans},\ref{smalltrans}), one may consider the
spinor ${\bar E}$ as a sort of `superspace' coordinate, and one can
combine the bosonic field $\phi (x)$ and its fermionic partner $\psi (x)$
into a superfield:
\begin{equation}
\delta_{\bar E} \phi (x) \; = \; \Phi (x, {\bar E}) - \Phi (x): \; \; 
\Phi (x, {\bar E}) \equiv \phi (x) \; + \; \sqrt{2} {\bar E} \psi (x).
\label{superfield}
\end{equation}
At this level, the introduction of superspace and superfields may appear
superfluous, but it gives deeper insights into the theory and facilitates
the derivation of many important results, such as the non-renormalization
theorems of supersymmetry, that we discuss next. In some sense, the next
generation of accelerators such as the LHC is `guaranteed' to discover
extra dimensions, either bosonic ones, as discussed here by 
Antoniadis~\cite{A}, or fermionic.

Many remarkable no-renormalization theorems can be proved in
supersymmetric field theories~\cite{noren}. First and foremost, they have 
no quadratic
divergences. One way to understand this is to compare the renormalizations
of bosonic and fermionic mass terms:
\begin{equation}
m_B^2 | \phi |^2 \; \leftrightarrow \; m_F {\bar \psi} \psi.
\label{noquad}
\end{equation}
We know well that fermion masses $m_F$ can only be renormalized
logarithmically. Since supersymmetry guarantees that $m_B = m_F$, it
follows that there can be no quadratic divergence in $m_B$. Going further,
chiral symmetry guarantees that the one-loop renormalization of a fermion
mass has the general multiplicative form:
\begin{equation}
\delta m_F \; = \; {\cal O}({\alpha \over \pi}) \; m_F \; {\rm ln} ({\mu_1 
\over \mu_2}),
\label{mFren}
\end{equation}
where $\mu_{1,2}$ are different renormalization scales. This means that if
$m_F$ (and hence also $m_B$) vanish at the tree level in a supersymmetric
theory, then both $m_F$ and $m_B$ remain zero after renormalization. This
is one example of the reduction in the number of logarithmic divergences
in a supersymetric theory.

In general, there is no intrinsic renormalization of any superpotential
parameters, including the Yukawa couplings $\lambda$, apart from overall
multiplicative factors due to wave-function renormalizations:
\begin{equation}
\Phi \; \to \; Z \Phi,
\label{Zfactors}
\end{equation}
which are universal for both the bosonic and fermionic components $\phi,
\psi$ in a given superfield $\Phi$. However, gauge couplings {\it are}
renormalized, though the $\beta$-function is changed:
\begin{equation}
\beta (g) \; \ne \; 0: \; \; - 11 N_c \; \to \; -9 N_c
\label{eleventonine}
\end{equation}
at one-loop order
in an $SU(N_c)$ supersymmetric gauge theory with no matter, as a result 
of the extra gaugino contributions.

There are even fewer divergences in theories with more supersymmetries.
For example, there is only a finite number of divergent diagrams in a
theory with $N = 2$ supersymmetries, which may be cancelled by imposing a
few simple relations on the spectrum of supermultiplets. Finally, there
are no divergences at all in theories with $N = 4$ supersymmetries, which
obey automatically the necessary finiteness conditions.

Many theorists from Dirac onwards have found the idea of a completely
finite theory attractive, so it is natural to ask whether theories with $N
\ge 2$ supersymmetries could be interesting as realistic field theories.
Unfortunately, the answer is `not immediately', because they do not allow
the violation of parity. To see why, consider the simplest possible
extended supersymmetric theory containing an $N = 2$ matter multiplet,
which contains both left- and right-handed fermions with helicities $\pm
1/2$. Suppose that the left-handed fermion with helicity $+1/2$ sits in a
representation $R$ of the gauge group. Now act on it with either of the
two supersymmetry charges $Q_{1,2}$: they each yield bosons, that each sit
in the same representation $R$. Now act on either of these with the other
supercharge, to obtain a right-handed fermion with helicity $-1/2$: this
must also sit in the same representation $R$ of the gauge group. Hence,
left- and right-handed fermions have the same interactions, and parity is
conserved. There is no way out using gauginos, because they are forced to
sit in adjoint representations of the gauge group, and hence also cannot
distinguish between right and left.

Thus, if we want to make a supersymmetric extension of the Standard Model,
it had better be with just $N = 1$ supersymmetry, and this is what we do
in the next Section.

\subsection{Building Supersymmetric Models}

Any supersymmetric model is based on a Lagrangian that contains a
supersymmetric part and a supersymmetry-breaking part:
\begin{equation}
{\cal L} \; = \; {\cal L}_{susy} \; + \; {\cal L}_{susy \times}.
\label{twobits}
\end{equation}
We discuss the supersymmetry-breaking part ${\cal L}_{susy \times}$ in the
next Lecture: here we concentrate on the supersymmetric part ${\cal
L}_{susy}$. The minimal supersymmetric extension of the Standard Model
(MSSM) has the same gauge interactions as the Standard Model, and Yukawa
interactions that are closely related. They are based on a superpotential
$W$ that is a cubic function of complex superfields corresponding to
left-handed fermion fields. Conventional left-handed lepton and quark
doublets are denoted $L, Q$, and right-handed fermions are introduced via
their conjugate fields, which are left-handed, $e_R \to E^c, u_R \to U^c,
d_R \to D^c$. In terms of these,
\begin{equation}
W \; = \; \Sigma_{L,E^c} \lambda_L L E^c H_1 \; + \; \Sigma_{Q, U^c} 
\lambda_U Q U^c H_2 \; + \; \Sigma_{Q, D^c} \lambda_D Q D^c H_1 \; 
+ \mu H_1 H_2.
\label{SMW}
\end{equation}
A few words of explanation are warranted. The first three terms in
(\ref{SMW}) yield masses for the charged leptons, charge-$(+2/3)$ quarks
and charge-$(-1/3)$ quarks respectively. All of the Yukawa couplings
$\lambda_{L,U,D}$ are $3 \times 3$ matrices in flavour space, whose
diagonalizations yield the mass eigenstates and Cabibbo-Kobayashi-Maskawa
mixing angles.

Note that two distinct Higgs doublets $H_{1,2}$ have been introduced, for
two important reasons. One reason is that the superpotential must be an
analytic polynomial: as we saw in (\ref{SMW}), it cannot contain both $H$ 
and $H^*$, whereas the
Standard Model uses both of these to give masses to all the quarks and
leptons with just a single Higgs doublet. The other reason is to cancel
the triangle anomalies that destroy the renormalizability of a gauge
theory. Ordinary Higgs boson doublets do not contribute to these
anomalies, but the fermions in Higgs supermultiplets do, and two doublets
are required to cancel each others' contributions. Once two Higgs
supermultiplets have been introduced, there is the possibility, even the
necessity, of a bilinear term $\mu H_1 H_2$ coupling them together.

Once the MSSM superpotential (\ref{SMW}) has been specified, the effective
potential is also fixed:
\begin{equation}
V \; = \; \Sigma_i |F^i|^2 \; + \; 
{1 \over 2} \Sigma_a (D^a)^2: \; \; 
F^*_i \equiv {\partial W \over \partial \phi^i}, \; 
D^a \equiv g_a \phi^*_i (T^a)^i_j \phi^j,
\label{SMV}
\end{equation}
according to the rules explained earlier in this Lecture, where the sums
run over the different chiral fields $i$ and the $SU(3), SU(2)$ and $U(1)$
gauge-group factors $a$.

There are important possible variations on the MSSM superpotential
(\ref{SMW}), which are impossible in the Standard Model, but are allowed
by the gauge symmetries of the MSSM supermultiplets. These are additional
superpotential terms that violate the quantity known as $R$ parity:
\begin{equation}
R \; \equiv \; (-1)^{3 B + L + 2 S},
\label{Rparity}
\end{equation}
where $B$ is baryon number, $L$ is lepton number, and $S$ is spin. It is
easy to check that $R = + 1$ for all the particles in the Standard Model,
and $R = - 1$ for all their spartners, which have identical values of $B$
and $L$, but differ in spin by half a unit. Clearly, $R$ would be
conserved if both $B$ and $L$ were conserved, but this is not automatic.
Consider the following superpotential terms:
\begin{equation}
\lambda_{ijk} L_i L_j E^c_k \; + \; \lambda^\prime_{ijk} L_i Q_j D^c_k \; 
+ \; \lambda^{\prime\prime}_{ijk} U^c_i D^c_j D^c_k \; + \; \epsilon_i H L_i,
\label{RVW}
\end{equation}
which are visibly $SU(3) \times SU(2) \times U(1)$ symmetric. The first
term in (\ref{RVW}) would violate $L$, causing for example ${\tilde \ell}
\to \ell + \ell$, the second would violate both $B$ and $L$, causing for
example ${\tilde q} \to q + \ell$, the third would violate $B$, causing
for example ${\tilde q} \to {\bar q} + {\bar q}$, and the last would
violate $L$ by causing $H \leftrightarrow L_i$ mixing. These interactions
would provide many exciting signatures for supersymmetry, such as dilepton 
events, jets plus leptons and multijet events. Such interactions are 
constrained by
direct searches, by the experimental limits on flavour-changing
interactions and other rare processes, and by cosmology: they would tend
to wipe out the baryon asymmetry of the Universe if they are too 
strong~\cite{CDEO}.
They would also cause the lightest supersymmetric particle to be unstable,
not necessarily a disaster in itself, but it would remove an excellent
candidate for the cold dark matter that apparently abounds throughout the
Universe. For simplicity, the conservation of $R$ parity will be assumed
in the rest of these Lectures.

We now look briefly at the construction of supersymmetric GUTs, of which
the minimal version is based on the group $SU(5)$~\cite{DG}. As in the
transition from the Standard Model to the MSSM, one simply extends the
conventional GUT multiplets to supermultiplets, so that matter particles
are assigned to ${\mathbf {\bar 5}}$ representations ${\bar F}$ and
${\mathbf {10}}$ representations $ T$, one doubles the electroweak Higgs
fields to include both ${ H, {\bar H}}$ in ${\mathbf 5}, \mathbf {\bar 
5}$
representations, and one postulates a ${\mathbf {24}}$ representation
${\Phi}$ to break the $SU(5)$ GUT symmetry down to $SU(3) \times SU(2)
\times U(1)$. The superpotential for the Higgs sector takes the general
form
\begin{equation}
W_5 \; = \; (\mu + {3 \over 2} \lambda M) H {\bar H} \; + \; \lambda H
\Phi {\bar H} \; + \; f(\Phi).
\label{W5}
\end{equation}
Here, $f(\Phi)$ is chosen so that the vacuum expectation value of $\Phi$
has the form
\begin{equation}
< 0 | \Phi | 0 > \; = \; M \times {\rm diag}(1,1,1,-{3 \over 2},-{3 \over 
2}).
\label{Phivev}
\end{equation}
The coefficient of the $H {\bar H}$ term has been chosen so that it almost
cancels with the term $\propto H < 0 | \Phi | 0 > {\bar H}$ coming from
the second term in (\ref{W5}), {\it for the last two components}. In this
way, the triplet components of $H, {\bar H}$ acquire large masses $\propto
M$, whilst the last two may acquire a vacuum expectation value: 
$<~0~|~H~|~0~> = {\rm column}(0,0,0,0,v), < 0 | {\bar H} | 0 > = {\rm 
column}(0,0,0,0,{\bar
v})$, once supersymmetry breaking and radiative corrections are taken into
account, as in the next Lecture.

In order that $v, {\bar v} \sim 100$~GeV, it is necessary that the
residual $H {\bar H}$ mixing term $\mu \lappeq 1$~TeV. Since, as we recall
shortly, $M \sim 10^{16}$~GeV, this means that the parameters of $W_5$
(\ref{W5}) must be tuned finely to one part in $10^{13}$. This fine-tuning
may appear very unreasonable, but it is technically natural, in the sense
that there are no big radiative corrections. Thanks to the supersymmetric
no-renormalization theorem for superpotential parameters, we know that
$\delta \lambda, \delta \mu = 0$, apart from wave-function renormalization
factors. Thus, if we adjust the input parameters of (\ref{W5}) so that
$\mu$ is small, it will stay small. However, this begs the more profound
question: how did $\mu$ get to be so small in the first place?

As already mentioned, a striking piece of circumstantial evidence in 
favour of the idea of
supersymmetric grand unification is provided by the measurements of
low-energy gauge couplings at LEP and elsewhere~\cite{GUTs}. The three 
gauge couplings
of the Standard Model are renormalized as follows:
\begin{equation}
{d g^2_a \over d t} \; = \; b_a {g_a^4 \over 16 \pi^2} + \dots,
\label{rengp}
\end{equation}
at one-loop order,
and the corresponding value of the electroweak mixing angle $\sin^2
\theta_W (m_Z)$ is given at the one-loop level by:
\begin{equation}
\sin^2 \theta_W (m_Z) \; = \; {g^{\prime^2} \over g_2^2 + g^{\prime^2}} \\
\; = \; {3 \over 5} {g^2_1 (m_Z) \over g_2^2(m_Z) + {3 \over 5} g^2_1 
(m_Z)} \\
\; = \; {1 \over 1 + 8 x} [ 3 x + {\alpha_{em}(m_Z) \over \alpha_3 (m_Z)}],
\label{sin2}
\end{equation}
where
\begin{equation}
x \; \equiv \; {1 \over 5} ( {b_2 - b_3 \over b_1 - b_2}).
\label{defx}
\end{equation}
One can distinguish the predictions of different GUTs by their different
values of the renormalization coefficients $b_i$, which are in turn
determined by the spectra of light particles around the electroweak scale.
In the cases of the Standard Model and the MSSM, these are:
\begin{eqnarray}
{4 \over 3} N_G - 11 \; \leftarrow & b_3 & \rightarrow  \; 2 N_G - 9 \;  
= \; - 3 \\
{1 \over 6} N_H + {4 \over 3} N_G - {22 \over 3} \; \leftarrow & b_2 & 
\rightarrow \; {1 \over 2} N_H + 2 N_G - 6 \; = \; + 1  \\
{1 \over 10} N_H + {4\over 3} N_G \; \leftarrow & b_1 & \rightarrow  \; {3 
\over 10} N_H + 2 N_G \; = \; {33 \over 5} \\
{23 \over 218} = 0.1055 \; \leftarrow & x & \rightarrow \; {1 \over 7} \;  .
\label{SMvsMSSM}
\end{eqnarray}
If we insert the best available values of the gauge couplings:
\begin{equation}
\alpha_{em} \; = \; {1 \over 128}; \; \alpha_3 (m_Z) \; 
= \; 0.119 \pm 0.003, \; \sin^2 \theta_W(m_Z) \; = \; 0.2315,
\label{alphas}
\end{equation}
we find the following value:
\begin{equation}
x \; = \; {1 \over 6.92 \pm 0.07}.
\label{valuex}
\end{equation}
We see that experiment strongly favours the inclusion of supersymmetric
particles in the renormalization-group equations, as required if the
effective low-energy theory is the MSSM (\ref{SMvsMSSM}), as in a simple
supersymmetric GUT such as the minimal $SU(5)$ model introduced above.

\section{TOWARDS REALISTIC MODELS}

\subsection{Supersymmetry Breaking}

This is clearly necessary: $m_e \ne m_{\tilde e}, m_\gamma \ne m_{\tilde
\gamma}$, etc. The Big Issue is whether the breaking of supersymmetry is
explicit, i.e., present already in the underlying Lagrangian of the
theory, or whether it is spontaneous, i.e., induced by a
non-supersymmetric vacuum state. There are in fact several reasons to
disfavour explicit supersymmetry breaking. It is ugly, it would be unlike
the way in which gauge symmetry is broken, and it would lead to
inconsistencies in supergravity theory.  For these reasons, theorists have
focused on spontaneous supersymmetry breaking.

If the vacuum is not to be supersymmetric, there must be some 
fermionic state $\chi$ that is coupled to the vacuum by the supersymmetry 
charge $Q$:
\begin{equation}
< 0 | Q | \chi > \; \equiv \; f_\chi^2 \; \ne \; 0.
\label{Goldstino}
\end{equation}
The fermion $\chi$ corresponds to a Goldstone boson in a spontaneously 
broken bosonic symmetry, and therefore is often termed a Goldstone fermion 
or a Goldstino.

There is just one small problem in globally supersymmetric models, i.e., 
those without gravity: spontaneous supersymmetry breaking necessarily 
entails a positive vacuum energy $E_0$. To see this, consider the vacuum 
expectation value of the basic supersymmetry anticommutator:
\begin{equation}
\{ Q, Q \} \propto \gamma_\mu P^\mu.
\label{anticomm}
\end{equation}
According to (\ref{Goldstino}), there is an intermediate state $\chi$, so 
that
\begin{equation}
< 0 | \{ Q, Q \} | 0 > \; = \; | < 0 | Q | \chi > |^2 = f_\chi^4 \; 
\propto \; < 0 | P_0 | 0 > = E_0,
\label{vacuumenergy}
\end{equation}
where we have used Lorentz invariance to set the spatial components $< 0 | 
P_i | 0 > = 0$. 
Spontaneous breaking of global supersymmetry (\ref{Goldstino}) requires
\begin{equation}
E_0 \; = \; f_\chi^4 \; \ne \; 0.
\label{oops}
\end{equation}
The next question is how to generate non-zero vacuum energy. Hints are 
provided by the effective potential in a globally supersymmetric theory:
\begin{equation}
V \; = \; \Sigma_i | {\partial W \over \partial \phi^i} |^2 \; 
+ \; {1 \over 2} \Sigma_\alpha g^2_\alpha | \phi^* T^\alpha \phi |^2.
\label{effpot}
\end{equation}
It is apparent from this expression that either the first `$F$ term' or 
the second `$D$ term' must be positive definite.

The option $D > 0$ requires constructing a model with a $U(1)$ gauge 
symmetry~\cite{FI}. The simplest example contains just one chiral (matter) 
supermultiplet with unit charge, for which the effective potential is:
\begin{equation}
V_D \; = \; {1 \over 2} ( \xi + g \phi^* \phi )^2.
\label{FIV}
\end{equation}
the extra constant term $\xi$ is not allowed in a non-Abelian theory, 
which is why one must use a $U(1)$ theory. We see immediately that the 
minimum of the effective potential (\ref{FIV}) is reached when $< 0 | \phi 
| 0 > = 0$, in which case $V_F = 1/2 \xi^2 > 0$ and supersymmetry is broken 
spontaneously. Indeed, it is easy to check that, in this vacuum:
\begin{equation}
m_\phi \; = \; g \xi, \; m_\psi \; = \; 0, \; m_V \; = \; m_{\tilde 
V} 
\; = \; 0,
\label{FIM}
\end{equation}
exhibiting explicitly the boson-fermion mass splitting in the $(\phi, 
\psi)$
supermultiplet. Unfortunately, this example cannot be implemented with the
$U(1)$ of electromagnetism in the Standard Model, because there are fields
with both signs of the hypercharge $Y$, enabling $V_D$ to vanish. So, one
needs a new $U(1)$ gauge group factor, and many new fields in order to
cancel triangle anomalies. For these reasons, $D$-breaking models did not
attract much attention for quite some time, though they have had a revival
in the context of string theory~\cite{DSW}.

The option $F > 0$ also requires additional chiral (matter) fields with 
somewhat `artificial' couplings~\cite{FO}: again, those of the Standard 
Model do not 
suffice. The simplest example uses three chiral supermultiplets $A, B, C$ 
with the superpotential
\begin{equation}
W \; = \; \alpha A B^2 \; + \; \beta C (B^2 - m^2).
\label{ORW}
\end{equation}
using the rules given in the previous Lecture, it is easy to calculate the 
corresponding $F$ terms:
\begin{equation}
F_A \; = \; \alpha B^2, \; F_B \; = \; 2 B (\alpha A + \beta C), \; F_C \; 
= \; \beta (B^2 - m^2),
\label{ORF}
\end{equation}
and hence the effective potential
\begin{equation}
V_F \; = \; \Sigma_i | F_i |^2 \; =
\; 4 |B(\alpha A + \beta C)|^2 \; + \; | \alpha B^2|^2 \; + 
\; | \beta (B^2 - m^2) |^2.
\label{ORV}
\end{equation}
Likewise, it is not difficult to check that the three different 
positive-semidefinite terms in (\ref{ORV}) cannot all vanish 
simultaneously. Hence, necessarily $V_F > 0$, and hence supersymmetry {\it 
must} be broken.

The principal outcome of this brief discussion is that there are no
satisfactory models of global superysmmetry breaking, so we will now look 
at the options in local supersymmetry, i.e., supergravity theory.

\subsection{Supergravity and Local Supersymmetry Breaking}

So far, we have considered global supersymmetry transformations, in which 
the infinitesimal transformation spinor $E$ is constant throughout space. 
Now we consider the possibility of a space-time-dependent field $E(x)$. 
Why?

This step of making symmetries local has become familiar with bosonic
symmetries, where it leads to gauge theories, so it is natural to try the 
analogous step with fermionic symmetries. Moreover, as we see shortly, it 
leads to an elegant mechanism for spontaneous supersymmetry breaking, 
again by analogy with gauge theories, the super-Higgs mechanism. Further, 
as we also see shortly, making supersymmetry local necessarily involves 
gravity, and even opens the prospect of unifying all the particle 
interactions and matter fields with extended supersymmetry 
transformations:
\begin{equation}
G (J = 2 ) \; \to \; {\tilde G} ( J = 3/2 ) \; \to \; V ( J = 1 ) \; \to 
\; q, \ell ( J = 
1/2 ) \; \to \; H ( J = 0 )
\label{TOE}
\end{equation}
in supergravity with $N > 1$ supercharges. In (\ref{TOE}), $G$ 
denotes the gravition, and ${\tilde G}$ the spin-3/2 
gravitino, which accompanies it in the graviton 
supermultiplet:
\begin{eqnarray}
\pmatrix{ G\cr
\tilde G} = 
\pmatrix{ 2 \cr 
{3\over 2}}
\label{Gsmultiplet}
\end{eqnarray}
Supergravity is in any case an 
essential ingredient in the discussion of gravitational interactions of 
supersymmetric particles, needed, for example, for any meaningful 
discussion of the cosmological constant.

To gain insight into why making supersymmetry local {\it necessarily} 
involves gravity, consider what happens if one applies both of the 
following pair of supersymmetry transformations:
\begin{eqnarray}
\delta_i \phi \; & = & \; \sqrt{2} {\bar E_i} \psi + \dots, \\
\delta_j \psi \; & = & \; - i \sqrt{2} \gamma_\mu \partial^\mu \phi E_j + 
\dots
\label{bothS}
\end{eqnarray}
on either of the fields $\phi, \psi$. One finds in each case:
\begin{equation}
[\delta_i, \delta_j] (\phi, \psi) = - 2 ({\bar E_j} \gamma_\mu E_j) i 
\partial_\mu (\phi, \psi).
\label{deconstruct}
\end{equation}
In either case, the effect is equivalent to a space-time translation, 
since $i \partial_\mu \leftrightarrow P_\mu$. Clearly, if the 
infinitesimal spinorial transformations $E_{i,j}$ are independent of $x$, 
this 
translation is global. However, if the $E_{i,j}$ depend on $x$, the net 
effect is equivalent to a local coordinate transformation, and we know 
that a theory invariant under these necessarily includes gravity.

To see further why gravity must be taken into account when making 
supersymmetry 
local, let us develop further the analogy with gauge theories. We consider 
the variation of a typical fermion kinetic term: $\delta ( i {\bar \psi}
\gamma^\mu \partial_\mu \psi )$. In a gauge theory, one makes a 
space-time-dependent phase transformation $\epsilon (x)$:
\begin{equation}
\psi (x) \; \to \; e^{i \epsilon (x)} \psi (x), 
\label{gauge}
\end{equation}
which leads to a term in the variation of the fermion kinetic term of the 
form:
\begin{equation}
- {\bar \psi} \gamma_\mu \psi \partial^\mu \epsilon (x).
\label{newvariation}
\end{equation}
This is cancelled in a gauge theory by the variation in the gauge 
interaction:
\begin{equation}
{\bar \psi} (x) \gamma_\mu \psi (x) A^\mu (x): \; \delta A^\mu (x) \; = \; 
\partial^\mu \epsilon (x).
\label{gaugecancel}
\end{equation}
In the supersymmetric case, when the supersymmetric variation 
becomes local:
\begin{equation}
\delta \psi (x) = - i \gamma_\mu \partial^\mu ( \phi (x) E 
(x)) + \dots, 
\label{susyvariation}
\end{equation}
the variation in the fermionic kinetic term includes a 
piece
\begin{equation}
\propto \; {\bar \psi} \gamma_\mu \gamma_\nu \partial^\nu \phi 
\partial^\mu E 
(x),
\label{Snewvariation}
\end{equation}
which is cancelled by introducing a new field $\psi_\mu$ 
with coupling
\begin{equation}
\kappa {\bar \psi} \gamma_\mu \gamma_\nu \partial^\nu \phi \psi_\mu (x): 
\; \delta \psi_\mu (x) \; = \; - {2 \over \kappa} \partial_\mu E (x).
\label{Sgaugecancel}
\end{equation}
The new field $\psi^\mu (x)$ may be regarded as a `gauge fermion': it 
represents the gravitino.

OK, so now we are convinced that making supersymmetry local necessarily 
involves gravity, and the analogy with gauge theory suggests the 
introduction of a gravitino field. Consider now the simplest possible 
Lagrangian for a gravitino and graviton~\cite{sugra}, which consists just 
of the 
Einstein Lagrangian for general relativity and a Rarita-Schwinger 
Lagrangian for a spin-3/2 field, made suitably invariant under general 
coordinate transformations by minimal substitution:
\begin{equation}
L \; = \; - {1 \over 2 \kappa^2} \sqrt{-g} R \; - \; {1 \over 2} 
\epsilon^{\mu \nu \rho \sigma} {\bar \psi_\mu} \gamma_5 \gamma_\nu {\cal 
D}_\rho \psi_\sigma,
\label{ERS}
\end{equation}
where $g \equiv det(g_{\mu \nu})$ with $g_{\mu \nu}$ the metric 
tensor:
\begin{equation}
g_{\mu \nu} \; \equiv \; \epsilon^m_\mu \epsilon^n_\nu \eta_{mn}
\label{vierbein}
\end{equation}
where $\epsilon^m_\mu$ is the vierbein, and
\begin{equation}
{\cal D}_\rho \; \equiv \; \partial_\rho + {1 \over 4} \omega^{mn}_\rho 
[\gamma_m, \gamma_n],
\label{GCD}
\end{equation}
with $\omega^{mn}_\rho$ the spin connection, is the generally-covariant 
derivative. It is a remarkable fact that the simple Lagrangian (\ref{ERS}) 
is locally supersymmetric~\cite{sugra}. To check this invariance, you need 
the 
following local supersymmetry transformation laws:
\begin{eqnarray}
\delta \epsilon^m_\mu \; & = & \; {\bar E} (x) \gamma^m \psi_\mu (x), \\
\delta \omega^{mn}_\mu \; & = & \; 0, \\
\delta \psi_\mu \; & = & \; {1 \over \kappa} {\cal D}_\mu E (x).
\label{localsusy}
\end{eqnarray}
Once again, we have been speaking prose all our lives without realizing! 

We shall discuss later the coupling of supergravity to matter. First, 
however, now is a good time to mention the remarkable phenomenon of 
spontaneous breaking of local supersymmetry: the super-Higgs 
effect~\cite{Polonyi, Fetal}. You 
recall that, in the conventional Higgs effect in spontaneously broken 
gauge theories, a massless Goldstone boson is `eaten' by a gauge boson to 
provide it with the third polarization state it needs to become massive: 
\begin{equation}
(2 \times V_{m = 0}) \; + \; (1 \times GB) \; = \; (3 \times V_{m \ne 0}).
\label{Higgs}
\end{equation}
In a locally supersymmetric theory, the two polarization states of the 
massless Goldstone fermion (Goldstino) are `eaten' by a massless 
gravitino, giving it the total of four polarization states it needs to 
become massive:
\begin{equation}
(2 \times \psi^\mu_{m = 0}) \; + \; (2 \times GF) \; = \; (4 \times 
\psi^\mu_{m \ne 0}).
\label{SHiggs}   
\end{equation}
This process clearly involves the breakdown of local 
supersymmetry, since the end result is to give the gravitino a different 
mass from the graviton: $m_G = 0 \ne m_{\tilde G} \ne 0$. It is indeed the 
only known consistent way of breaking local supersymmetry, just as the 
Higgs mechanism is the only consistent way of breaking gauge symmetry. We 
shall not go here through all the details of the super-Higgs effect, but 
there is one noteworthy feature: this local breaking of supersymmetry can 
be achieved with zero vacuum energy:
\begin{equation}
< 0 | V | 0 > \; = \; 0 \; \leftrightarrow \; \Lambda \; = \; 0.
\label{zerovacen}
\end{equation}
As we discuss shortly in more detail, there is no inconsistency between 
local supersymmetry breaking and a 
vanishing cosmological constant $\Lambda$, unlike the case of global 
superymmetry breaking that we discussed earlier.

\subsection{Effective Low-Energy Theory}

The coupling of matter particles to supergravity is more complicated than 
the globally supersymmetric case discussed in the previous lecture. 
Therefore, it is not developed here in detail. Instead, a few key results 
are presented without proof, and then we study the general form of the 
effective low-energy theory~\cite{LEtheory} derivable from a supergravity 
theory.

The superpotential of global supersymmetry is upgraded in $N = 
1$ supergravity to a `K\"ahler potential', which describes the geometry of 
the internal space parameterized by the scalar fields in the chiral 
supermultiplets. This K\"ahler potential is a Hermitean function $G(\phi, 
\phi^*)$ of the chiral fields and their complex conjugates, and plays 
several r\^oles. It is an order parameter for supersymmetry breaking:
\begin{equation}
m_{\tilde G} \; \equiv \; m_{3/2} = e^{G\over 2}.
\label{Kgravitino}
\end{equation}
It also determines the kinetic terms for the chiral fields:
\begin{equation}
L_K \; = \; G^j_i \partial^\mu \phi^*_j \partial_\mu \phi^i
\label{Kkinetic}
\end{equation}
via the K\"ahler metric
\begin{equation}
G^j_i \; \equiv \; {\partial^2 G \over \partial \phi^i \partial_j^*},
\label{Kmetric}
\end{equation}
as well as the effective potential:
\begin{equation}
V \; = \; e^G [ G_i (G'')^{-1 i}_j G^j - 3]: \; \; G_i \; \equiv \; 
{\partial G \over \partial \phi^i}.
\label{Kpotential}
\end{equation}
The first term in (\ref{Kpotential}) corresponds to the effective 
potential in a globally supersymmetric theory. 

The second term in (\ref{Kpotential}) is novel: it permits the
reconciliation of supersymmetry breaking: $m_{\tilde G}^2 = e^G \ne 0$
with a vanishing effective potential: $V = 0$, as a result of a
cancellation between the first and second terms in (\ref{Kpotential}).  
This is certainly good news, but there is also bad news. For general forms
of $G$, there are certain values of the fields where $V$ is {\it
negative}, with values $- {\cal O}(m_P^4)$. This would be a catastrophe
for cosmology, since our Universe would surely fall down one of these
holes. Fortunately, there is a particular class of K\"ahler potentials,
known as no-scale supergravities~\cite{noscale}, where the effective
potential is positive semidefinite. Fortunately but not fortuitously, this
is the class of supergravity that emerges from string theory~\cite{W}.

Just as the K\"ahler potential determines the geometry and kinetic terms
for chiral fields, there is a corresponding function that describes the 
geometry and determines the kinetic terms for gauge fields. For generic 
choices of this function, there are also non-vanishing 
supersymmetry-breaking masses for the gauginos:
\begin{equation}
m_{1/2} \; \propto \; m_{\tilde G} \; \equiv \; m_{3/2}.
\label{gauginomass}
\end{equation}
It is not inevitable that the masses of the $SU(3), SU(2)$ and $U(1)$ 
gauginos be universal, but this emerges naturally if the geometry is not 
too complicated. 

Expanding the effective potential (\ref{Kpotential}), one also finds in
general terms $\propto |\phi|^2$, that are interpreted as 
supersymmetry-breaking scalar masses:
\begin{equation}
m_{0} \; \propto \; m_{\tilde G} \; \equiv \; m_{3/2}.
\label{scalarmass}
\end{equation}
In this case, there is no particularly good theoretical motivation for 
universality: indeed, this is broken in many string models. There are, 
however, phenomenological reasons to think that the supersymmetry-breaking 
scalar masses for sparticles with the same charge, e.g., all the sleptons, 
should be universal, in order to suppress flavour-changing neutral 
interactions mediated by virtual sparticles~\cite{FCNI}.

Generic forms of the effective potential (\ref{Kpotential}) also yield 
trilinear supersymmetry-breaking interactions among the scalar particles:
\begin{equation}
A_\lambda \lambda \phi^3: \; \; A_\lambda \; \propto \; m_{\tilde G} \; 
\equiv \; m_{3/2},
\label{Aterm}
\end{equation}
where here $\phi$ denotes a scalar component of a generic chiral 
supermultiplet.
If the supersymmetric theory also includes bilinear interactions $\mu 
\phi^2$, as is 
the case in the minimal supersymmetric extension of the Standard Model 
(MSSM), one also expects an analogous bilinear supersymmetry-breaking term
$B_\mu \mu \phi^2$ among the scalar components.

Thus, the final form of the effective low-energy theory suggested by 
spontaneous supersymmetry breaking in supergravity is:
\begin{equation}
- {1 \over 2} \Sigma_a m_{1/2_a} {\tilde V}_\alpha {\tilde 
V}_\alpha - \Sigma_i m^2_{0_i} |\phi^i|^2 - (\Sigma_\lambda A_\lambda 
\lambda \phi^3 + \Sigma_\mu B_\mu \mu \phi^2 + {\rm Herm. Conj.}), 
\label{efftheory}
\end{equation}
which contains many free parameters and phases. The breaking of 
supersymmetry in the effective low-energy theory (\ref{efftheory}) is 
explicit but `soft', in the sense that the renormalization of the 
parameters $m_{1/2_a}, m_{0_i}, A_\lambda$ and $B_\mu$ is 
logarithmic. Of course, these parameters are not considered to be 
fundamental, and the underlying mechanism of supersymmetry breaking is 
thought to be spontaneous, for the reasons described at the beginning of 
this lecture.

The logarithmic renormalization of the parameters means that one can
calculate their low-energy values in terms of high-energy inputs from a
supergravity or superstring theory, using standard renormalization-group
equations~\cite{RGEs}. In the case of the low-energy gaugino masses $M_a$, 
the 
renormalization is 
multiplicative and identical with that of the corresponding gauge coupling 
$\alpha_a$ at the one-loop level:
\begin{equation}
{M_a \over m_{1/2_a}} \; = \; {\alpha_a \over \alpha_{GUT}}
\label{m12renn}
\end{equation}
where we assume GUT unification of the gauge couplings at the input 
supergravity scale. In the case of the scalar masses, there is both 
multiplicative renormalization and renormalization related to the gaugino 
masses:
\begin{equation}
{ \partial m^2_{0_i} \over \partial t} \; = \; {1 \over 16 \pi^2} [ 
\lambda^2 (m_0^2 + A_\lambda^2) - g_a^2 M_a^2]
\label{m0renn}
\end{equation}
at the one-loop level, where $t \equiv {\rm ln} ( Q^2 / m_{GUT}^2)$, and 
the ${\cal O}(1)$ group-theoretical 
coefficients have been omitted. In the case of the first two generations, 
the first terms in (\ref{m0renn}) are negligible, and one may integrate 
(\ref{m0renn}) trivially to obtain effective low-energy parameters
\begin{equation}
m^2_{0_i} \; = \; m_0^2 + C_i m_{1/2}^2,
\label{m0rend}
\end{equation}
where universal inputs are assumed, and the coefficients $C_i$ are 
calculable in any given model. The first terms in 
(\ref{m0renn}) are, 
however, important for the third generation and for the Higgs bosons of 
the MSSM, as we now see.

Notice that the signs of the first terms in (\ref{m0renn}) are positive,
and that of the last term negative. This means that the last term tends to
{\it increase} $m^2_{0_i}$ as the renormalization scale $Q$ {\it
decreases}, an effect seen in Fig.~\ref{fig:EWSB}. The positive signs of
the first terms mean that they tend to {\it decrease} $m^2_{0_i}$ as 
$Q$ {\it
decreases}, an effect seen for a Higgs squared-mass in 
Fig.~\ref{fig:EWSB}. Specifically, the negative effect on $H_u$ seen in 
Fig.~\ref{fig:EWSB} is due to its large Yukawa coupling to the 
$t$ quark: $\lambda_t \sim g_{2,3}$. The exciting aspect of this 
observation is that spontaneous electroweak symmetry breaking is 
possible~\cite{RGEs} 
when $m_H^2 (Q) < 0$, as occurs in Fig.~\ref{fig:EWSB}. Thus the 
spontaneous breaking of supersymmetry, which normally provides $m_0^2 > 
0$, and renormalization, which then drive $m_H^2 (Q) < 0$, conspire to 
make spontaneous electroweak symmetry breaking possible. Typically, this 
occurs at a renormalization scale that is exponentially smaller than the 
input supergravity scale:
\begin{equation}
{m_W \over m_P} \; = \; exp( { - {\cal O}(1) \over \alpha_t}): \; \; 
\alpha_t \equiv {\lambda_t^2 \over 4 \pi}.
\label{EWSBhierarchy}
\end{equation}
Typical dynamical calculations find that $m_W \sim 100$~GeV emerges 
naturally if $m_t \sim 60$ to $200$~GeV, and this was in fact one of the 
first suggestions that $m_t$ might be as high as was subsequently 
observed.

\begin{figure}%[t]
\centerline{\includegraphics[height=3in]{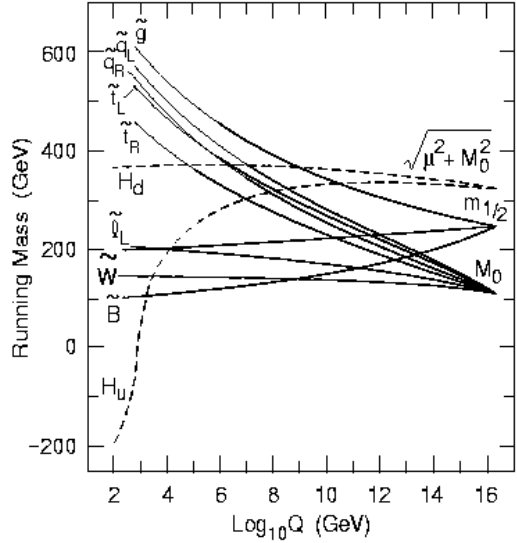}}
\caption[]{The renormalization-group evolution of the soft 
supersymmetry-breaking parameters in the MSSM, showing the 
increase in the squark and slepton masses as the renormalization 
scale decreases, whilst the Higgs squared-mass may become 
negative, triggering electroweak symmetry breaking.}
\label{fig:EWSB}
\end{figure}

To 
conclude 
this section, let us briefly review the 
reasons why soft
supersymmetry breaking might be universal, at least in some respects.  
There are important constraints on the mass differences of squarks and
sleptons with the same internal quantum numbers, coming from
flavour-changing neutral interactions~\cite{FCNI}. These are suppressed in 
the
Standard Model by the Glashow-Iliopoulos-Maiani mechanism~\cite{GIM}, 
which limits
them to magnitudes $\propto \Delta m_q^2 / m_W^2$ for small squared-mass
differences $\Delta m_q^2$. Depending on the process considered, it is
either necessary or desirable that sparticle exchange contributions, which
would have expected magnitudes $ \sim \Delta m_{\tilde q}^2 / m_{\tilde
q}^2$, be suppressed by a comparable factor. In particular, one would like
\begin{equation}
m_0^2 ( {\rm first~generation} ) - m_0^2 ( {\rm second~generation} ) 
\; \sim \; \delta m_q^2 \times {m_{\tilde q}^2 \over m_W^2}.
\label{SFCNC}
\end{equation}
The limits on third-generation sparticle masses from flavour-changing 
neutral interactions are less severe, and the first/second-generation 
degeneracy could be relaxed if $m_{\tilde q}^2 \gg m_W^2$, but models with 
physical values of $m_0^2$ degenerate to ${\cal O} (m_q^2)$ are certainly 
preferred. This is possible in models with a universal K\"ahler geometry 
for the scalar fields $\phi^i$. For example:
\begin{equation}
G \; = \; |\phi^i|^2 \; \rightarrow \; {\partial^2 G \over \partial \phi^i 
\partial \phi_j^*} \; = \; \delta^j_i,
\label{trivialK}
\end{equation}
resulting in universal $m_{0_i}^2$, and there are other examples such as 
certain no-scale modelds. However, this restriction is not respected in 
many low-energy effective theories derived from string models.

The desirability of degeneracy between sparticles of different generations
help encourage some people to study models in which this property would
emerge naturally, such as models of gauge-mediated supersymmetry breaking
or extra dimensions~\cite{A}. However, for the rest of these lectures we
shall mainly stick to familiar old supergravity.

\subsection{Sparticle Masses and Mixing}

We now progress to a more complete discussion of sparticle masses and
mixing. \\
~ \\
\noindent
{\bf Sfermions} : Each flavour of charged lepton or quark has both
left- and right-handed components $f_{L,R}$, and these have separate
spin-0 boson superpartners $\tilde f_{L,R}$. These have different
isospins $I = {1\over 2},~0$, but may mix as soon as the electroweak
gauge symmetry is broken. Thus, for each flavour we should consider a
$2\times 2$ mixing matrix for the 
$\tilde f_{L,R}$, which takes the following general form:
\beq
M^2_{\tilde f} \equiv \left( \matrix{m^2_{\tilde f_{LL}} & m^2_{\tilde
f_{LR}} \cr \cr m^2_{\tilde f_{LR}} & m^2_{\tilde f_{RR}}}\right)
\label{threeten}
\eeq
The diagonal terms may be written in the form
\beq
m^2_{\tilde f_{LL,RR}} = m^2_{\tilde f_{L,R}} + m^{D^2}_{\tilde
f_{L,R}} + m^2_f
\label{threeeleven}
\eeq
where $m_f$ is the mass of the corresponding fermion, $\tilde
m^2_{\tilde f_{L,R}}$ is the soft supersymmetry-breaking mass discussed
in the previous section, and $m^{D^2}_{\tilde f_{L,R}}$ is a
contribution due to the quartic $D$ terms in the effective potential:
\beq
m^{D^2}_{\tilde f_{L,R}} = m^2_Z~\cos 2\beta~~(I_3 + \sin^2\theta_WQ_{em})
\label{threetwelve}
\eeq
where the term $\propto I_3$ is non-zero only for the $\tilde f_L$.
Finally, the off-diagonal mixing term takes the general form
\beq
m^{2}_{\tilde f_{L,R}} = m_f \left(A_f +
\mu^{\tan\beta}_{\cot\beta}\right)~~{\rm for}~~f =
^{e,\mu,\tau,d,s,b}_{u,c,t}
\label{threethirteen}
\eeq
It is clear that $\tilde f_{L,R}$ mixing is likely to be important for
the $\tilde t$, and it may also be important for the $\tilde b_{L,R}$ and
$\tilde\tau_{L,R}$ if $\tan\beta$ is large. 

We also see from (\ref{threeeleven}) that the diagonal entries for the
$\tilde t_{L,R}$ would be different from those of the $\tilde u_{L,R}$ and
$\tilde c_{L,R}$, even if their soft supersymmetry-breaking masses were
universal, because of the $m^2_f$ contribution. In fact, we also expect
non-universal renormalization of $m^2_{\tilde t_{LL,RR}}$ (and also
$m^2_{\tilde b_{LL,RR}}$ and $m^2_{\tilde \tau_{LL,RR}}$ if $\tan\beta$ is
large), because of Yukawa effects analogous to those discussed in the
previous section for the renormalization of the soft Higgs masses. For
these reasons, the $\tilde t_{L,R}$ are not usually assumed to be
degenerate with the other squark flavours. Indeed, one of the $\tilde t$
could well be the lightest squark, perhaps even lighter than the $t$ quark
itself~\cite{ERu}. \\

\noindent
{\bf Charginos}: These are the supersymmetric partners of the
$W^\pm$ and $H^\pm$, which mix through a $2\times 2$ matrix
\beq
-{1\over 2} ~(\tilde W^-, \tilde H^-)~~M_C ~~\left(\matrix{\tilde
W^+\cr\tilde H^+}\right)~~+~~{\rm herm.conj.}
\label{threeforteen}
\eeq
where
\beq
M_C \equiv \left(\matrix{M_2 & \sqrt{2} m_W\sin\beta \cr \sqrt{2}
m_W\cos\beta & \mu}\right)
\label{threefifteen}
\eeq
Here $M_2$ is the unmixed $SU(2)$ gaugino mass and $\mu$ is the Higgs
mixing parameter introduced in (\ref{SMW}). Fig.~\ref{fig:lsp}
displays (among other lines to be discussed later) the contour
$m_{\chi^\pm}$ = 91 GeV for the lighter of the two chargino mass
eigenstates~\cite{EFGOS}. \\

\begin{figure}
\begin{center}
%\vskip 0.75in
\vspace*{-0.75in}
\hspace*{-.40in}
\begin{minipage}{8in}
\epsfig{file=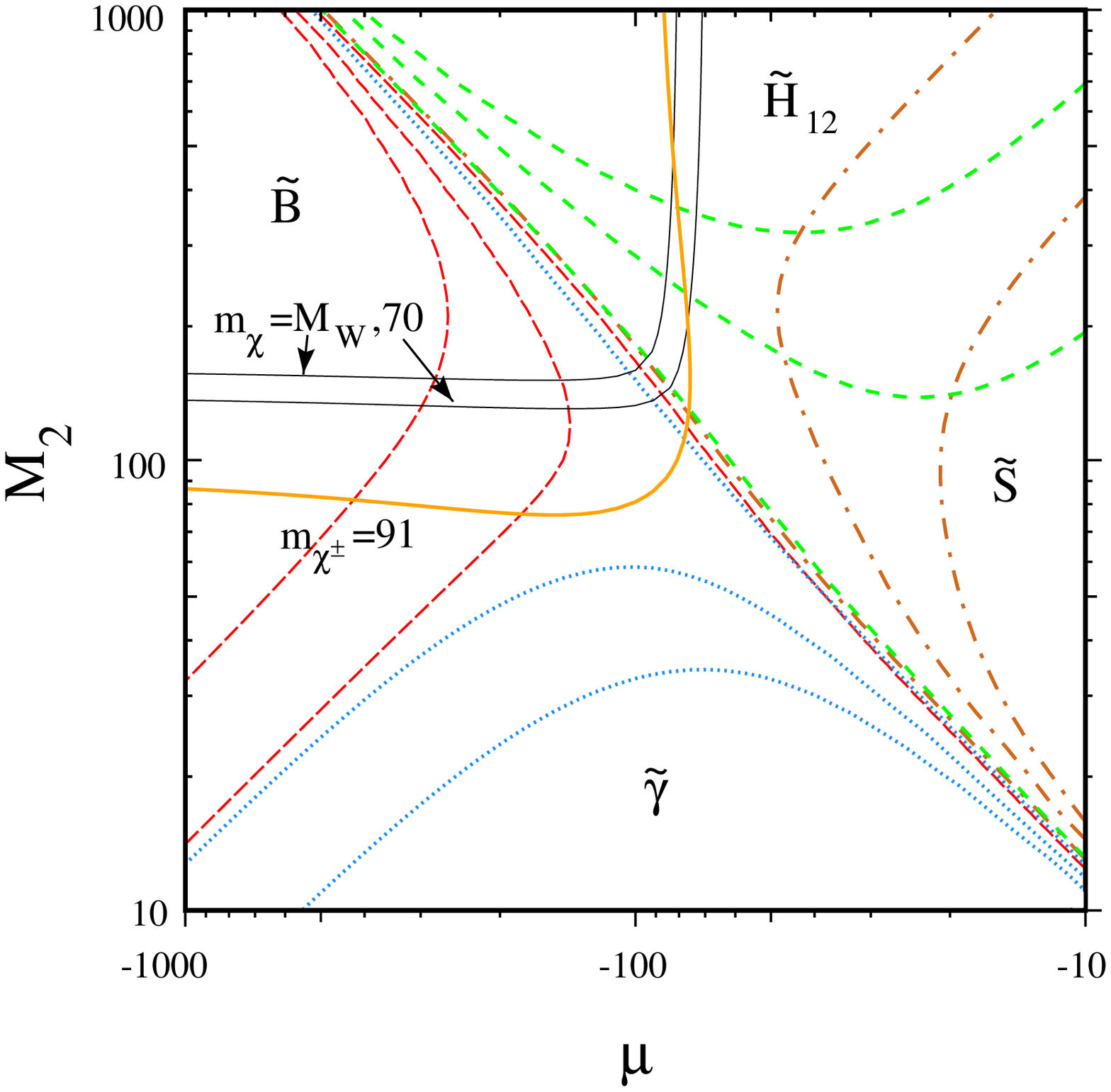,height=5in}
\hspace*{-0.17in}
\epsfig{file=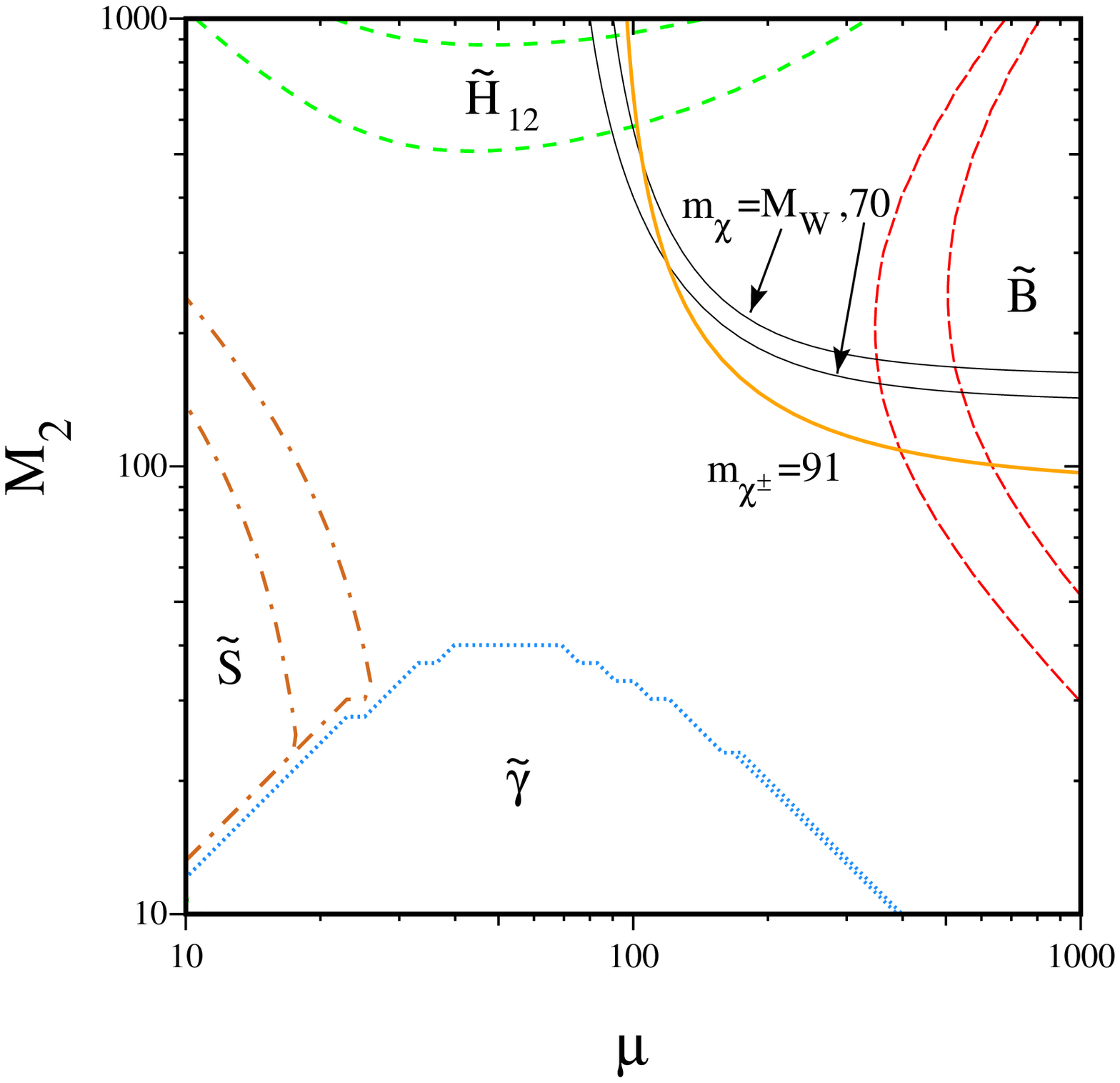,height=5in} \hfill
\end{minipage}
\end{center}
\vspace*{-0.75in}
\caption[]{
The $(\mu , M_2)$ plane characterizing charginos and neutralinos,
for (a) $\mu < 0$ and (b) $\mu > 0$, including contours of $m_\chi$ and
$m_{\chi^\pm}$, and of neutralino purity~\cite{EFGOS}.}
\label{fig:lsp}
\end{figure}

\noindent
{\bf Neutralinos}: These are characterized by a $4\times 4$ mass
mixing matrix~\cite{EHNOS}, which takes the following form in the $(\tilde
W^3, \tilde
B, \tilde H^0_2, \tilde H^0_1)$ basis :
\beq
m_N = \left( \matrix{
M_2 & 0 & {-g_2v_2\over\sqrt{2}} & {g_2v_1\over\sqrt{2}}\cr\cr
0 & M_1 & {g^\prime v_2\over\sqrt{2}} & {-g^\prime 
v_1\over\sqrt{2}}\cr\cr
{-g_2 v_2\over\sqrt{2}} & {g^\prime v_2\over\sqrt{2}} & 0 & \mu \cr\cr
{g_2v_1\over\sqrt{2}} & {-g^\prime v_1\over\sqrt{2}} & \mu & 0}\right)
\label{threesixteen}
\eeq
Note that this has a structure similar to $M_C$ (\ref{threefifteen}), but 
with its entries replaced by $2\times 2$ submatrices. As has already been
mentioned, one conventionally assumes that the $SU(2)$ and $U(1)$ gaugino
masses $M_{1,2}$ are universal at the GUT or supergravity scale, so that
\beq
M_1 \simeq M_2~~{\alpha_1\over\alpha_2}
\label{threeseventeen}
\eeq
so the relevant parameters of (\ref{threesixteen}) are generally taken to
be $M_2 = (\alpha_2/ \alpha_{GUT}) m_{1/2}$, $\mu$ and
$\tan\beta$.

Figure 20 also displays contours of the mass of the lightest neutralino
$\chi$, as well as contours of its gaugino and Higgsino
contents~\cite{EFGOS}. In the
limit $M_2\rightarrow 0$, $\chi$ would be approximately a photino and it
would be approximately a Higgsino in the limit $\mu\rightarrow 0$.
Unfortunately, these idealized limits are excluded by unsuccessful LEP
and other searches for neutralinos and charginos, as discussed in
more detail in the next Lecture.

\subsection{The Lightest Supersymmetric Particle}

This is expected to be stable in the MSSM,  and hence should be present in
the Universe today as a cosmological relic from the Big
Bang~\cite{Goldberg,EHNOS}. Its stability
arises because there is a multiplicatively-conserved quantum number called
$R$ parity, that takes the values +1 for all conventional particles and -1
for all sparticles~\cite{Fayet}. The conservation of $R$ parity can be
related to that of
baryon number $B$ and lepton number $L$, since
\beq
R = (-1)^{3B+L+2S}
\label{threeeighteen}
\eeq
where $S$ is the spin. Note that $R$ parity could be violated either
spontaneously if $<0\vert\tilde\nu\vert 0> \not= 0$ or
explicitly if one
of the supplementary couplings (\ref{RVW}) is
present. There
could also be a coupling $HL$, but this can be defined away be choosing a
field basis such that $\bar H$ is defined as the superfield with a
bilinear coupling to $H$. Note that $R$ parity is not violated by the
simplest models for neutrino masses, which have $\Delta L = 0, \pm 2$,
nor by simple GUTs, which violate
combinations of $B$ and $L$ that leave $R$ invariant.
There are three important consequences of $R$ conservation: 
\begin{enumerate}
\item sparticles are always produced in pairs, e.g., $\bar
pp\rightarrow\tilde q \tilde g X$, $e^+e^-\rightarrow \tilde\mu^+ +
\tilde\mu^-$, 
\item heavier sparticles decay to lighter ones, e.g., $\tilde q \rightarrow
q\tilde g, \tilde\mu\rightarrow\mu\tilde\gamma$, and 
\item the lightest sparticles is stable, 
\end{enumerate}
because it has no legal decay
mode.

This feature constrains strongly the possible nature of the lightest
supersymmetric sparticle. If it had either electric charge or strong
interactions, it would surely have dissipated its energy and condensed
into galactic disks along with conventional matter. There it would surely
have bound electromagnetically or via the strong interactions to
conventional nuclei, forming anomalous heavy isotopes that should have
been detected. There are upper limits on the possible abundances of such
bound relics, as compared to conventional nucleons:
\beq
{n({\rm relic})\over n(p)} \lappeq 10^{-15}~~{\rm to}~~10^{-29}
\label{threenineteen}
\eeq
for 1 GeV $\lappeq m_{\rm relic} \lappeq$ 1 TeV. These are far below the
calculated abundances of such stable relics:
\beq
{n({\rm relic})\over n(p)} \gappeq 10^{-6}~~(10^{-10})
\label{threetwenty}
\eeq
for relic particles with electromagnetic (strong) interactions. We may
conclude~\cite{EHNOS} that any supersymmetric relic is probably
electromagnetically neutral with only weak interactions, and could in
particular not be a gluino. Whether the lightest hadron containing
a gluino is charged or neutral, it would surely bind to some nuclei.
Even if one pleads for some level of fractionation, it is difficult
to see how such gluino nuclei could avoid the stringent bounds
established for anomalous isotopes of many species.

Plausible scandidates of different spins are the sneutrinos $\tilde\nu$ of
spin 0, the lightest neutralino $\chi$ of spin 1/2, and the gravitino
$\tilde G$ of spin 3/2. The sneutrinos have been ruled out by the
combination of LEP experiments and direct searches for cosmological
relics. The gravitino cannot be ruled out, but we concentrate on the
neutralino possibility for the rest of these Lectures.

A very attractive feature of the neutralino candidature for the lightest
supersymmetric particle is that it has a relic density of interest to
astrophysicists and cosmologists: $\Omega_\chi h^2 = {\cal O}(0.1)$ over
generic domains of the MSSM parameter space~\cite{EHNOS}, as discussed in
the next Lecture.  In these domains, the lightest neutralino $\chi$ could
constitute the cold dark matter favoured by theories of cosmological
structure formation.

\subsection{Supersymmetric Higgs Bosons}

As was discussed in Lecture 2, one expects two complex Higgs doublets
$H_2\equiv (H^+_2 , H^0_2)~,~~H_1\equiv (H^+_1 , H^0_1)$ in the MSSM,
with a total of 8 real degrees of freedom. Of these, 3 are eaten via the
Higgs mechanism to become the longitudinal polarization states of the
$W^\pm$ and $Z^0$, leaving 5 physical Higgs bosons to be discovered by
experiment. Three of these are neutral: the lighter CP-even neutral $h$,
the heavier CP-even neutral $H$, the CP-odd neutral $A$, and charged
bosons $H^\pm$. The quartic potential is completely determined by the $D$
terms
\beq
V_4 = {g^2 + g^{\prime 2}\over 8}~~\left( \vert H^0_1\vert^2 - \vert
H^0_2\vert^2 \right)
\label{threefour}
\eeq
for the neutral components, whilst the quadratic terms may be
parametrized at the tree level by
\beq
m^2_{H_1}~\vert H_1\vert^2 + m^2_{H_2}~\vert H_2\vert^2 +
(m^2_3~H_1H_2 + {\rm herm.conj.})
\label{threefive}
\eeq
where $m^2_3 = B_\mu\mu$. One combination of the three parameters
$(m^2_{H_1},m^2_{H_2},m^2_3)$ is fixed by the Higgs vacuum expectation $v
= \sqrt{v^2_1+v^2_2}$ = 246 GeV, and the other two combinations may be
rephrased as $(m_A,\tan\beta)$. These characterize all Higgs masses and
couplings in the MSSM at the tree level. Looking back at
(\ref{threefour}), we see that the gauge coupling strength of the
quartic interactions suggests a relatively low mass for
at least the lightest MSSM Higgs boson $h$, and this is indeed the case,
with $m_h \leq m_Z$ at the tree level:
\beq
m^2_h = m^2_Z~\cos^2 2\beta
\label{threesix}
\eeq
This raised considerable hope that the lightest MSSM Higgs boson could be
discovered at LEP, with its prospective reach to $m_H \sim$ 100 GeV.

However, radiative corrections to the Higgs masses are calculable in a
supersymmetric model (this was, in some sense, the whole point of
introducing supersymmetry!), and they turn out to be non-negligible for
$m_t \sim$ 175 GeV~\cite{susyHiggs}. Indeed, the leading one-loop
corrections to $m^2_h$ depend quartically on $m_t$:
\beq
\Delta m^2_h = {3m^4_t\over 4\pi^2v^2}~~\ln~~\left({m_{\tilde t_1} m_{\tilde t_2}\over
m^2_t}\right) + {3m^4_t \hat A^2_t\over 8\pi^2 v^2}~~\left[2h(m^2_{\tilde t_1}, m^2_{\tilde
t_2})+ \hat A^2_t ~~f(m^2_{\tilde t_1}, m^2_{\tilde t_2})\right] + \ldots
\label{threeseven}
\eeq
where $m_{\tilde t_{1,2}}$ are the physical masses of the two stop
squarks $\tilde t_{1,2}$ to be discussed in more detail shortly, $\hat A_t
\equiv A_t - \mu \cot\beta$, and
\beq
h(a,b) \equiv {1\over a-b}~\ln \left({a\over b}\right)~,~~f(a,b) =
{1\over (a-b)^2}~\left[2 - {a+b\over a-b}~\ln\left({a\over
b}\right)\right]
\label{threeeight}
\eeq
Non-leading one-loop corrections to the MSSM Higgs masses are also known,
as are corrections to coupling vertices, two-loop corrections and
renormalization-group resummations~\cite{FeynHiggs}. For $m_{\tilde
t_{1,2}} \lappeq$ 1
TeV and a plausible range of $A_t$, one finds
\beq
m_h \lappeq 130~{\rm GeV}
\label{threenine}
\eeq
as seen in Fig. 14. There we see the sensitivity of $m_h$ to $(m_A,
\tan\beta)$, and we also see how $m_A, m_H$ and $m_{H^\pm}$ approach each
other for large $m_A$.

\begin{figure}%[t]
\centerline{\includegraphics[height=3in]{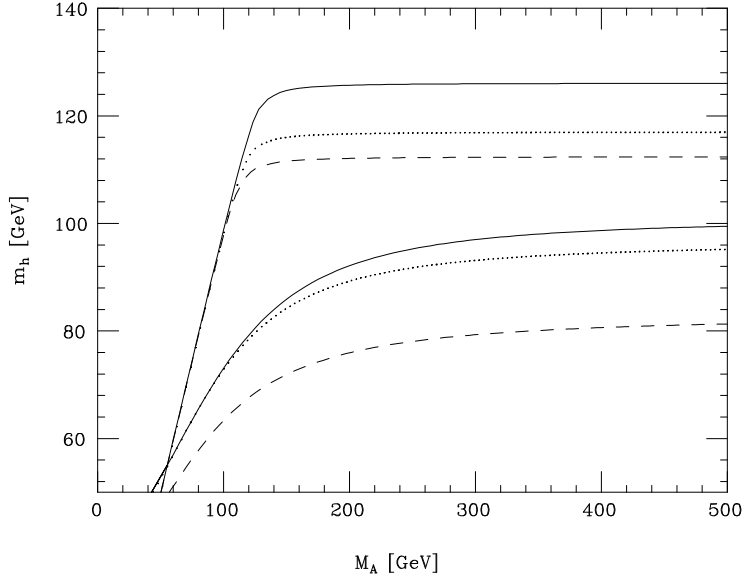}}
\caption[]{The lightest Higgs boson mass in the MSSM, for different values of
$\tan\beta$ and the CP-odd Higgs boson mass $M_A$.}
\label{fig:EllisScotfig14}
\end{figure}

\section{PHENOMENOLOGY}

\subsection{Constraints on the MSSM}

Important experimental constraints on the MSSM parameter space are
provided by direct searches at LEP and the Tevatron collider, as compiled 
in Fig.~\ref{fig:CMSSM}. One of these is the
limit $m_{\chi^\pm} \gappeq$ 103.5 GeV provided by chargino searches at 
LEP~\cite{LEPsusy},
where the third significant figure depends on other CMSSM parameters. LEP
has also provided lower limits on slepton masses, of which the strongest
is $m_{\tilde e}\gappeq$ 99 GeV \cite{LEPSUSYWG_0101}, again depending
only sightly on the other CMSSM parameters, as long as $m_{\tilde e} -
m_\chi \gappeq$ 10 GeV. The most important constraints on the $u, d, s,
c, b$ squarks and gluinos are provided by the Tevatron collider: for
equal masses
$m_{\tilde q} = m_{\tilde g} \gappeq$ 300 GeV. In the case of the $\tilde
t$, LEP provides the most stringent limit when $m_{\tilde t} - m_\chi$ is
small, and the Tevatron for larger $m_{\tilde t} - m_\chi$~\cite{LEPsusy}.

\begin{figure}
\vskip 0.5in
\vspace*{-0.75in}
%\hspace*{-.70in}
\begin{minipage}{8in}
\epsfig{file=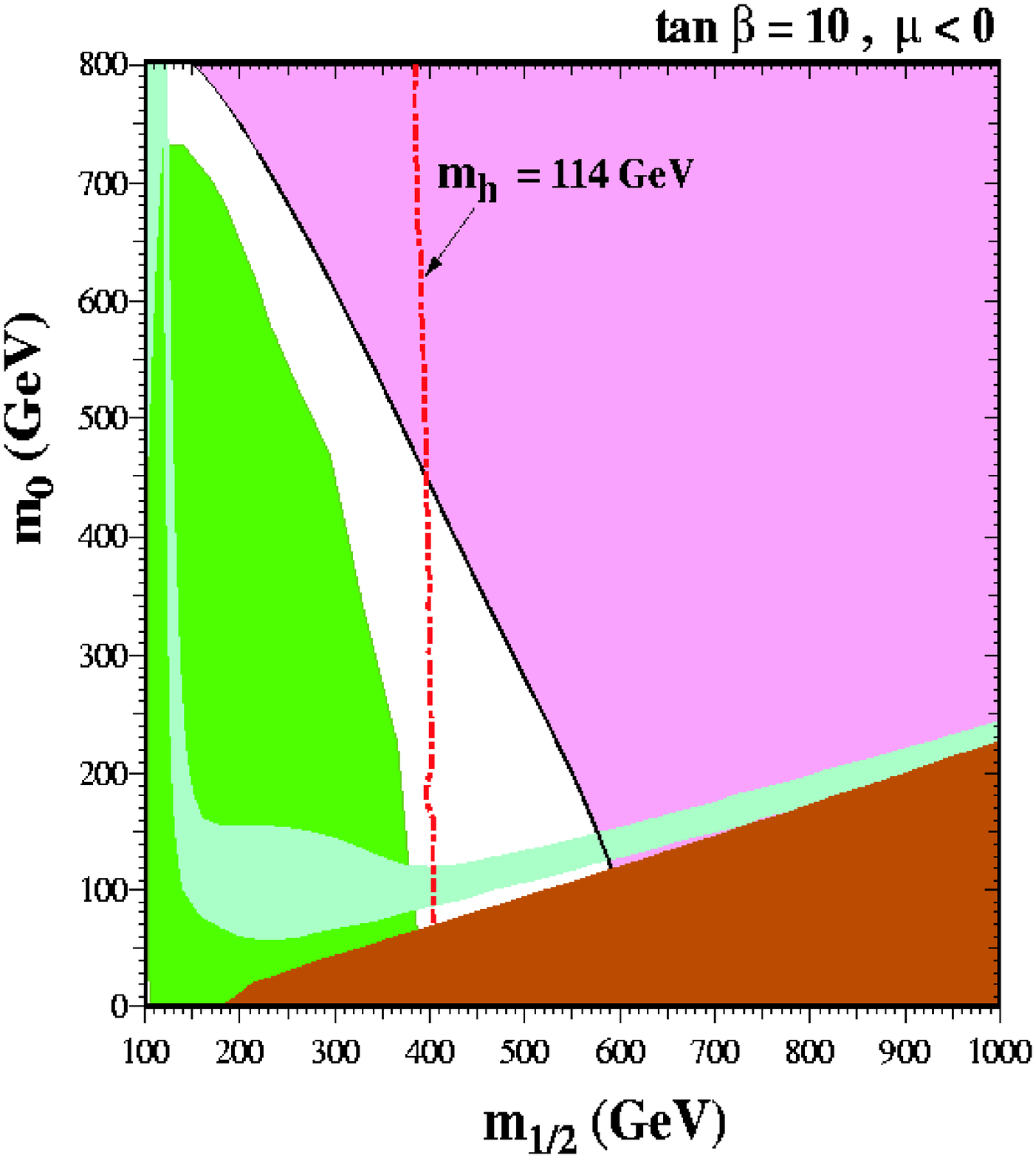,height=3.3in}
%\hspace*{-0.17in}
\epsfig{file=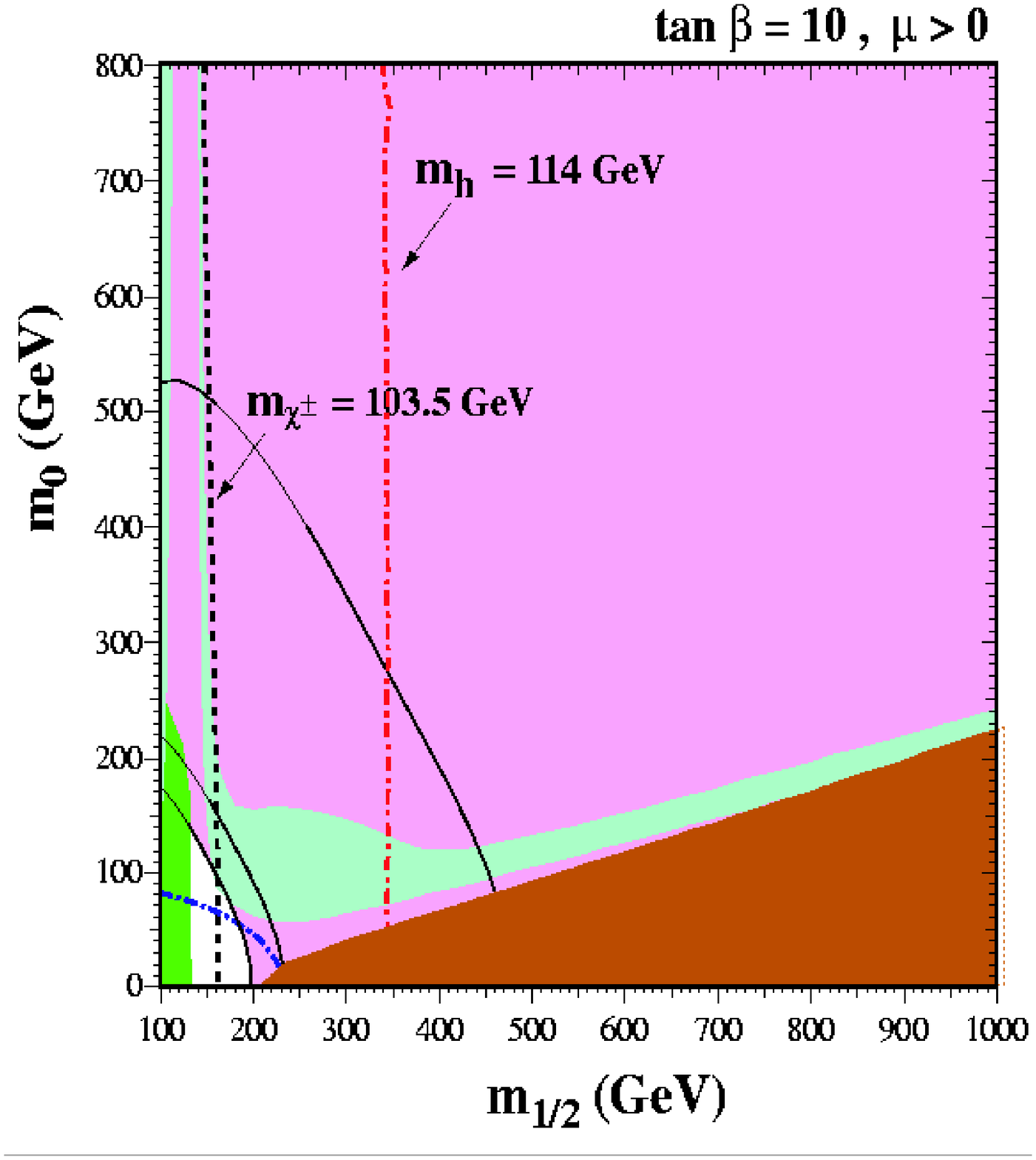,height=3.3in} \hfill
\end{minipage}
%\vspace*{-3in}
%\hspace*{-.70in}
\begin{minipage}{8in}
%\hskip -1.40in
%\vskip -.75in
\epsfig{file=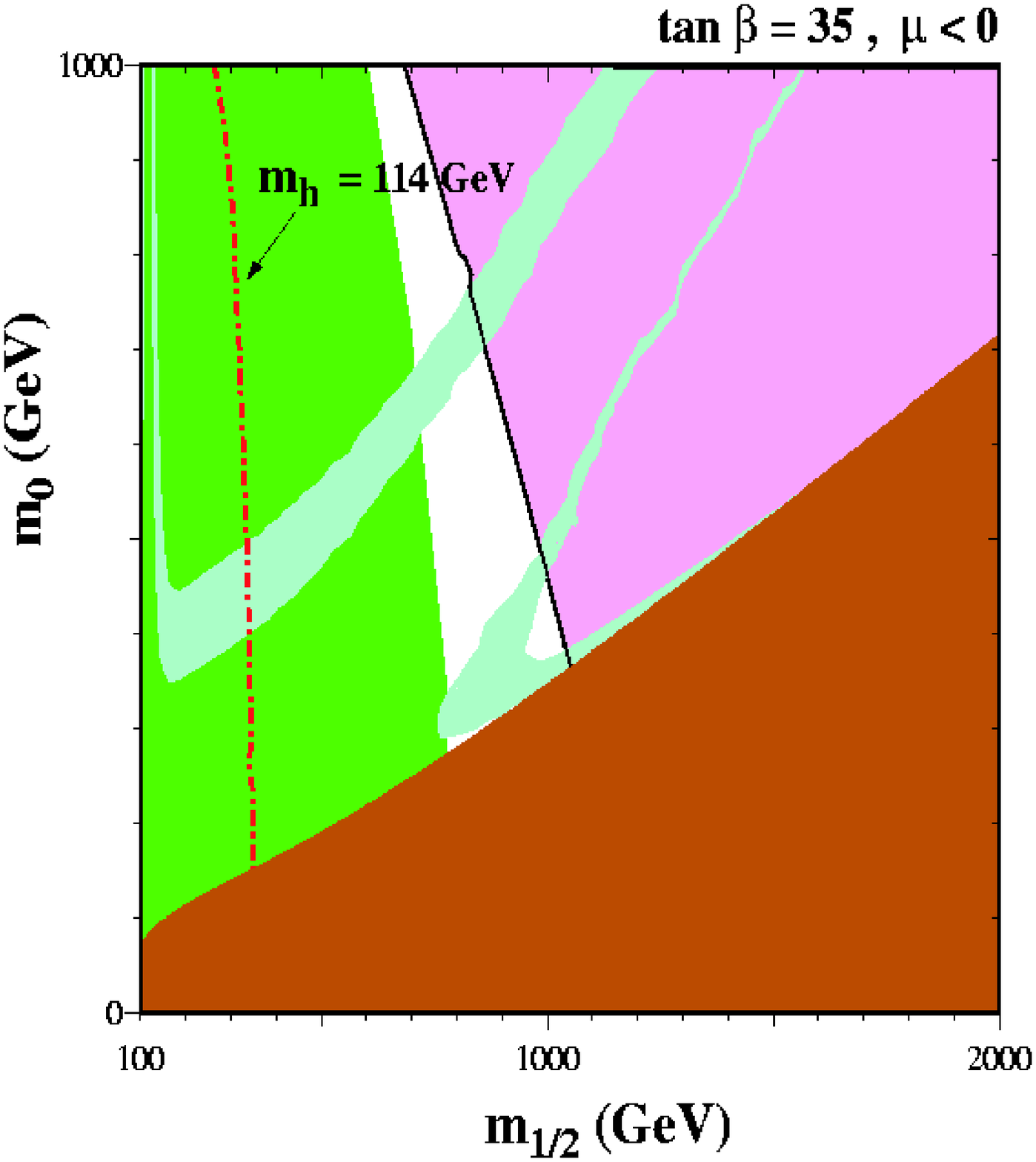,height=3.3in}
%\hspace*{-0.2in}
\epsfig{file=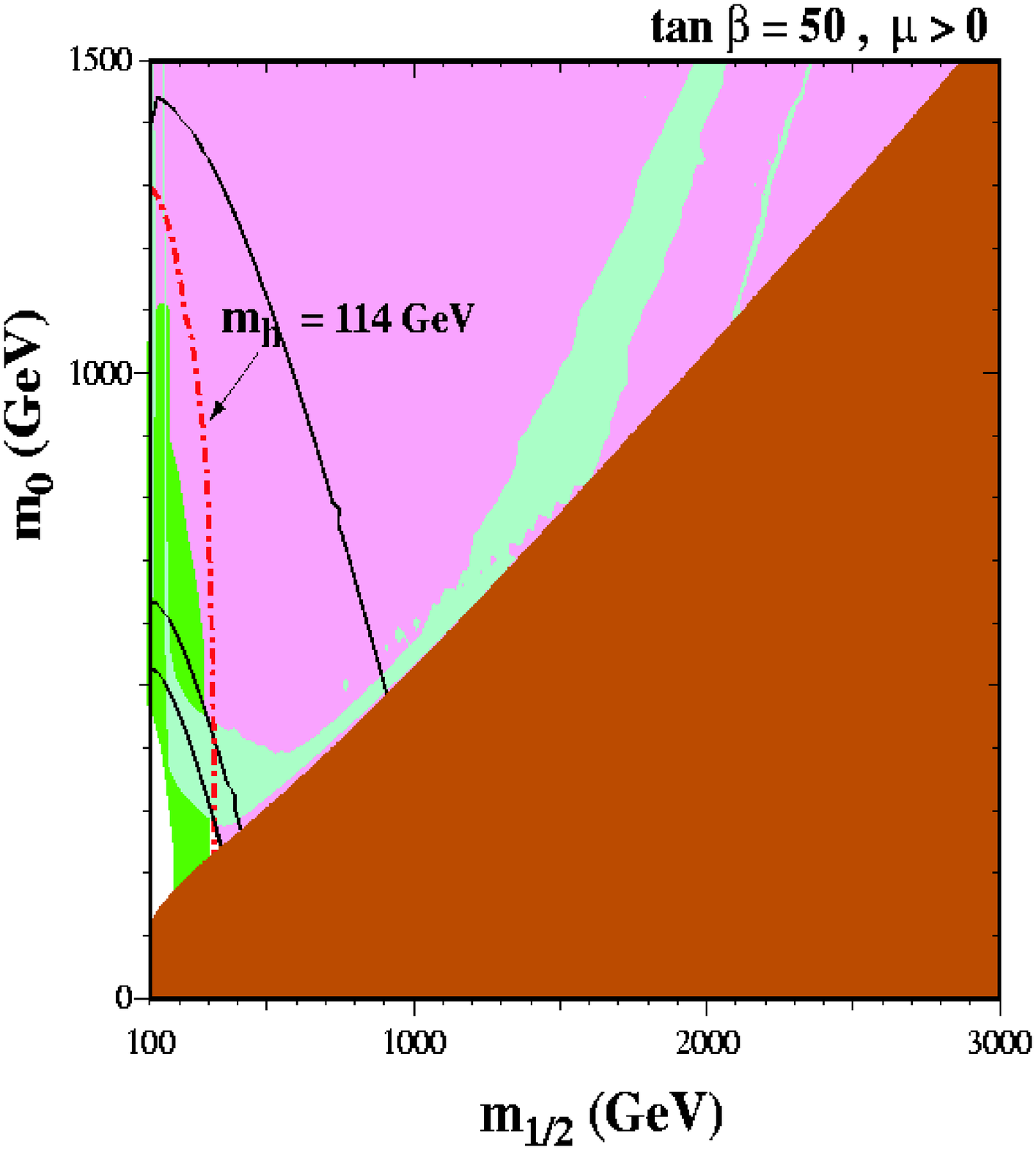,height=3.3in} \hfill
\end{minipage}
%\vskip 2.5in
\caption{
{Compilations of phenomenological constraints on the CMSSM for
(a) $\tan \beta = 10, \mu < 0$,  (b) $\tan \beta = 10, \mu > 0$, (c)
$\tan \beta = 35, \mu < 0$ and (d)  $\tan \beta = 50, \mu > 0$, assuming
$A_0 = 0, m_t = 175$~GeV and $m_b(m_b)^{\overline {MS}}_{SM} = 4.25$~GeV
\cite{EFGOSi}.  The near-vertical lines are the LEP limits
$m_{\chi^\pm} = 103.5$~GeV (dashed and black)~\cite{LEPsusy}, shown in
(b) only, and
$m_h = 114.1$~GeV (dotted and red)~\cite{LEPHWG}. 
Also, in the lower left corner of (b), we show the $m_{\tilde e} = 99$
GeV contour \protect\cite{LEPSUSYWG_0101}.  In the dark (brick red)
shaded regions, the LSP is the charged
${\tilde
\tau}_1$, so this region is excluded. The light (turquoise) shaded areas
are the cosmologically preferred regions with
\protect\mbox{$0.1\leq\ohsq\leq 0.3$}~\cite{EFGOSi}. The medium (dark
green) shaded regions that are most prominent in panels (a) and (c) are
excluded by $b \to s \gamma$~\cite{bsg}. The shaded (pink) regions in the 
upper right regions delineate the $\pm 2 \, \sigma$ ranges of $g_\mu -
2$. For $\mu > 0$, the $\pm 1 \, \sigma$ contours are also shown as solid 
black lines.
}}
\label{fig:CMSSM}
\end{figure}

Another important constraint is provided by the LEP lower limit on the
Higgs mass: $m_H > $ 114.1 GeV \cite{LEPHWG}. This holds in the
Standard Model, for the lightest Higgs boson $h$ in the general MSSM for
$\tan\beta
\lappeq 8$, and almost always in the CMSSM for all $\tan\beta$, at least
as long as CP is conserved~\footnote{The lower bound on the lightest MSSM
Higgs boson may be relaxed significantly if CP violation feeds into the
MSSM Higgs sector~\cite{CEPW}.}. Since $m_h$ is sensitive to sparticle
masses, particularly $m_{\tilde t}$, via loop corrections:
\beq
\delta m^2_h \propto {m^4_t\over m^2_W}~\ln\left({m^2_{\tilde t}\over
m^2_t}\right)~ + \ldots
\label{nine}
\eeq
the Higgs limit also imposes important constraints on the CMSSM parameters,
principally $m_{1/2}$~\cite{EGNO} as seen in Fig.~\ref{fig:CMSSM}. The 
constraints are here
evaluated using {\tt FeynHiggs}~\cite{FeynHiggs}, which is estimated to 
have a residual uncertainty of a couple of GeV in $m_h$.

Also shown in Fig.~\ref{fig:CMSSM} is the constraint imposed by
measurements of $b\rightarrow s\gamma$~\cite{bsg}. These agree with the
Standard Model, and therefore provide bounds on MSSM particles, 
such as the chargino and charged Higgs
masses, in particular. Typically, the $b\rightarrow s\gamma$
constraint is more important for $\mu < 0$, as seen in
Fig.~\ref{fig:CMSSM}a and c, but it is also relevant for $\mu > 0$, 
particularly when $\tan\beta$ is large as seen in Fig.~\ref{fig:CMSSM}d.

The final experimental constraint we consider is that due to the
measurement of the anamolous magnetic moment of the muon.  The BNL E821
experiment reported last year a new measurement of
$a_\mu\equiv {1\over 2} (g_\mu -2)$ which deviated by 2.6 standard
deviations from the best Standard Model prediction available at that
time~\cite{BNL}. The largest contribution to the errors in the comparison
with theory was thought to be the statistical error of the experiment,
which will soon be significantly reduced, as many more data have already
been recorded. However, it has recently been realized that the sign of
the most important pseudoscalar-meson pole part of the light-by-light 
scattering contribution
\cite{lightbylight} to the Standard Model prediction should be reversed,
which reduces the apparent experimental discrepancy to about 1.6 standard
deviations. The next-largest error is thought to be that due to
strong-interaction uncertainties in the Standard Model prediction, for
which recent estimates converge to about 7$\times 10^{-10}$~\cite{DHSNTY}.

As many authors have pointed out~\cite{susygmu}, a discrepancy between
theory and the BNL experiment could well be explained by supersymmetry. As
seen in Fig.~\ref{fig:CMSSM}, this is particularly easy if $\mu > 0$.
With the change in sign of the meson-pole contributions to
light-by-light scattering, good consistency is also possible for $\mu <
0$ so long as either $m_{1/2}$ or $m_0$ are taken sufficiently large.
We show in Fig.~\ref{fig:CMSSM} as medium (pink) shaded the new $2 \,
\sigma$  allowed region: $-6 < \delta a_\mu \times 10^{10} <  58 $.

The new regions preferred by the $g-2$ experimental data shown in 
Fig.~\ref{fig:CMSSM} differ considerably from the older
ones~\cite{susygmu} which were based on the range $11 < \delta a_\mu
\times 10^{10} <  75 $.  First of all, the older bound completely 
excluded $\mu < 0$ at the $2 \, \sigma$ level.  As one can see this is no
longer true: $\mu < 0 $ is allowed so long as either (or both) $m_{1/2}$
and $m_0$ are large. Thus, for $\mu < 0$, one is forced into either
the $\chi-{\tilde \tau}$ coannihilation region or the funnel region
produced by the s-channel annihilation via the heavy Higgses $H$ and $A$, 
as described below.
Second, whereas the older limits produced definite upper bounds on the
sparticle masses (which were accepted with delight by future collider
builders), the new bounds, which are consistent with $a_{\mu} = 0$, 
allow arbitrarily high sparticle masses.  Now only the very low mass
corner of the $(m_{1/2}, m_0)$ plane is excluded.

Fig.~\ref{fig:CMSSM} also displays the regions where the supersymmetric 
relic density $\rho_\chi = \Omega_\chi \rho_{critical}$ falls within the 
preferred range
\beq
0.1 < \Omega_\chi h^2 < 0.3
\label{ten}
\eeq
The upper limit is rigorous, since astrophysics and cosmology tell us that
the total matter density $\Omega_m \lappeq 0.4$, and the Hubble expansion
rate $h \sim 1/\sqrt{2}$ to within about 10 \% (in units of 100 km/s/Mpc). On
the other hand, the lower limit in (\ref{ten}) is optional, since there
could be other important contributions to the overall matter density.

As is seen in Fig.~\ref{fig:CMSSM}, there are generic regions of the CMSSM
parameter space where the relic density falls within the preferred range
(\ref{ten}). What goes into the calculation of the relic density? It is
controlled by the annihilation cross section~\cite{EHNOS}:
\beq
\rho_\chi = m_\chi n_\chi \, , \quad n_\chi \sim {1\over
\sigma_{ann}(\chi\chi\rightarrow\ldots)}\, ,
\label{eleven}
\eeq
where the typical annihilation cross section $\sigma_{ann} \sim 
1/m_\chi^2$.
For this reason, the relic density typically increases with the relic
mass, and this combined with the upper bound in (\ref{ten}) then leads to
the common expectation that $m_\chi \lappeq$ O(200) GeV. 

However, there are various ways in which the generic upper bound on
$m_\chi$ can be increased along filaments in the $(m_{1/2},m_0)$ plane.
For example, if the next-to-lightest sparticle (NLSP) is not much heavier
than $\chi$: $\Delta m/m_\chi \lappeq 0.1$, the relic density may be
suppressed by coannihilation: $\sigma (\chi + $NLSP$ \rightarrow \ldots
)$~\cite{coann}.  In this way, the allowed CMSSM region may acquire a
`tail' extending to larger sparticle masses. An example of this 
possibility is the
case where the NLSP is the lighter stau: $\tilde\tau_1$ and
$m_{\tilde\tau_1} \sim m_\chi$, as seen in Figs.~\ref{fig:CMSSM}(a) and
(b) and extended to larger $m_{1/2}$ in
Fig.~\ref{fig:coann}(a)~\cite{ourcoann}. Another example is
coannihilation when the NLSP is the lighter stop
\cite{stopco}: $\tilde t_1$ and
$m_{\tilde t_1}
\sim m_\chi$, which may be important in the general MSSM or in the CMSSM
when
$A$ is large, as seen in Fig.~\ref{fig:coann}(b)~\cite{EOS}. In
the cases studied, the upper limit on $m_\chi$ is not affected by stop
coannihilation. Another mechanism for extending the allowed CMSSM region
to large
$m_\chi$ is rapid annihilation via a direct-channel pole when $m_\chi
\sim {1\over 2} m_{Higgs, Z}$~\cite{funnel,EFGOSi}. This may yield a
`funnel' extending to large $m_{1/2}$ and $m_0$ at large $\tan\beta$, as
seen in panels (c) and (d) of  Fig.~\ref{fig:CMSSM}~\cite{EFGOSi}. Yet
another allowed region at large
$m_{1/2}$ and $m_0$ is the `focus-point' region~\cite{focus}, which is
adjacent to the boundary of the region where electroweak symmetry breaking
is possible, as seen in Fig.~\ref{fig:focus}.

\begin{figure}
\vskip 0.5in
\vspace*{-0.75in}
%\hspace*{-.70in}
\begin{minipage}{8in}
\epsfig{file=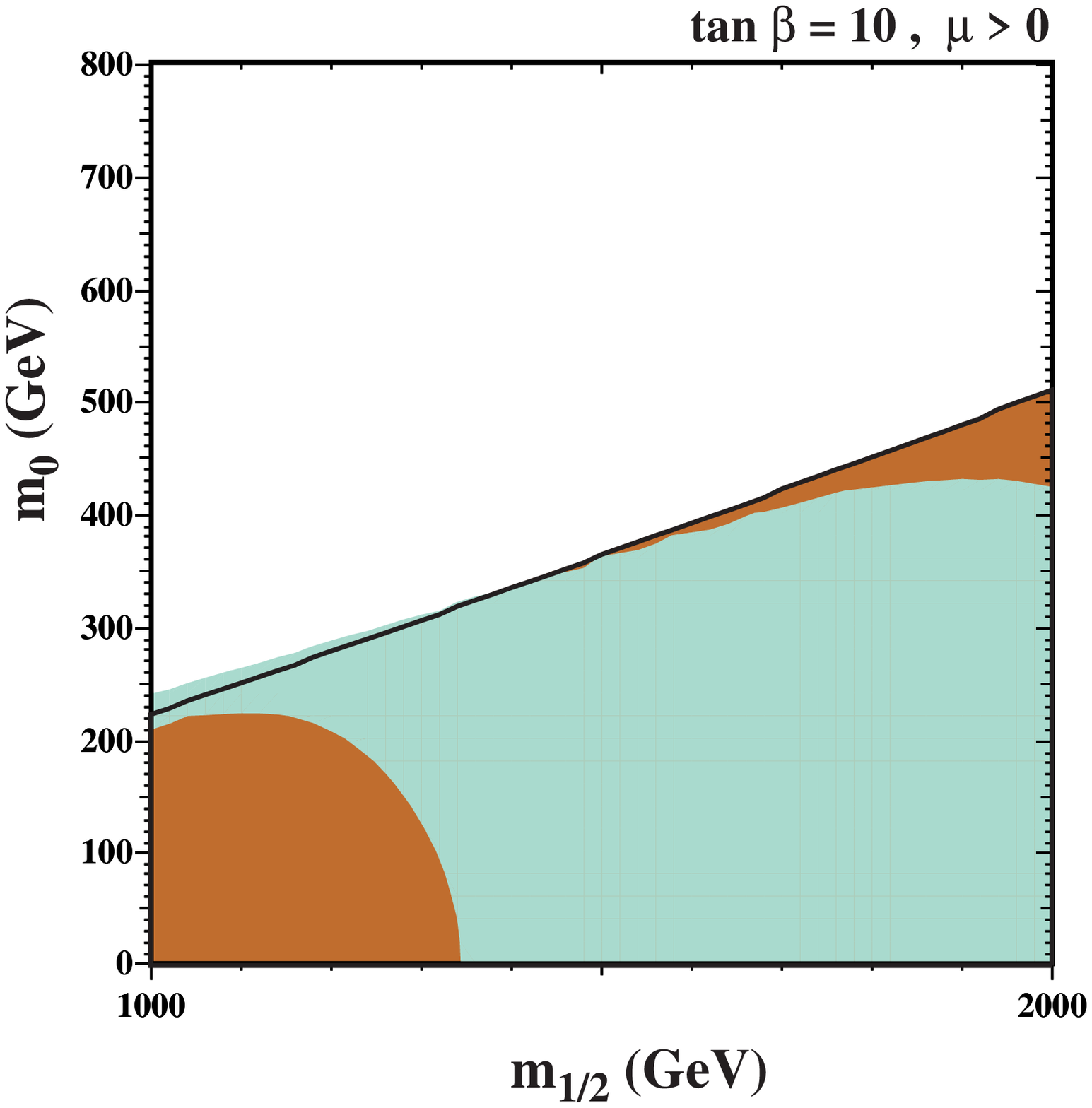,height=3.3in}
\hspace*{-0.17in}
\epsfig{file=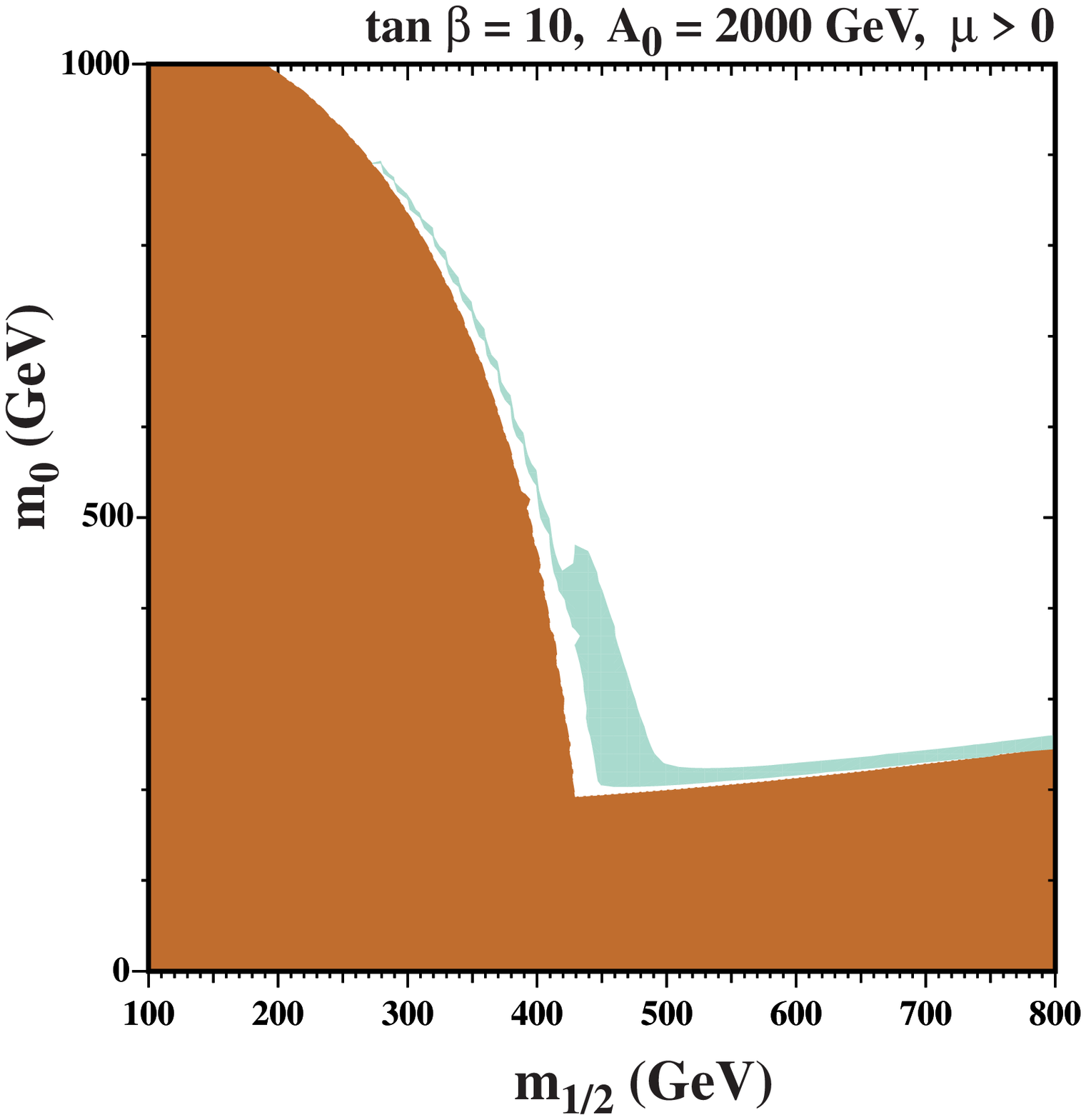,height=3.3in} \hfill
\end{minipage}
\caption[]{(a) The large-$m_{1/2}$ `tail' of the $\chi - {\tilde 
\tau_1}$ 
coannihilation region 
for $\tan \beta = 10$, $A = 0$ and $\mu < 0$~\cite{ourcoann}, superimposed 
on the disallowed dark (brick red) shaded region where $m_{\tilde
\tau_1} < m_\chi$, and (b) the $\chi - {\tilde t_1}$ coannihilation region
for $\tan \beta = 10$, $A = 2000$~GeV and $\mu > 0$~\cite{EOS}, exhibiting 
a large-$m_0$ `tail'.}
\label{fig:coann}
\end{figure}

\begin{figure}
%\vspace*{-0.75in}
\hspace*{-.40in}
\begin{minipage}{8in}
\epsfig{file=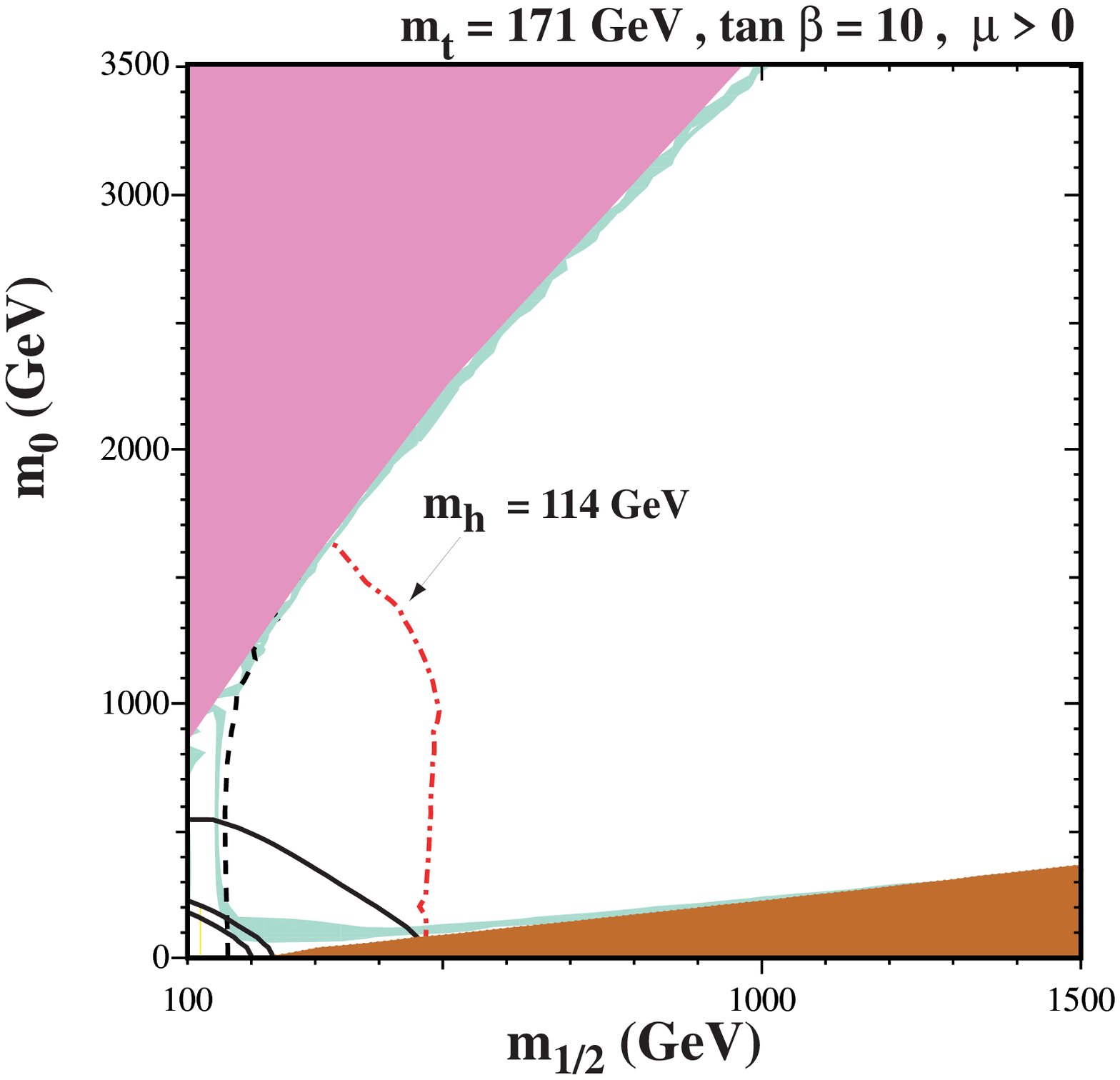,height=3.3in}
\hspace*{-0.17in}
\epsfig{file=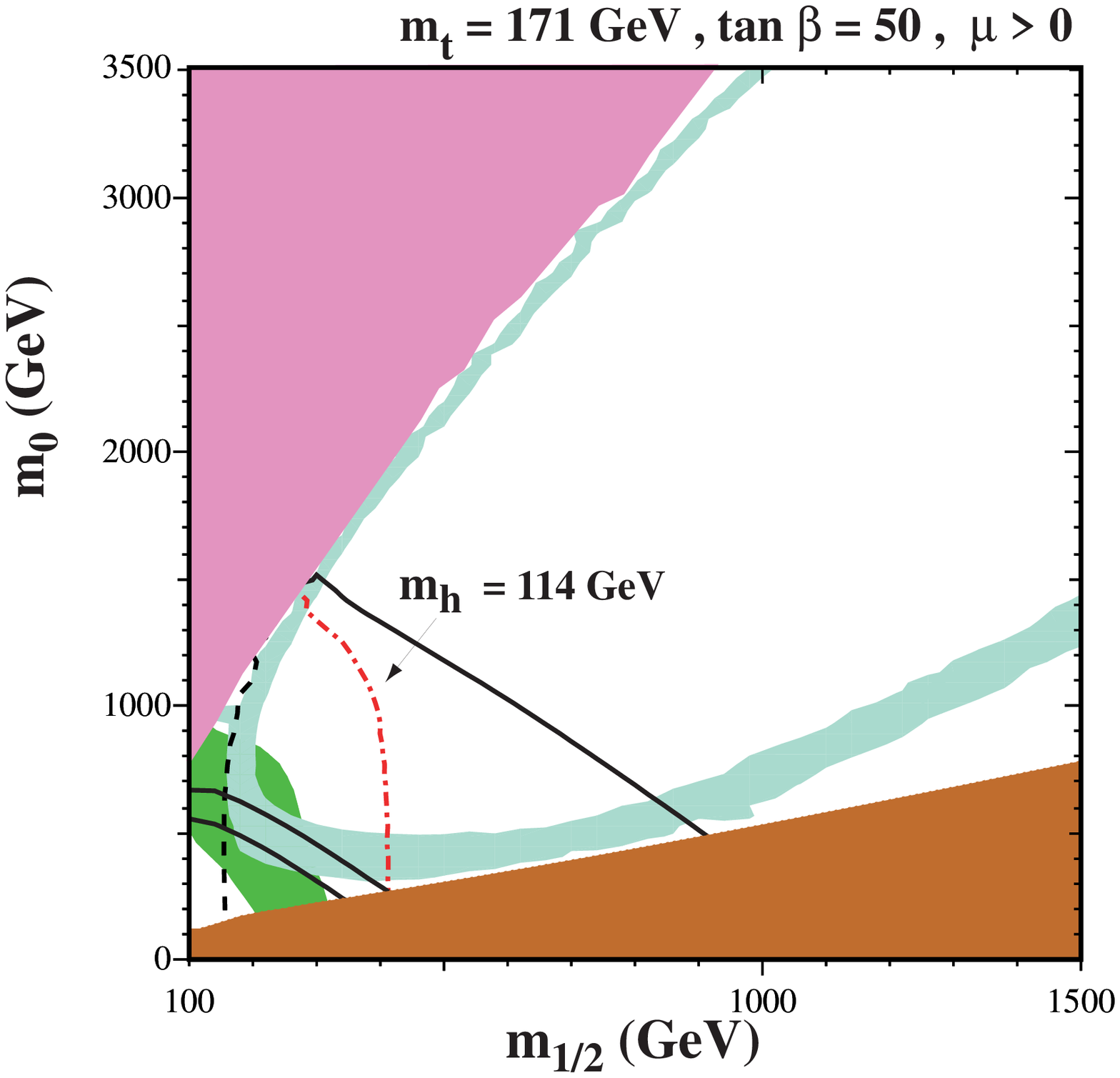,height=3.3in} \hfill
\end{minipage}
\caption[]{An expanded view of the $m_{1/2} - m_0$ parameter plane
showing the focus-point regions \protect\cite{focus} at large $m_0$ for 
(a) $tan \beta
= 10$, and (b) $\tan \beta = 50$. In the shaded (mauve) region in the 
upper
left corner, there are no solutions with proper electroweak symmetry 
breaking, so these are
excluded in the CMSSM.  Note that we have chosen $m_t = 171$ GeV, in 
which case the focus-point region is at lower $m_0$ than when $m_t = 175$ 
GeV, as assumed in the other figures. The position of this region
is very sensitive to $m_t$. The black contours (both dashed and solid)
are as in Fig.~\protect\ref{fig:CMSSM}, we do not shade the preferred
$g-2$ region. }
\label{fig:focus}
\end{figure}

\subsection{Fine Tuning}

The filaments extending the preferred CMSSM parameter space are clearly
exceptional, in some sense, so it is important to understand the sensitivity
of the relic density to input parameters, unknown higher-order effects, 
etc. One proposal is the relic-density fine-tuning measure~\cite{EO}
\beq
\Delta^\Omega \equiv \sqrt{\sum_i ~~\left({\partial\ln (\Omega_\chi h^2)\over
\partial
\ln a_i}\right)^2 }
\label{twelve}
\eeq
where the sum runs over the input parameters, which might include
(relatively) poorly-known Standard Model quantities such as $m_t$ and
$m_b$, as well as the CMSSM parameters $m_0, m_{1/2}$, etc. As seen in
Fig.~\ref{fig:overall}, the sensitivity $\Delta^\Omega$ (\ref{twelve}) is 
relatively small
in the `bulk' region at low $m_{1/2}$, $m_0$, and $\tan\beta$. However, it
is somewhat higher in the $\chi - \tilde\tau_1$ coannihilation `tail', and
at large $\tan\beta$ in general. The sensitivity measure $\Delta^\Omega$
(\ref{twelve}) is particularly high in the rapid-annihilation `funnel' and
in the `focus-point' region. This explains why published relic-density
calculations may differ in these regions~\cite{otherOmega}, whereas they 
agree well when
$\Delta^\Omega$ is small: differences may arise because of small
differences in the treatments of the inputs.

\begin{figure}
\vskip 0.5in
\vspace*{-0.75in}
\hspace*{-.20in}
\begin{minipage}{8in}
\epsfig{file=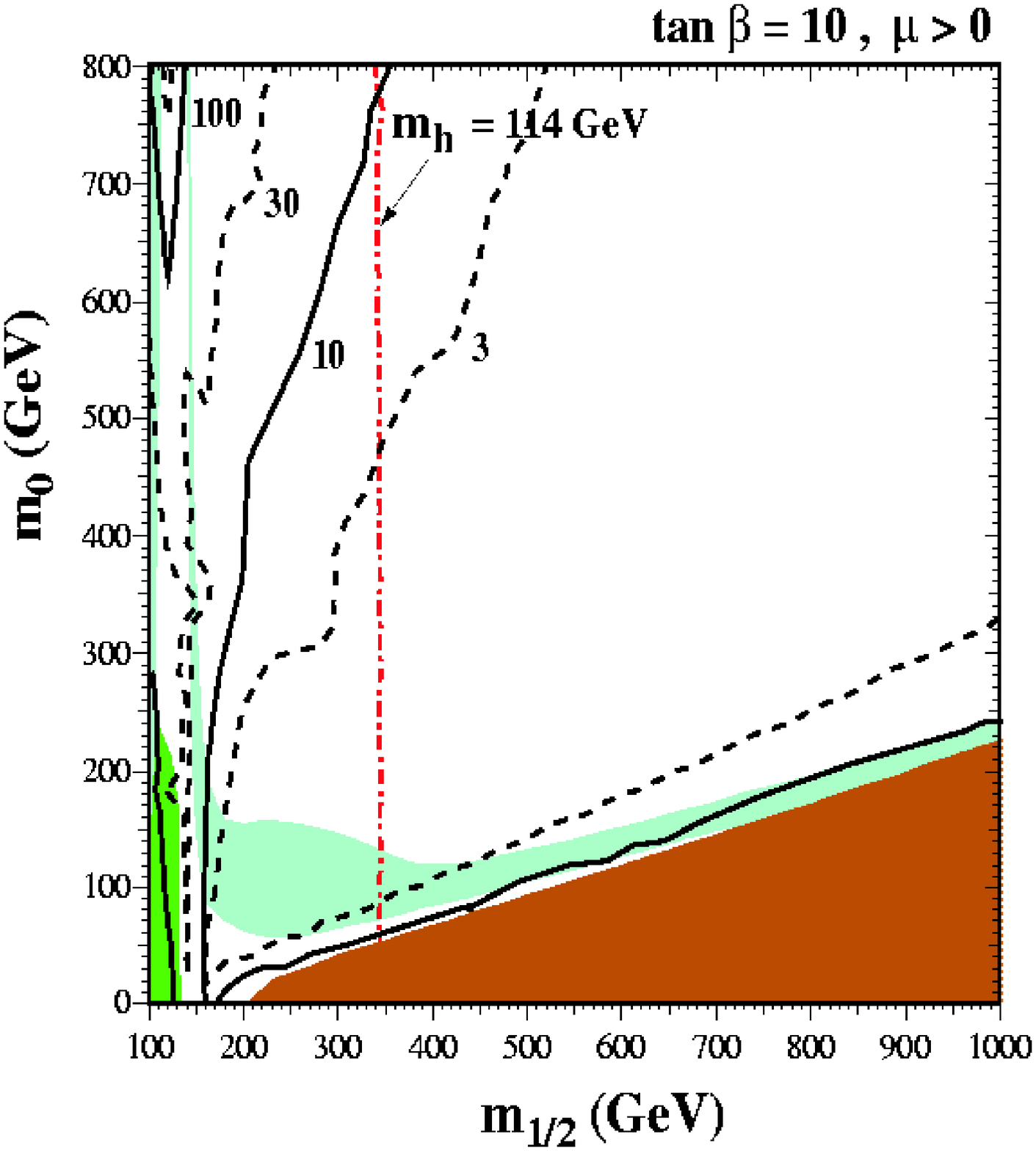,height=3.3in}
\hspace*{+0.10in}
\epsfig{file=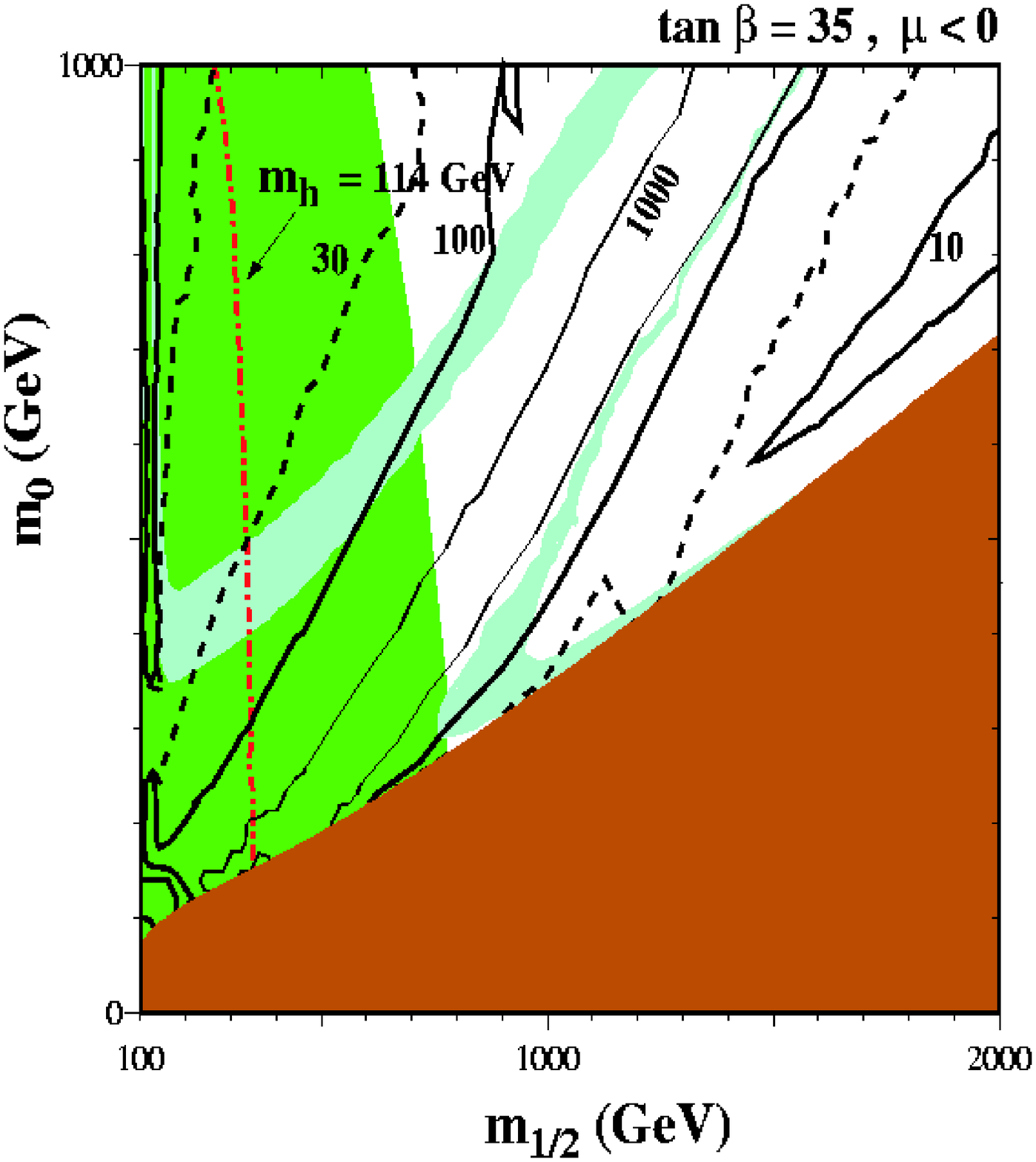,height=3.3in} \hfill
\end{minipage}
%\vspace*{-3in}
%\hspace*{-.70in}
\hspace*{-.20in}
\begin{minipage}{8in}
%\hskip -1.40in
%\vskip -.75in
\epsfig{file=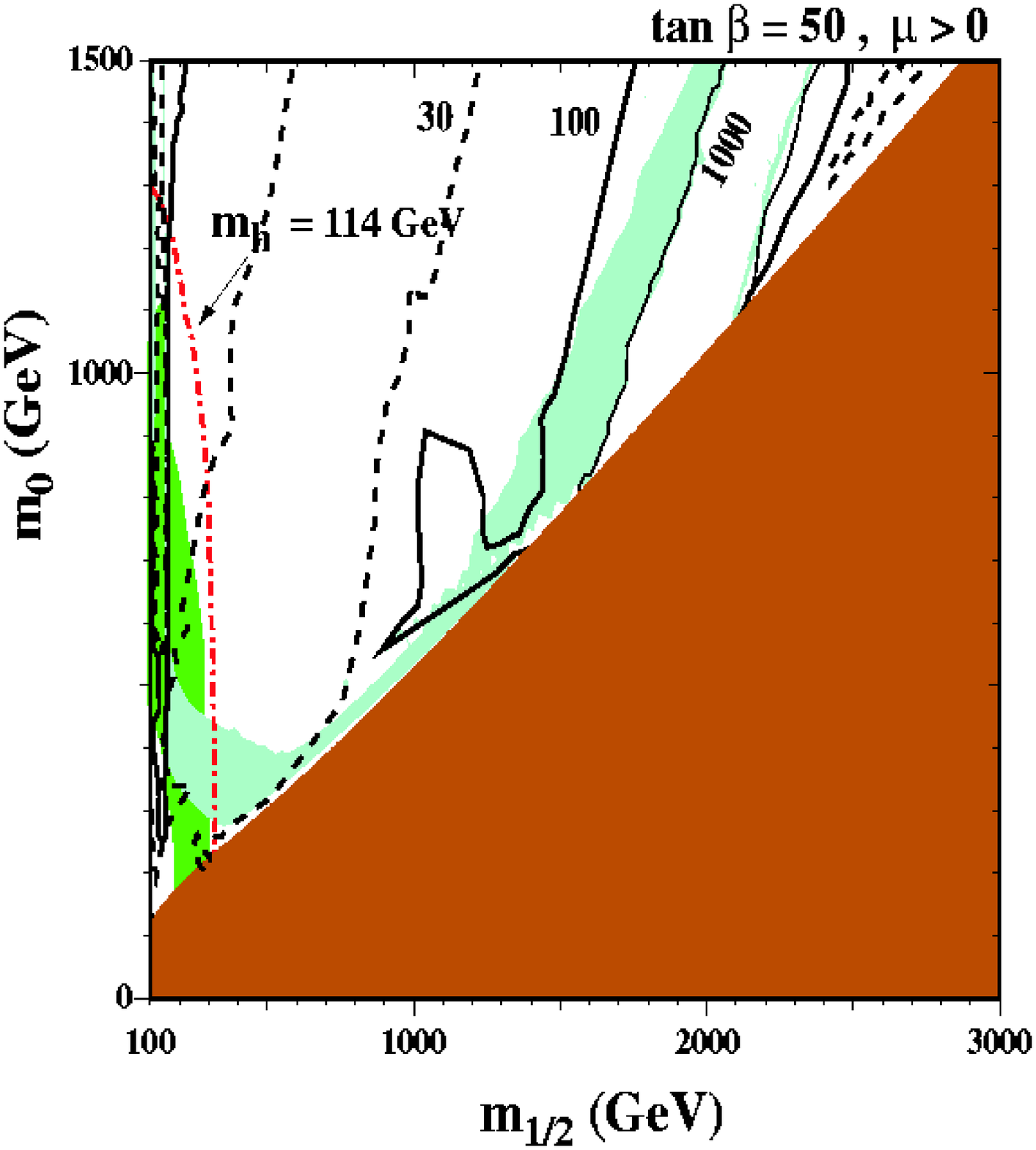,height=3.3in}
%\hspace*{-0.10in}
\epsfig{file=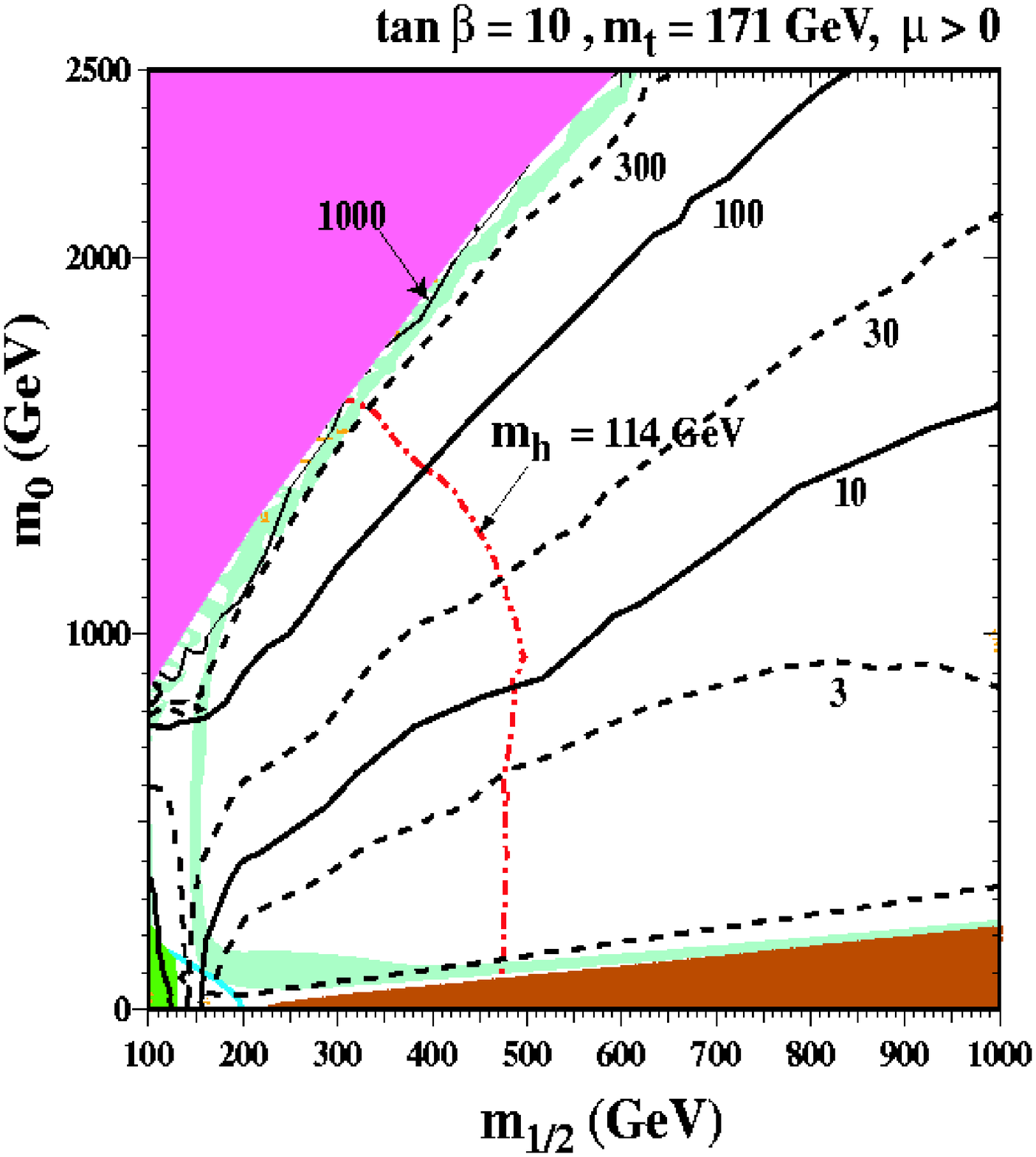,height=3.3in} \hfill
\end{minipage}
%\vskip 2.5in 
\caption{\label{fig:overall}
{Contours of the total sensitivity $\Delta^\Omega$ (\ref{twelve}) of 
the relic density in the
$(m_{1/2}, m_0)$ planes for (a) $\tan \beta = 10, \mu > 0, m_t =
175$~GeV, (b) $\tan \beta = 35, \mu < 0, m_t = 175$~GeV, (c)
$\tan \beta = 50, \mu > 0, m_t = 175$~GeV, and (d) $\tan \beta =
10, \mu > 0, m_t = 171$~GeV, all for $A_0 = 0$. The light (turquoise)
shaded areas are the cosmologically preferred regions with
\protect\mbox{$0.1\leq\ohsq\leq 0.3$}. In the dark (brick red) shaded
regions, the LSP is the charged ${\tilde \tau}_1$, so these regions are
excluded. In panel (d), the medium shaded (mauve) region is excluded by
the electroweak vacuum conditions. }}
\end{figure}

\begin{figure}
\vskip 0.5in
\vspace*{-0.75in}
\hspace*{-.20in}
\begin{minipage}{8in}
\epsfig{file=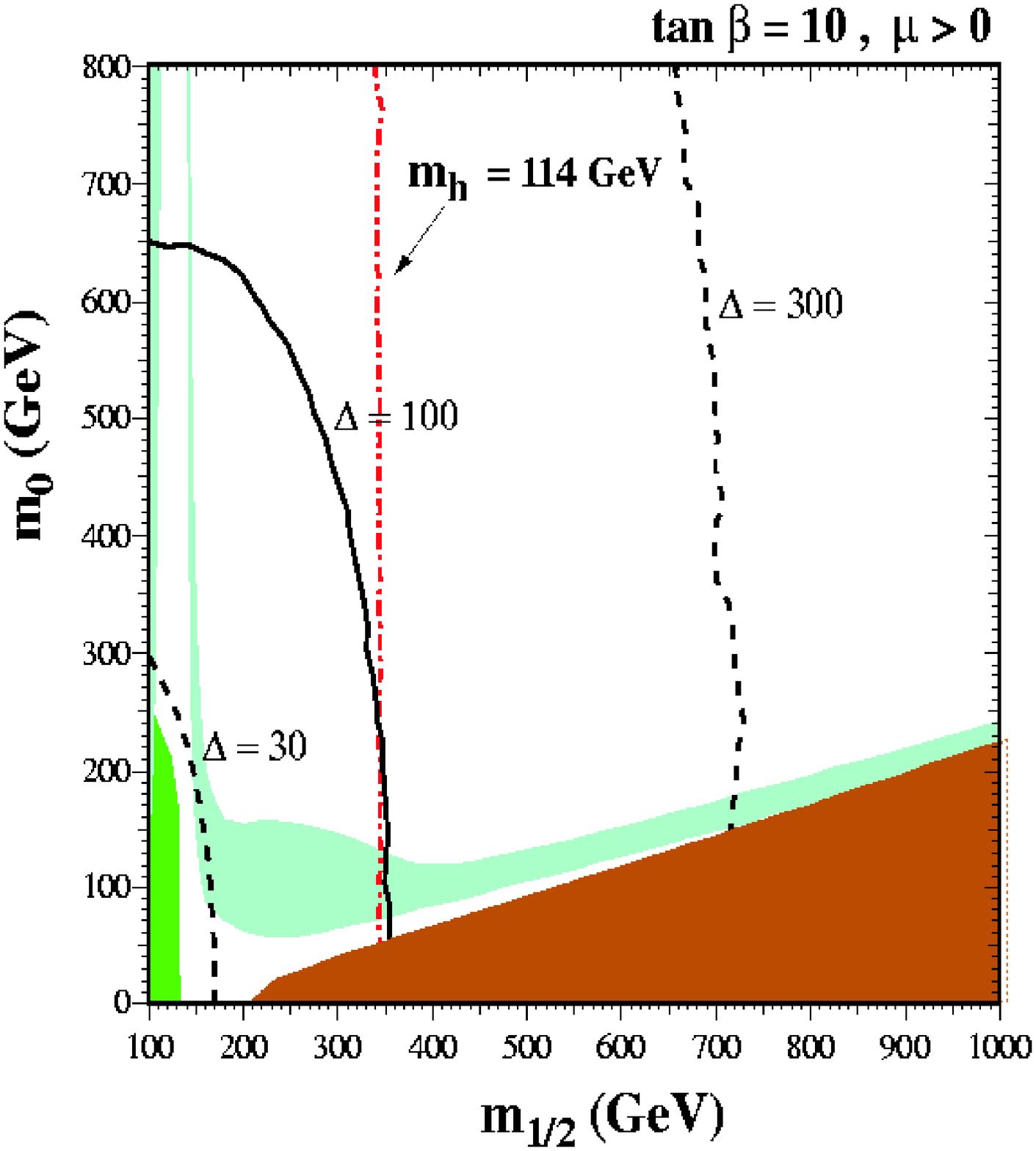,height=3.3in}
\hspace*{+0.10in}
\epsfig{file=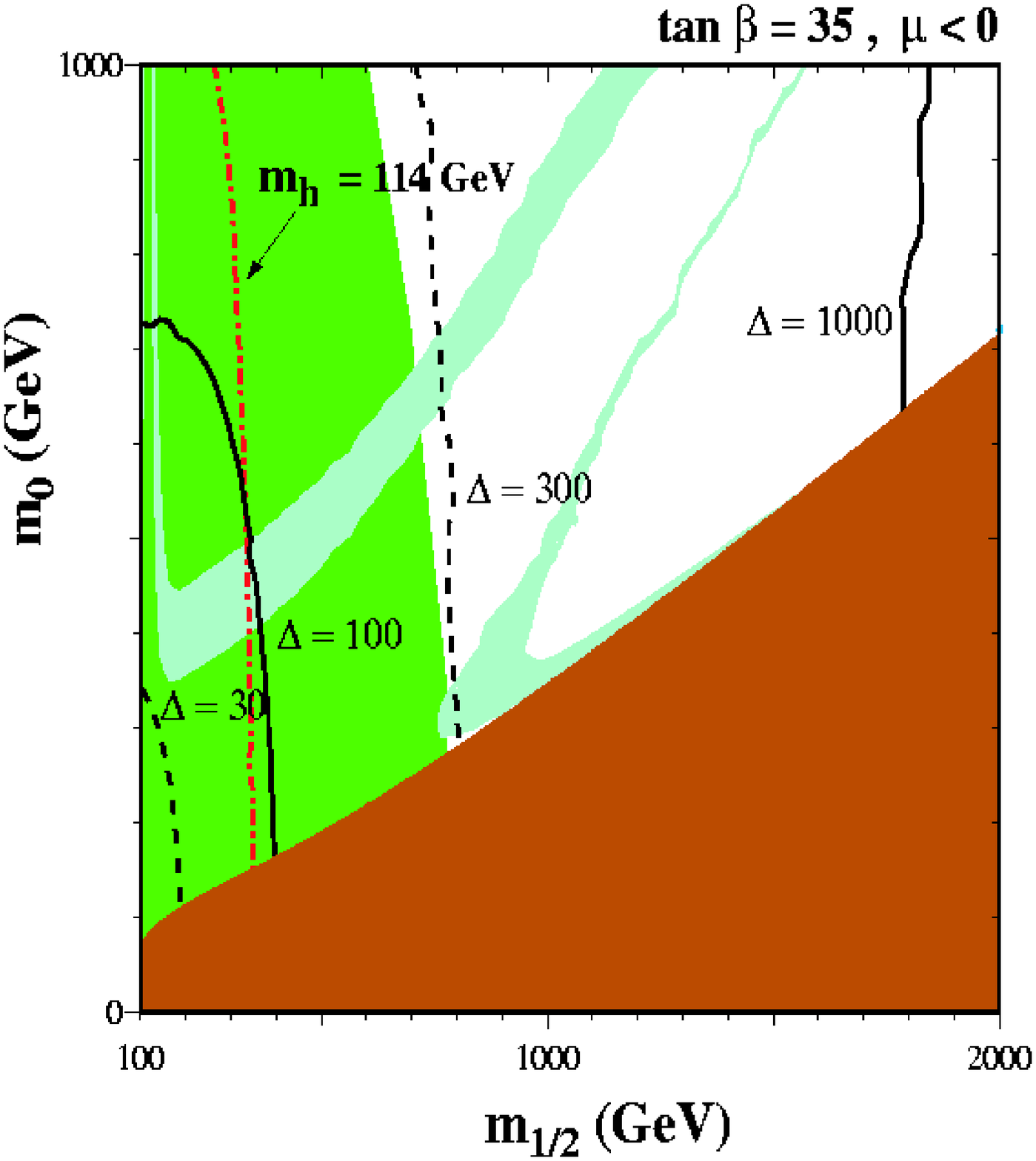,height=3.3in} \hfill
\end{minipage}
%\vspace*{-3in}
%\hspace*{-.70in}
\hspace*{-.20in}
\begin{minipage}{8in}
%\hskip -1.40in
%\vskip -.75in
\epsfig{file=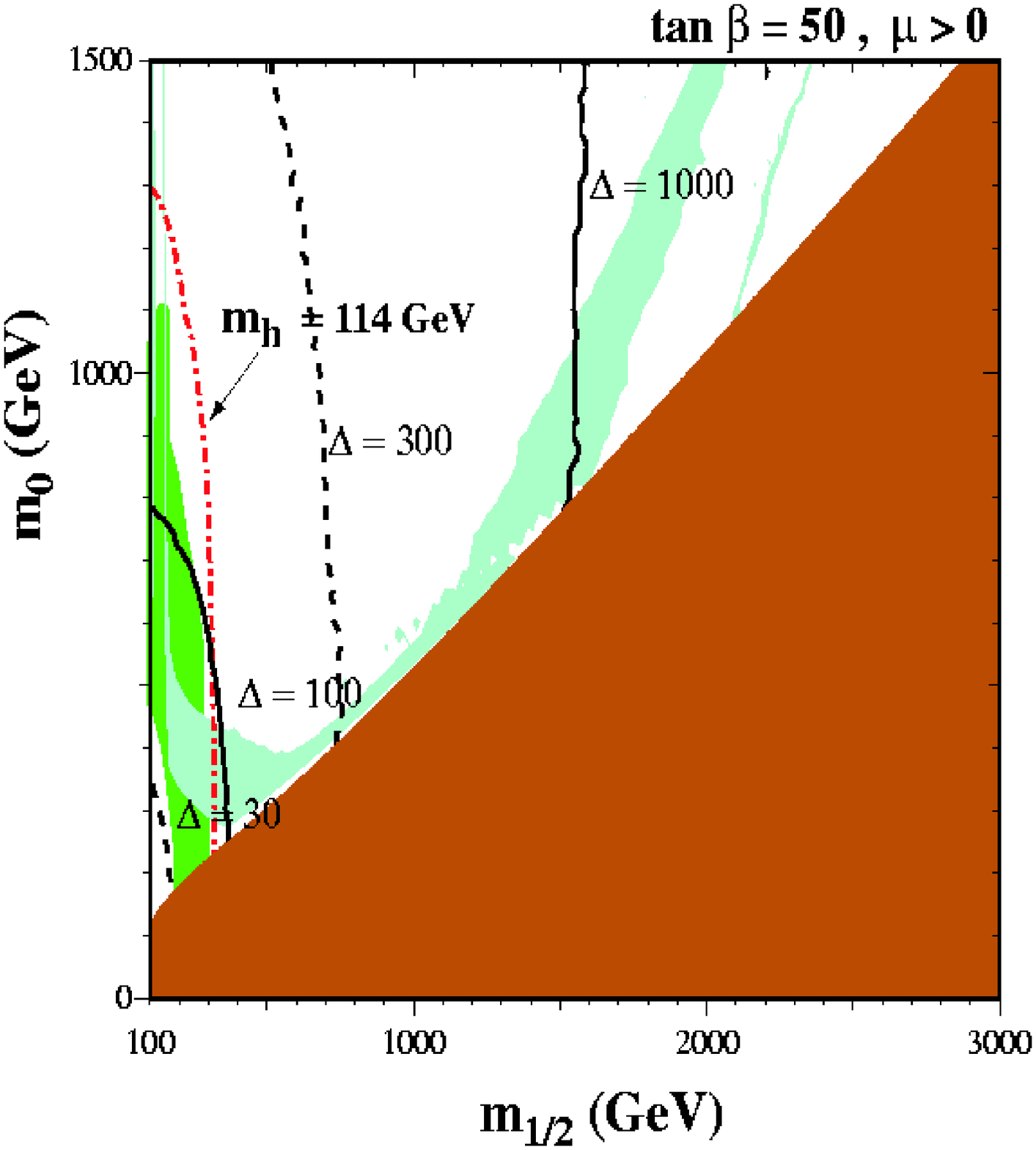,height=3.3in}
\hspace*{-0.10in}
\epsfig{file=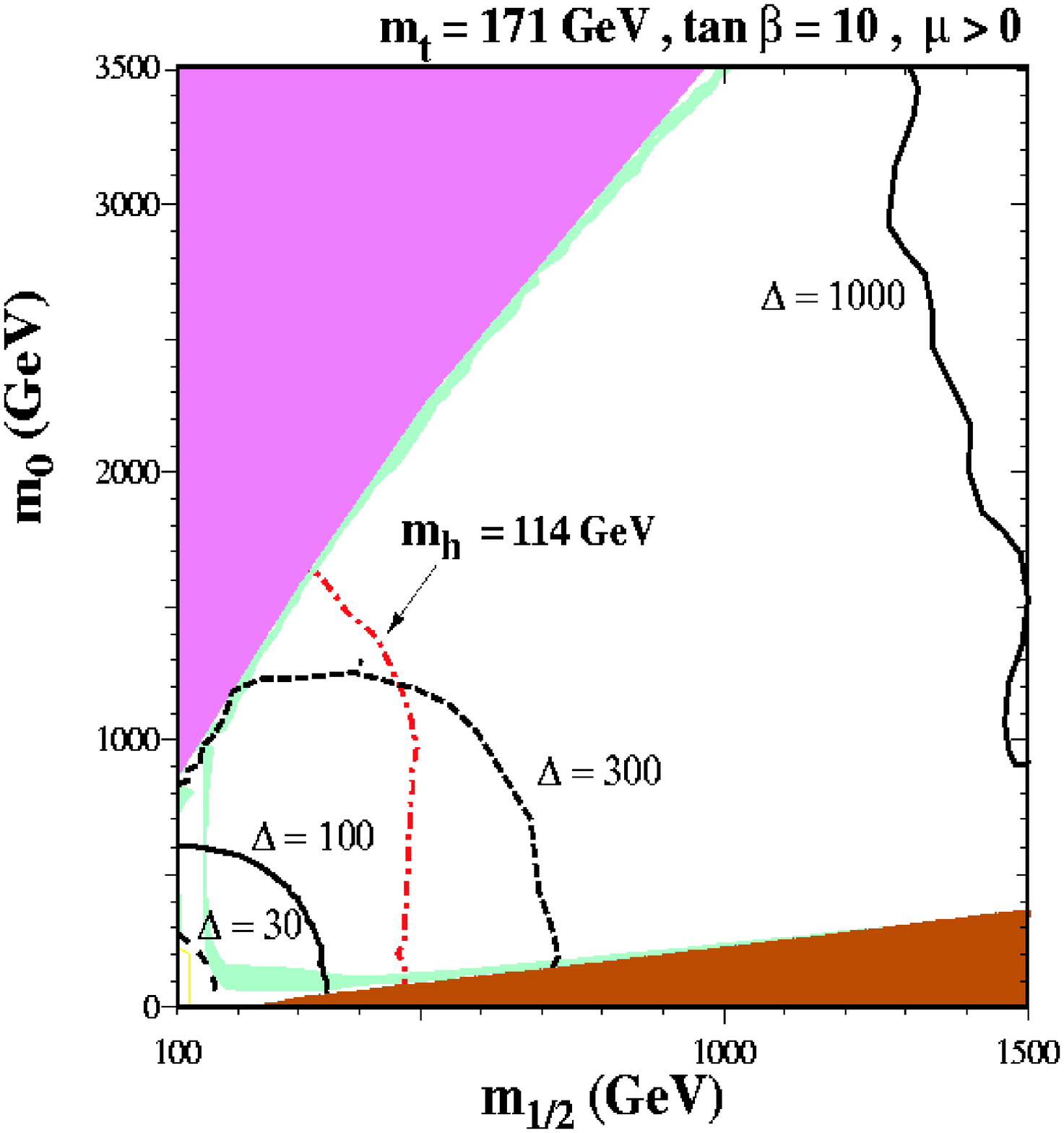,height=3.3in} \hfill
\end{minipage}
%\vskip 2.5in 
\caption{\label{fig:EWFT}
{Contours of the electroweak fine-tuning measure $\Delta$ 
(\ref{thirteen}) in the
$(m_{1/2}, m_0)$ planes for (a) $\tan \beta = 10, \mu > 0, m_t =
175$~GeV, (b) $\tan \beta = 35, \mu < 0, m_t = 175$~GeV, (c)
$\tan \beta = 50, \mu > 0, m_t = 175$~GeV, and (d) $\tan \beta =
10, \mu > 0, m_t = 171$~GeV, all for $A_0 = 0$. The light (turquoise)
shaded areas are the cosmologically preferred regions with
\protect\mbox{$0.1\leq\ohsq\leq 0.3$}. In the dark (brick red) shaded
regions, the LSP is the charged ${\tilde \tau}_1$, so this region is
excluded. In panel (d), the medium shaded (mauve) region is excluded by
the electroweak vacuum conditions. }}
\end{figure}  

It is important to note that the relic-density fine-tuning measure
(\ref{twelve}) is distinct from the traditional measure of the fine-tuning
of the electroweak scale~\cite{EENZ}:
\beq
\Delta = \sqrt{\sum_i ~~\Delta_i^{\hspace{0.05in} 2}}\, , \quad \Delta_i \equiv
{\partial \ln
m_W\over \partial \ln a_i}
\label{thirteen}
\eeq
Sample contours of the electroweak fine-tuning measure are shown 
(\ref{thirteen}) are shown in Figs.~\ref{fig:EWFT}.
This electroweak fine tuning is logically different from
the cosmological fine tuning, and values
of $\Delta$ are not necessarily related to values of
$\Delta^\Omega$, as is apparent when comparing the contours in 
Figs.~\ref{fig:overall} and \ref{fig:EWFT}.  Electroweak fine-tuning is 
sometimes used as a 
criterion
for restricting the CMSSM parameters. However, the interpretation of 
$\Delta$ (\ref{thirteen}) is unclear. How large a value of $\Delta$ is
tolerable? Different physicists may well have different pain thresholds.
Moreover, correlations between input parameters may reduce its value in
specific models, and the regions allowed by the different constraints can 
become very
different when we relax some of the CMSSM assumptions,
e.g. the universality between the input Higgs masses and those of the squarks
and sleptons, a subject beyond the scope of these Lectures.

\subsection{Prospects for Observing Supersymmetry at Accelerators}

As an aid to the assessment of the prospects for detecting sparticles at
different accelerators, benchmark sets of supersymmetric parameters have
often been found useful~\cite{benchmarks}, since they provide a focus for
concentrated discussion. A set of proposed post-LEP benchmark scenarios in
the CMSSM~\cite{benchmark} are illustrated schematically in
Fig.~\ref{fig:Bench}. They take into account the direct searches for
sparticles and Higgs bosons, $b\rightarrow s\gamma$ and the preferred
cosmological density range (\ref{ten}). The proposed
benchmark points are consistent with $g_\mu -2$ at the
$2 \, \sigma$ level, but this was not imposed as an absolute requirement. 

\begin{figure}
%\vspace*{-0.75in}
%\hspace*{.40in}
\begin{center}
%\begin{minipage}{8in}
\epsfig{file=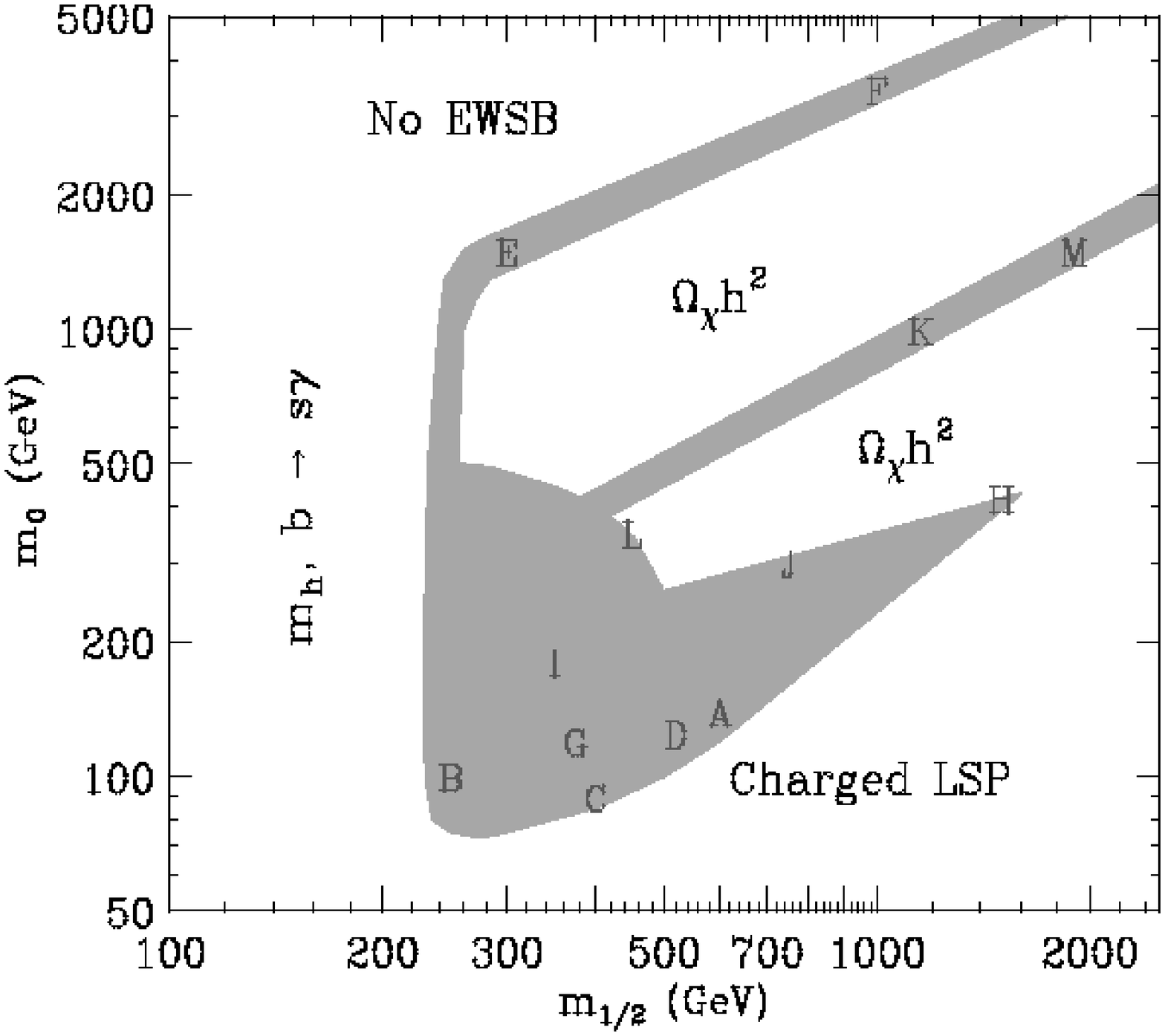,height=3.5in}
\end{center}
%\end{minipage}
\caption[]{Overview of the CMSSM benchmark points proposed 
in~\cite{benchmark}.
They were chosen to be compatible with the indicated experimental
constraints, as well as have a relic density in the preferred range
(\ref{ten}). The points have different values of $\tan \beta$, and are 
intended to illustrate the range of available possibilities.}
\label{fig:Bench}
\end{figure}

The proposed points were chosen not to provide an `unbiased' statistical
sampling of the CMSSM parameter space, whatever that means in the absence
of a plausible {\it a priori} measure, but rather are intended to
illustrate the different possibilities that are still allowed by the
present constraints~\cite{benchmark}~\footnote{This study is restricted 
to $A = 0$, for which $\stop_1 - \chi$
coannihilation is less important, so this effect has not influenced 
the selection of benchmark points.}. Five of the
chosen points are in the
`bulk' region at small $m_{1/2}$ and $m_0$, four are spread along the
coannihilation `tail' at larger $m_{1/2}$ for various values of
$\tan\beta$, two are in the `focus-point' region at large $m_0$, and two
are in rapid-annihilation `funnels' at large $m_{1/2}$ and $m_0$. The
proposed points range over the allowed values of $\tan\beta$ between 5 and
50.  Most of them have $\mu > 0$, as favoured by $g_\mu - 2$, but there
are two points with $\mu < 0$. 

Various derived quantities in these supersymmetric benchmark scenarios,
including the relic density, $g_\mu - 2, b \rightarrow s\gamma$,
electroweak fine-tuning $\Delta$ and the relic-density sensitivity
$\Delta^\Omega$, are given in~\cite{benchmark}. These enable the reader to
see at a glance which models would be excluded by which refinement of the
experimental value of $g_\mu - 2$. Likewise, if you find some amount of
fine-tuning uncomfortably large, then you are free to discard the
corresponding models.

The LHC collaborations have analyzed their reach for sparticle detection
in both generic studies and specific benchmark scenarios proposed
previously~\cite{susyLHC}. Based on these studies,
Fig.~\ref{fig:Manhattan} displays estimates of how many different
sparticles may be seen at the LHC in each of the newly-proposed benchmark
scenarios~\cite{benchmark}. The lightest Higgs boson is always found, and
squarks and gluinos are usually found, though there are some scenarios
where no sparticles are found at the LHC. The LHC often misses heavier
weakly-interacting sparticles such as charginos, neutralinos, sleptons and
the other Higgs bosons.

It was initially thought that the discovery of supersymmetry at the LHC
was `guaranteed' if the BNL measurement $g_\mu -2$ was within $2 \, 
\sigma$
of the true value, but this is no longer the case with the new sign of the
pole contributions to light-by-light scattering. This is the case, in 
particular, because arbitrarily large values of $m_{1/2}$ and $m_0$ are 
now compatible with the data at the $2 \, \sigma$ level \cite{bench2}.

\begin{figure}
%\vspace*{-0.75in}
%\hspace*{.40in}
\begin{center}
%\begin{minipage}{8in}
\epsfig{file=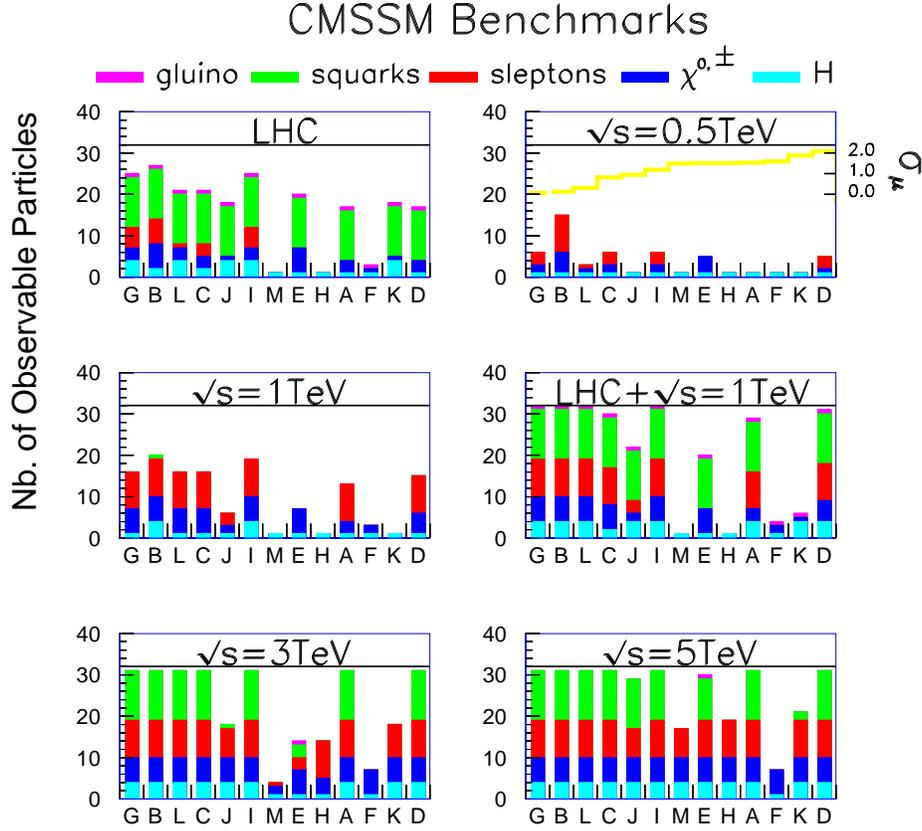,height=5in}
\end{center}
%\end{minipage}
\caption[]{Summary of the prospective sensitivities of the LHC and
linear
colliders at 
different $\sqrt{s}$ energies to CMSSM particle production 
in the proposed benchmark scenarios G, B, ..., which are
ordered by their distance from the central value of $g_\mu - 2$, as
indicated by the pale (yellow) line in the second panel. We see clearly
the complementarity
between an $e^+ e^-$ collider~\cite{LC,CLIC} (or $\mu^+ 
\mu^-$ collider~\cite{MC}) and the LHC in the TeV
range of energies~\cite{benchmark}, with the former
excelling for non-strongly-interacting particles, and the LHC for
strongly-interacting sparticles and their cascade decays. CLIC \cite{CLIC}
provides unparallelled physics reach for non-strongly-interacting
sparticles, extending beyond the TeV scale. We
recall that mass and coupling measurements at $e^+ e^-$ colliders
are usually much cleaner and more precise than at
hadron-hadron colliders such as the LHC. Note, in particular, that it is
not known how to distinguish the light squark flavours at the LHC. }
\label{fig:Manhattan} 
\end{figure}

The physics capabilities of linear $e^+e^-$ colliders are amply documented
in various design studies~\cite{LC}. Not only is the lightest MSSM Higgs
boson observed, but its major decay modes can be measured with high
accuracy. Moreover, if sparticles are light enough to be produced, their
masses and other properties can be measured very precisely, enabling
models of supersymmetry breaking to be tested~\cite{Zerwas}.

As seen in Fig.~\ref{fig:Manhattan}, the sparticles visible at an $e^+e^-$
collider largely complement those visible at the
LHC~\cite{benchmark,bench2}. In most of benchmark scenarios proposed, a
1-TeV linear collider would be able to discover and measure precisely
several weakly-interacting sparticles that are invisible or difficult to
detect at the LHC. However, there are some benchmark scenarios where the
linear collider (as well as the LHC) fails to discover supersymmetry.
Only a linear collider with a higher centre-of-mass energy appears sure
to cover all the allowed CMSSM parameter space, as seen in the lower
panels of Fig.~\ref{fig:Manhattan}, which illustrate the physics reach of
a higher-energy lepton collider, such as CLIC~\cite{CLIC} or a multi-TeV
muon collider~\cite{MC}.

\subsection{Prospects for Other Experiments}

\subsubsection{Detection of Cold Dark Matter}

Fig.~\ref{fig:DM} shows rates for the elastic spin-independent scattering
of supersymmetric relics~\cite{EFFMO}, including the
projected sensitivities for CDMS
II~\cite{Schnee:1998gf} and CRESST~\cite{Bravin:1999fc} (solid) and
GENIUS~\cite{GENIUS} (dashed).
Also shown are the cross sections 
calculated in the proposed benchmark scenarios discussed in the previous
section, which are considerably below the DAMA \cite{DAMA} range
($10^{-5} - 10^{-6}$~pb), but may be within reach of future projects.
The prospects for detecting elastic spin-independent scattering are less 
bright, as also shown in Fig.~\ref{fig:DM}.
Indirect searches for supersymmetric dark matter via the products of
annihilations in the galactic halo or inside the Sun also have prospects
in some of the benchmark scenarios~\cite{EFFMO}, as seen in 
Fig.~\ref{fig:indirectDM}.

\begin{figure}
\vskip 0.75in
\vspace*{-0.75in}
\hspace*{-.40in}
\begin{minipage}{8in}
\epsfig{file=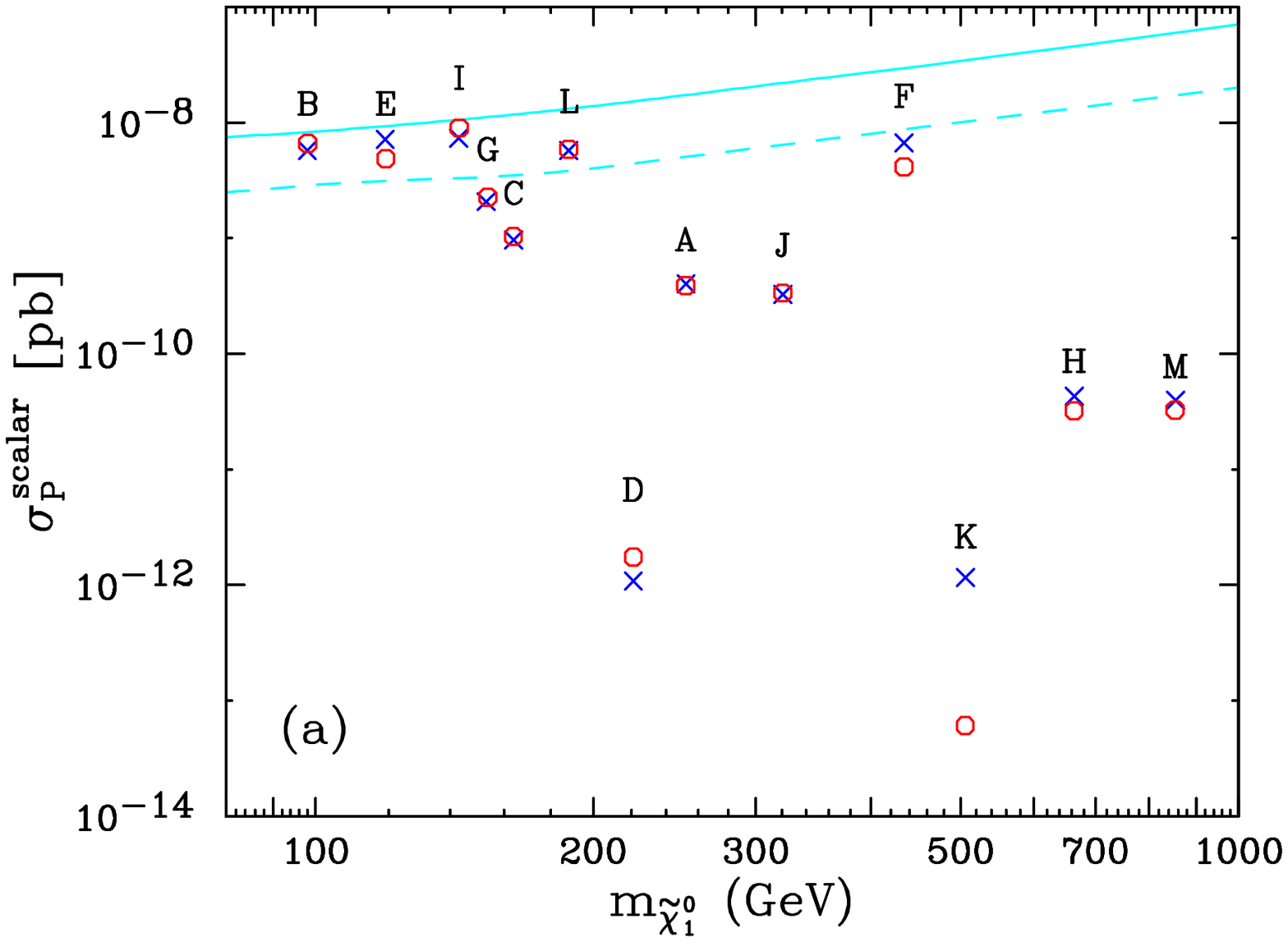,height=2.5in}
\hspace*{-0.17in}
\epsfig{file=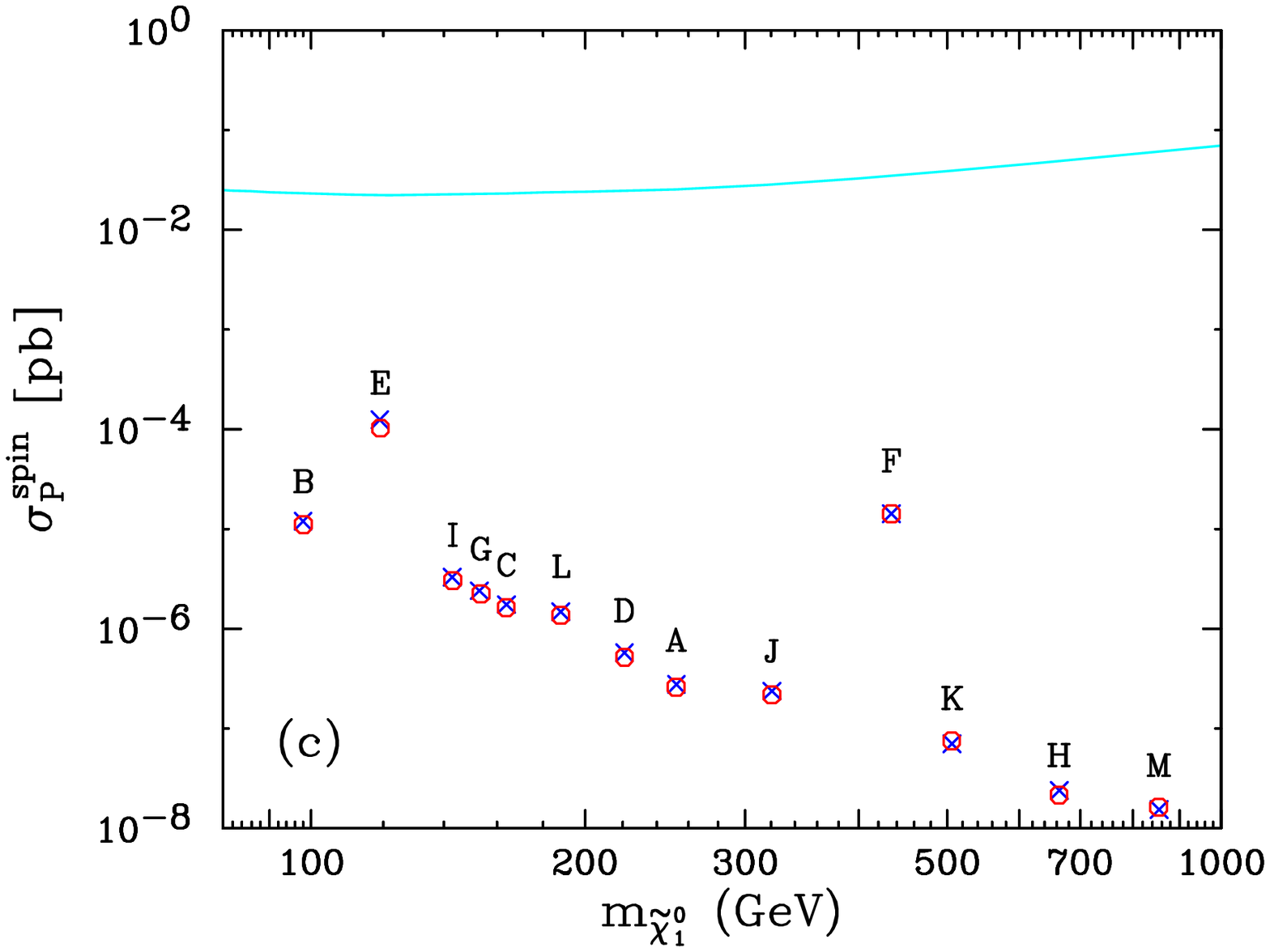,height=2.5in} \hfill
\end{minipage}
\caption[]{Left panel: elastic spin-independent scattering  
of supersymmetric relics on protons calculated in 
benchmark scenarios~\cite{EFFMO}, compared with the 
projected sensitivities for CDMS
II~\cite{Schnee:1998gf} and CRESST~\cite{Bravin:1999fc} (solid) and
GENIUS~\cite{GENIUS} (dashed).
The predictions of the {\tt SSARD} code (blue
crosses) and {\tt Neutdriver}\cite{neut} (red circles) for
neutralino-nucleon scattering are compared.
The labels A, B, ...,L correspond to the benchmark points as shown in 
Fig.~\protect\ref{fig:Bench}. Right panel: prospects for detecting 
elastic spin-independent scattering in the benchmark scenarios, which are 
less bright.}
\label{fig:DM}
\end{figure}  

\begin{figure}
\vskip 0.75in
\vspace*{-0.75in}
\hspace*{-.40in}
\begin{minipage}{8in}
\epsfig{file=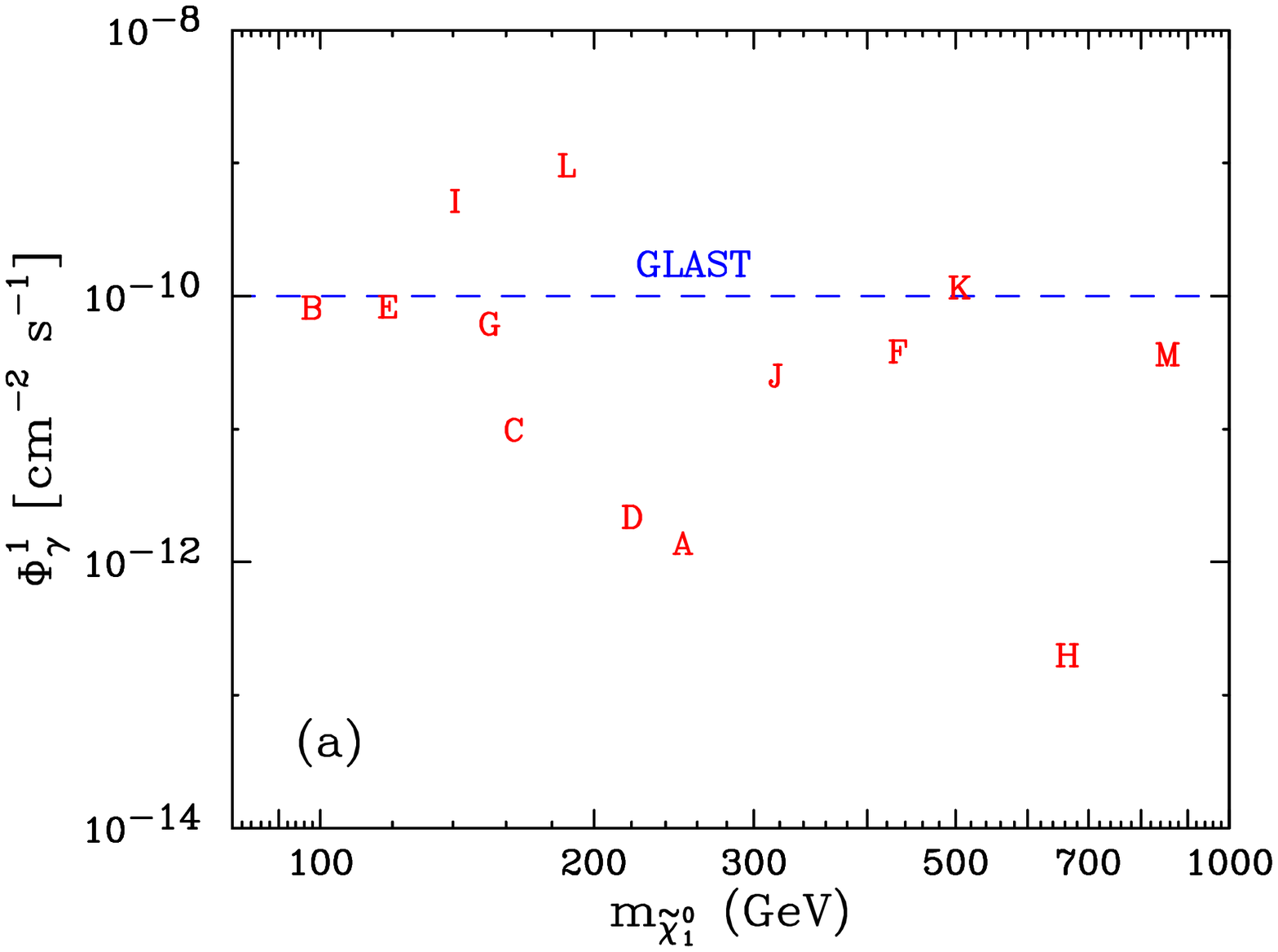,height=2.5in}
\hspace*{-0.17in}
\epsfig{file=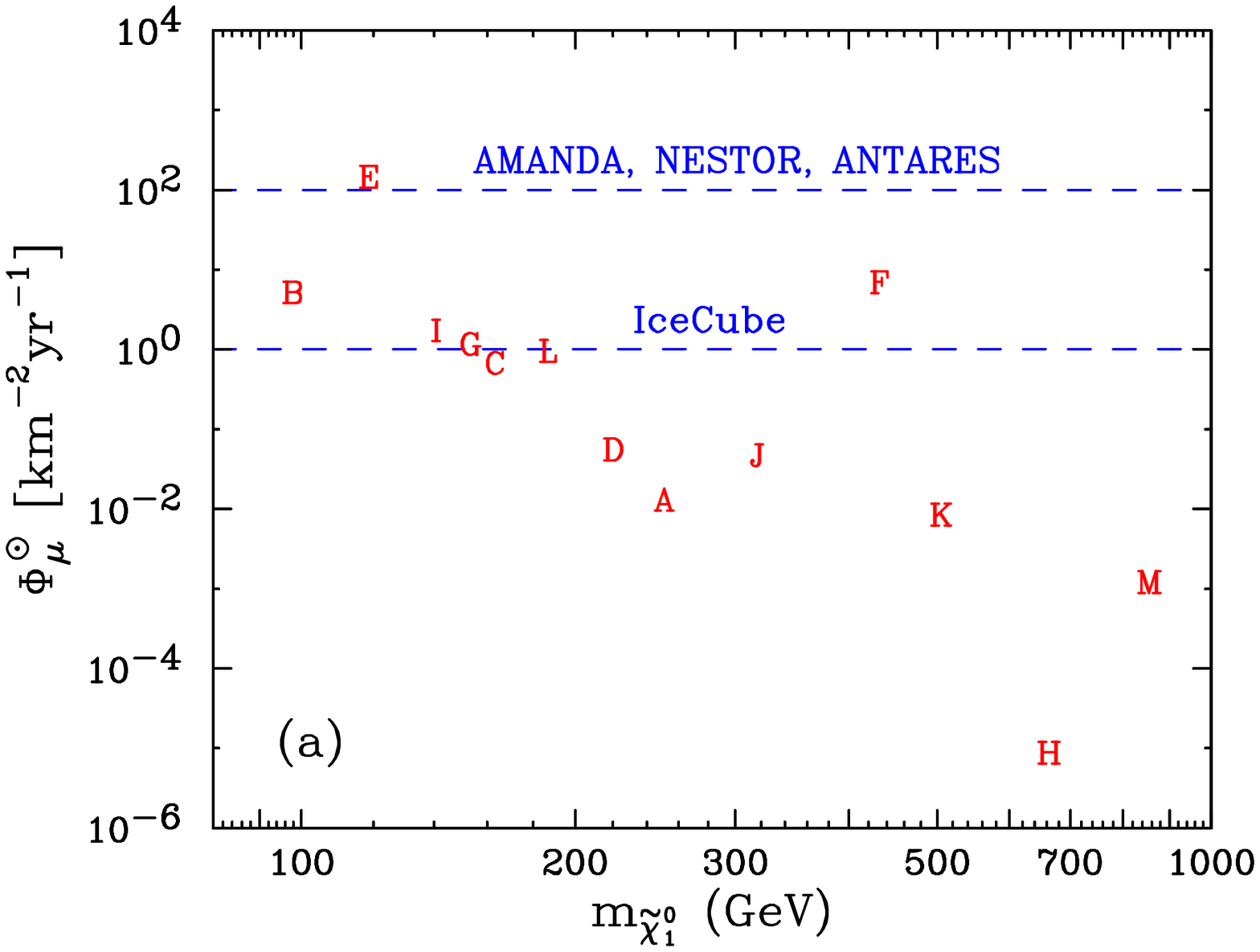,height=2.5in} \hfill
\end{minipage}
\caption[]{Left panel: prospects for detecting photons with energies above 
1~GeV 
from annihilations in the centre of the galaxy, assuming a moderate 
enhancement there of the overall halo density, and right panel: prospects 
for detecting 
muons from energetic solar neutrinos produced by relic annihilations in 
the Sun, as calculated~\cite{EFFMO} in the benchmark scenarios using {\tt 
Neutdriver}\cite{neut}.}
\label{fig:indirectDM}
\end{figure}  

\subsubsection{Proton Decay}

This could be within reach, with $\tau (p\rightarrow e^+\pi^0)$ via a
dimension-six operator possibly $\sim 10^{35}y$ if $m_{GUT} \sim 10^{16}$
GeV as expected in a minimal supersymmetric GUT. Such a model also
suggests that $\tau (p\rightarrow \bar\nu K^+) < 10^{32} y$ via
dimension-five operators~\cite{dim5}, unless measures are taken to
suppress them~\cite{fsu5}. This provides motivation for a next-generation
megaton experiment that could detect proton decay as well as explore new
horizons in neutrino physics~\cite{UNO}.

\subsection{Conclusions}

We have compiled in this Lecture the various experimental constraints 
on the MSSM, particularly in its constrained CMSSM version. These have 
been compared and combined with the cosmological constraint on the relic 
dark matter density. As we have shown, there is good overall compatibility 
between these various constraints. To exemplify the possible types of
supersymmetric phenomenology compatible with all these constraints, a set 
of benchmark scenarios have been proposed.

We have discussed the fine-tuning of parameters required for supersymmetry 
to have escaped detection so far. There are regions of parameter space 
where the neutralino relic density is rather sensitive to the exact values 
of the input parameters, and to the details of the calculations based on 
them. However, there are generic domains of parameter space where 
supersymmetric dark matter is quite natural. The fine-tuning price of the 
electroweak supersymmetry-breaking scale has been increased by the 
experimental constraints due to LEP, in particular, but its significance 
remains debatable.

As illustrated by these benchmark scenarios, future colliders such as the
LHC and a TeV-scale linear $e^+e^-$ collider have good prospects of
discovering supersymmetry and making detailed measurements. There are also
significant prospects for discovering supersymmetry via searches for cold
dark matter particles, and searches for proton decay also have interesting
prospects in supersymmetric GUT models.

One may be disappointed that supersymmetry has not already been
discovered, but one should not be disheartened. Most of the energy range
where supersymmetry is expected to appear has yet to be explored. Future
accelerators will be able to complete the search for supersymmetry, but
they may be scooped by non-accelerator experiments. In a few years' time,
we expect to be writing about the discovery of supersymmetry, not just
constraints on its existence.

\end{document}